\documentclass[showpacs,preprintnumbers,amsmath,amssymb,APSl,prd,nofootinbib,superscriptaddress,12pt]{revtex4-2}
\usepackage{booktabs}
\usepackage{floatrow}
\usepackage{adjustbox}
\usepackage{siunitx}
\usepackage{array}
\usepackage{multirow}
\usepackage{graphicx}
\usepackage{caption}
\usepackage{float}
\usepackage{subcaption}
\usepackage{xcolor}
\usepackage{hyperref}
\usepackage{amsfonts}
\usepackage{bm}
\usepackage[a4paper, margin=1.6cm]{geometry}
\usepackage{mathrsfs}
\usepackage{soul}
\usepackage{makecell,multirow}
\usepackage{rotating}
\usepackage[utf8]{inputenc}
\usepackage{amsmath}
\usepackage{makecell}

\usepackage{amssymb}
\usepackage{tensor}
\usepackage{graphicx}
\usepackage{bigints}
\usepackage{bbm}
\usepackage{graphicx}
\usepackage{subcaption}
\usepackage{epsf}
\usepackage{bm}
\usepackage{amsmath}
\usepackage{amsfonts}
\usepackage{amssymb}
\usepackage{graphicx}
\usepackage{tabularx}
\usepackage{multirow}
\usepackage{color}%
\providecommand{\U}[1]{\protect\rule{.1in}{.1in}}

\newcommand{\ie}{\begin{equation}}
\newcommand{\fe}{\end{equation}}

\newcommand{\mincir}{\raise
-3.truept\hbox{\rlap{\hbox{$\sim$}}\raise4.truept\hbox{$<$}\ }}
\newcommand{\magcir}{\raise
-3.truept\hbox{\rlap{\hbox{$\sim$}}\raise4.truept\hbox{$>$}\ }}

\providecommand{\U}[1]{\protect\rule{.1in}{.1in}}

\usepackage{tikz,xcolor,hyperref}

\usepackage{hyperref}             
\hypersetup{
    colorlinks=true,              
    breaklinks=true,              
    citecolor=blue,               
    linkcolor=[rgb]{0,0.5,0.9},   
    urlcolor=red,                 
    filecolor=green               
}

\definecolor{lime}{HTML}{A6CE39}
\DeclareRobustCommand{\orcidicon}{%
	\begin{tikzpicture}
	\draw[lime, fill=lime] (0,0) 
	circle [radius=0.16] 
	node[white] {{\fontfamily{qag}\selectfont \tiny ID}};
	\draw[white, fill=white] (-0.0625,0.095) 
	circle [radius=0.007];
	\end{tikzpicture}
	\hspace{-2mm}
}

\foreach \x in {A, ..., Z}{%
	\expandafter\xdef\csname orcid\x\endcsname{\noexpand\href{https://orcid.org/\csname orcidauthor\x\endcsname}{\noexpand\orcidicon}}
}

\begin{document}

\title{\Large{Gravitational wave signatures and periodic orbits of a charged black hole in a Hernquist dark matter halo}}

\author{N. Heidari\orcidA{}}
\email{heidari.n@gmail.com}

\affiliation{Departamento de Física, Universidade Federal de Campina Grande, Caixa Postal 10071, 58429-900 Campina Grande, Paraíba, Brazil.}
\affiliation{Center for Theoretical Physics, Khazar University, 41 Mehseti Street, Baku, AZ-1096, Azerbaijan.}
\affiliation{School of Physics, Damghan University, Damghan, 3671641167, Iran.}


\author{A. A. Ara\'{u}jo Filho\orcidB{}}
\email{dilto@fisica.ufc.br}
\affiliation{Departamento de Física, Universidade Federal de Campina Grande, Caixa Postal 10071, 58429-900 Campina Grande, Paraíba, Brazil.}
\affiliation{Center for Theoretical Physics, Khazar University, 41 Mehseti Street, Baku, AZ-1096, Azerbaijan.}
\affiliation{Departamento de Física, Universidade Federal da Paraíba, Caixa Postal 5008, 58051--970, João Pessoa, Paraíba,  Brazil.}


\author{Iarley P. Lobo\orcidD{}}
\email{lobofisica@gmail.com}
\affiliation{Departamento de Física, Universidade Federal de Campina Grande, Caixa Postal 10071, 58429-900 Campina Grande, Paraíba, Brazil.}
\affiliation{Department of Chemistry and Physics, Federal University of Para\'iba, Rodovia BR 079 - km 12, 58397-000 Areia-PB,  Brazil.}



\begin{abstract}

In this work, we study the motion of massive test particles and the gravitational wave emission associated with periodic trajectories around a magnetically charged black hole immersed in a \textit{Hernquist} dark matter halo. We begin by analyzing the effective potential and the conditions for stable motion, with particular attention to the marginally bound radius and the innermost stable circular orbit. Our results show that the dark matter parameters, namely the halo density and scale radius, enlarge the allowed region and generally shift the relevant characteristic radii and angular momenta toward larger values. In contrast, the magnetic charge partially counterbalances this behavior. We then examine periodic trajectories through the rational number $q$, which characterizes the relation between the azimuthal and radial frequencies, and construct representative zoom--whirl configurations together with their precessing counterparts. Finally, we investigate the imprints of dark matter and magnetic monopole charge on the gravitational wave polarizations in the extreme mass--ratio regime.To explore the observational relevance of the predicted gravitational wave signals, we compute their characteristic strains in the frequency domain and compare them with the LISA sensitivity curve.

\end{abstract}
\maketitle

\pagebreak

\tableofcontents


\section{Introduction}

Periodic orbits around black holes constitute a valuable framework for establishing the relation between the geodesic structure of the spacetime and the gravitational wave signal produced by orbiting bodies. In particular, they provide an effective description of strong--field orbital dynamics, including zoom--whirl behavior, and make it possible to determine how the underlying geometry influences the morphology of the emitted waveforms. Consequently, gravitational waves associated with periodic trajectories have been the subject of increasing interest in recent years and can trigger the search for new physics \cite{Addazi:2021xuf}.

This problem has been investigated in a variety of black hole backgrounds, including regular black holes in both static \cite{Gong:2025mne} and axisymmetric \cite{Kumar:2024our} configurations, the Dastan--Destounis--Suvorov--Kokkotas black hole spacetime \cite{Hua:2026kvw}, the Schwarzschild--Bertotti--Robinson black hole \cite{Xamidov:2026kqs}, effective field theory scenarios \cite{Alloqulov:2025dqi}, dyonic ModMax black holes \cite{Alloqulov:2025dqi}, non--commutative-inspired black holes surrounded by quintessence \cite{Ahmed:2025azu}, the $\gamma$ metric \cite{Zhang:2025wni}, quantum-corrected black holes \cite{Chen:2025aqh,Ahmed:2025shr}, Einstein--Æther black holes \cite{Lu:2025cxx}, black holes in bumblebee gravity \cite{Shi:2026zxx,Liu:2025swi}, Kalb--Ramond black holes \cite{Junior:2024tmi,Xia:2025yzg}, charged black holes with scalar hair \cite{Deng:2025wzz}, renormalization-group-improved Kerr black holes \cite{Li:2025sfe,kumar2024exploring}, and black holes without a Cauchy horizon \cite{Wang:2025hla}.
More recently, a new solution describing a Schwarzschild black hole immersed in a Hernquist dark matter halo was proposed in Ref.~\cite{Jha:2025xjf}, followed by its charged extension in Ref.~\cite{jha2025probing}. The uncharged geometry has already been examined in several contexts, including black hole shadows \cite{Luo:2025xjb}, the presence of a cloud of strings \cite{Ahmed:2025ttq}, accretion-disk properties \cite{Nieto:2025apz,Ban:2026tyh}, quasinormal modes \cite{Feng:2025iao,Qi:2026zrr}, motion of spining particles \cite{tan2025motion}, and thermodynamic fluctuations together with radiation properties \cite{Jumaniyozov:2025xxh}. In parallel, periodic orbits and the corresponding gravitational wave emission have also started to be explored in black hole spacetimes surrounded by dark matter halos \cite{Haroon:2025rzx,Hassanabadi:2026wku,Li:2025eln,Alloqulov:2025ucf}.

However, to the best of our knowledge, the charged \textit{Hernquist} black hole has not yet been explored from this perspective. In this work, we investigate the motion of massive test particles and the associated gravitational wave emission around a magnetically charged black hole immersed in a \textit{Hernquist} dark matter halo. Our analysis focuses on the effective potential, the marginally bound orbit, the innermost stable circular orbit, and the periodic zoom--whirl trajectories supported by this geometry, with special attention to the effects of the magnetic charge and dark matter halo parameters, as well as to the comparison with the corresponding uncharged case.


\section{The model}

In this section, we introduce a black hole solution describing a magnetically charged object generated within nonlinear electrodynamics and surrounded by a \textit{Hernquist} dark matter halo, hereafter referred to as the \textit{MHDM} black hole. The dynamical framework governing this combined setup is determined by the action presented in Ref.~\cite{jha2025probing}
\ie
\label{action}
S=\int \mathrm{d}^4x\sqrt{-\mathrm{g}}\bigg[\frac{\mathcal{R}}{2\kappa}-\frac{2\mathcal{L}(F)}{\kappa}+\mathcal{L}_{dm}\bigg].
\fe
Here, $\mathcal{R}$ denotes the Ricci scalar, while $\mathcal{L}(F)$ represents the nonlinear electrodynamics Lagrangian, written as a function of the electromagnetic invariant $F$, which is defined as $ F=\frac{1}{4}F_{\mu\nu} F^{\mu\nu}$.
In this expression, $F_{\mu\nu}=\partial_{\mu}A_{\nu}-\partial_{\nu}A_{\mu}$ denotes the electromagnetic tensor associated with the four--potential $A_{\mu}$ in nonlinear electrodynamics, whereas $\mathcal{L}_{dm}$ accounts for the dark matter sector through the \textit{Hernquist} halo Lagrangian density. The nonlinear electrodynamics model is required to recover the standard Maxwell theory in the weak--field regime, namely $\mathcal{L}(F)\to F$. By performing the variation of the action in Eq.~(\ref{action}) with respect to the metric tensor $\mathrm{g}{\mu\nu}$, one arrives at the corresponding gravitational field equations \cite{jha2025probing}
\ie
\mathcal{R}_{\mu \nu }-\frac{1}{2}\mathrm{g}_{\mu \nu }\mathcal{R}=\kappa  T^{\text{H}}_{\mu\nu}+2\left(\frac{\partial \mathcal{L}}{\partial F}F_{\mu \lambda}F_{\nu}^{\lambda}-\mathrm{g}_{\mu\nu}\mathcal{L}(F)\right).
\label{fe}
\fe
The field strength tensor satisfies two fundamental relations, namely the equation of motion and the corresponding Bianchi identity, which can be written, respectively, as follows:
$ \nabla_\mu\left(\frac{\partial \mathcal{L}(F)}{\partial F} F^{\nu \mu}\right) = 0$, and $\nabla_\mu\left(* F^{\nu \mu}\right)=0$.
Having established the governing equations, we next introduce the necessary elements for their explicit solution. Among the various profiles proposed to model extended dark matter distributions, we adopt the Hernquist profile here as an analytically tractable phenomenological benchmark with a finite total halo mass where the dark matter distribution is characterized by the density profile \cite{hernquist1990analytical,jha2025probing,mo2010galaxy}, i.e., 
${\rho}_{H}(r)={\tilde{\rho}_\mathrm{s}}\left(\frac{r}{{\tilde{r}_\mathrm{s}}}\right)^{-1}\left[1+\frac{r}{{\tilde{r}_\mathrm{s}}}\right]^{-3}$,
with ${\tilde{\rho}}_\mathrm{s}$ being the characteristic density of the halo, whereas ${\tilde{r}_\mathrm{s}}$ determines its characteristic radial scale. Then, a black hole embedded within a \textit{Hernquist} dark matter halo—hereafter referred to as an \textit{HDM} BH—is described by the following line element $\mathrm{d}s^2 = -\mathcal{A}(r) \mathrm{d}t^2 + \mathcal{A}(r)^{-1} \mathrm{d}r^2 + r^2 (\mathrm{d}\theta^2 + \sin^2 \theta \mathrm{d}\phi^2)$,
where 
$\mathcal{A}(r)=1-\frac{4\pi {\tilde{\rho}}_\mathrm{s} {\tilde{r}_\mathrm{s}}^3}{r+{\tilde{r}_\mathrm{s}}}$.  
For the \textit{Hernquist} halo, the matter sector is described by an anisotropic energy--momentum tensor of the form $T_{\mu}^{\nu\,(H)}=\mathrm{diag}(-\rho,\,p_r,\,p_\theta,\,p_\phi)$, where $\rho$ denotes the energy density, while $p_r$ and $p_{\theta,\phi}$ correspond to the radial and angular pressure components, respectively. These quantities are expressed as
$-\rho=p_r=-\frac{{\tilde{\rho}}_\mathrm{s} {\tilde{r}_\mathrm{s}} ^4}{2 r^2 ({\tilde{r}_\mathrm{s}} +r)^2}$ and $p_\theta=p_\phi=\frac{{\tilde{\rho}_\mathrm{s}}  {\tilde{r}_\mathrm{s}} ^4}{2 r^2 ({\tilde{r}_\mathrm{s}} +r)^2}$. 
One may then adopt the following ansatz for the electromagnetic field strength tensor: 
$F_{\mu \nu}=\left(\delta_\mu^\theta \delta_\nu^{\phi}-\delta_\nu^\theta \delta_\mu^{\phi}\right) B(r, \theta)=\left(\delta_\mu^\theta \delta_\nu^{\phi}-\delta_\nu^\theta \delta_\mu^{\phi}\right) g(r) \sin \theta$.

In this manner, we find that $g(r)$ must be a constant, so that $g(r)=\tilde{g}$. This constant $\tilde{g}$ is interpreted as the magnetic monopole charge. In turn, we obtain
$F_{\theta\phi}=-F_{\phi\theta}=\tilde{g}\sin \theta$  and $F=\frac{\tilde{g}^2}{2r^4}$.
In this work, we adopt the following form for the nonlinear electrodynamics Lagrangian density:
\begin{equation}
\mathcal{L}(F)=\frac{2 \sqrt{\tilde{g}} F^{5 / 4}}{s(\sqrt{2}+2 g \sqrt{F})^{3 / 2}}.
\end{equation}
Here, $s$ denotes a free constant whose value will be specified at a later stage. To construct a static and spherically symmetric black hole configuration motivated by nonlinear electrodynamics and embedded in a \textit{Hernquist} dark matter halo, we consider the following metric ansatz \cite{jha2025probing}:
\ie
\mathrm{d}s^2= -A(r)\mathrm{d}t^2+B(r)\mathrm{d}r^2+r^2 \mathrm{d}\theta^2+r^2 \sin^2\theta \mathrm{d}\phi^2.
\label{trial}
\fe 
With all the necessary ingredients now specified, we proceed to the explicit resolution of the field equations, which take the form
\begin{eqnarray}
&&-\frac{r B'(r)+(B(r)-1) B(r)}{r^2 B(r)^2}=-\kappa\rho-2\mathcal{L},\label{00}\\
&&\frac{r A'(r)-A(r) B(r)+A(r)}{r^2 A(r) B(r)}=\kappa p_r-2\mathcal{L},\label{11}\\\nonumber
&&\frac{-r B(r) A'(r)^2+A(r) \left(2 B(r) \left(r A''(r)+A'(r)\right)-r A'(r) B'(r)\right)-2 A(r)^2 B'(r)}{4 r A(r)^2 B(r)^2}\\
&& =\kappa p_{\theta}-2\left(\frac{r}{2}\frac{\partial \mathcal{L}}{\partial r}+\mathcal{L}\right),\label{22}
\end{eqnarray}
By combining Eqs.~(\ref{00}) and (\ref{11}), we obtains the relation connecting the metric functions $A(r)$ and $B(r)$, namely
$B(r) A'(r)+A(r) B'(r)=0$.
As a result, one finds $B(r)=1/A(r)$. Substituting this condition back into Eq.~(\ref{00}) then gives the metric function in the required form \cite{jha2025probing}
\ie
\mathrm{d}s^2=-A(r)\mathrm{d}t^2+\frac{1}{A(r)}\mathrm{d}r^2+r^2 \mathrm{d}\theta^2+r^2 \sin^2\theta \mathrm{d}\phi^2,\label{final}
\fe
where
\ie\label{eq:lapse}
A(r)=1-\frac{2M}{\sqrt{r^2+\tilde{g}^2}}-\frac{4\pi {\tilde{\rho}_\mathrm{s}}{\tilde{r}_\mathrm{s}}^3}{r+{\tilde{r}_\mathrm{s}}}.
\fe  

It is worth clarifying that, although the model considered here is a composite NED Hernquist configuration, in which the magnetic NED sector gives a Bardeen--like contribution, the curvature invariants of the complete spacetime remain singular at $r=0$, as shown in the original construction of the solution \cite{jha2025probing}.
The gravitational wave emission generated by periodic orbits in the spacetime described by the metric function in Eq.~(\ref{eq:lapse}), as a composite configuration describing the combined effects of a magnetic NED source and an extended Hernquist dark matter environment will be investigated in the following sections. {Hereafter, we apply the defined dimentionless parameters $r_s=\tilde{r}_\mathrm{s}/M$, $\rho_s=\tilde{\rho}_\mathrm{s}M^2$ and $g_s=\tilde{g}_\mathrm{s}/M$, in our calculations.} 


\section{Timelike geodesics}

\begin{figure}[ht!]
    \centering
    \includegraphics[width=80mm]{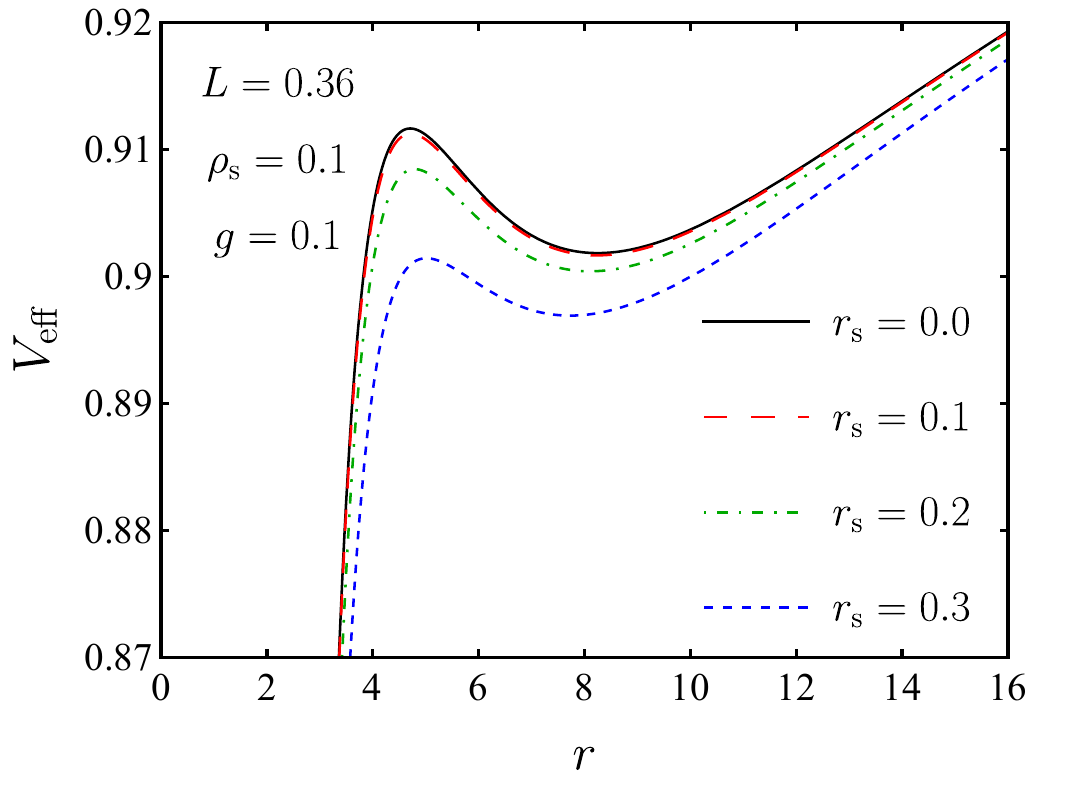}
    \includegraphics[width=80mm]{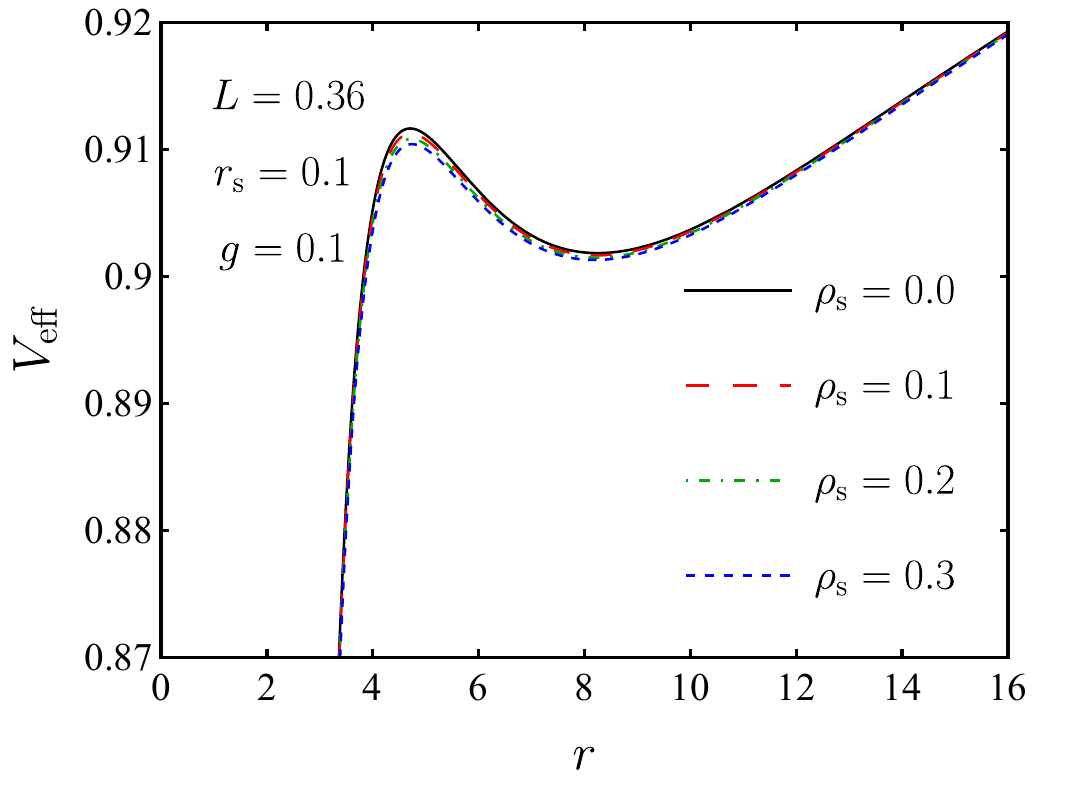}
    \includegraphics[width=80mm]{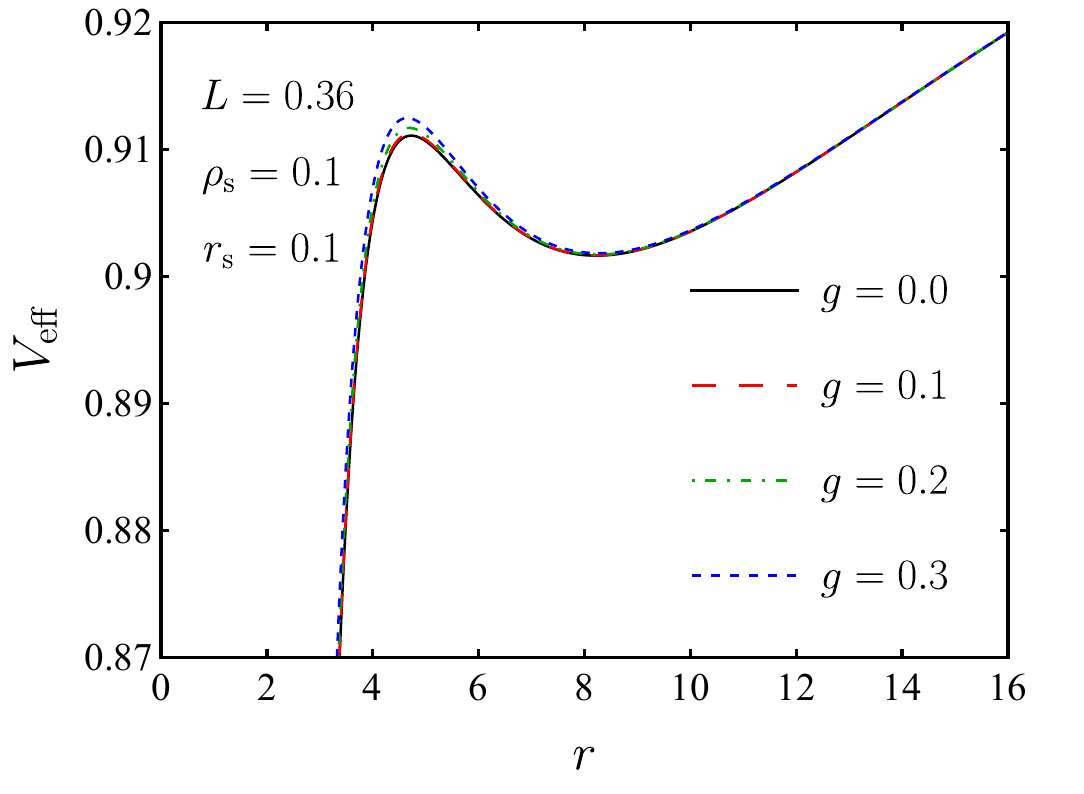}
    \includegraphics[width=80mm]{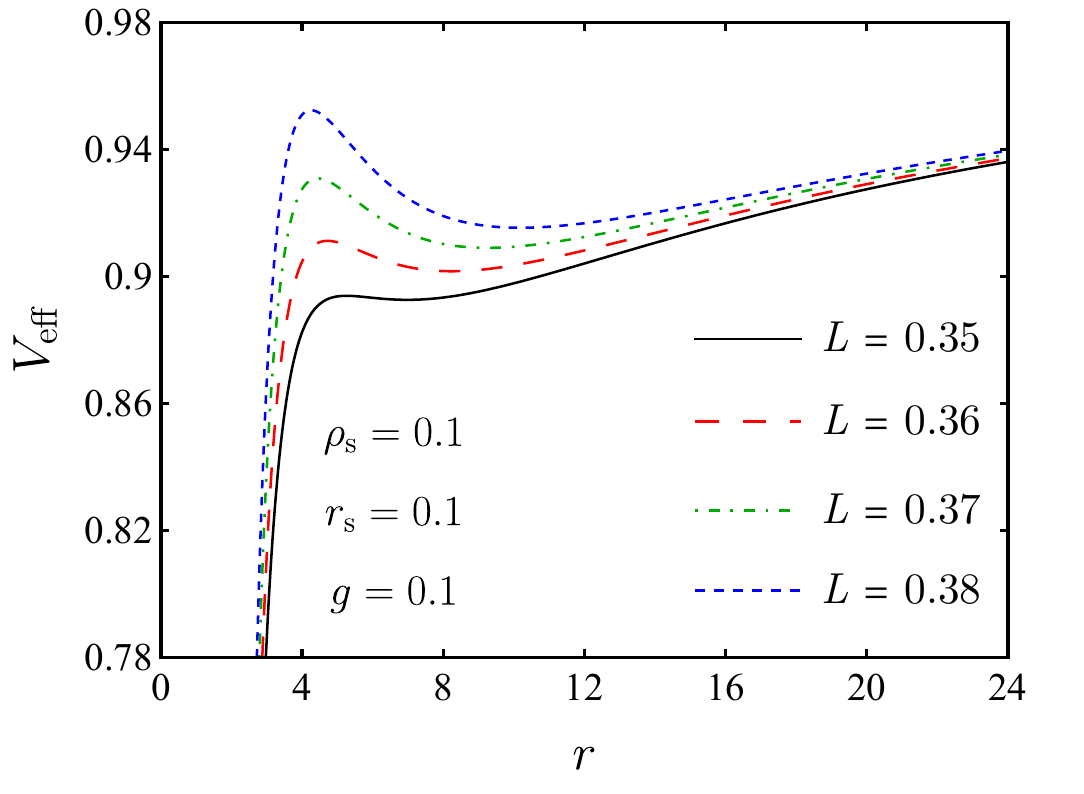}
    \protect\caption{Dependence of the radial behavior of the effective potential on the \textit{Hernquist} halo parameters; scale density $\rho_\mathrm{s}$, scale radius $r_\text{s}$, and magnetic monopole charge $g$, along with the orbital angular momentum $L$.}
    \label{fig:Veff}
\end{figure}

The trajectory of a massive test particle in a static, spherically symmetric spacetime is governed by the geodesic equations. Applying the Lagrangian formalism, it is governed by
\begin{equation}
    2\mathcal{L} = g_{\mu\nu} \dot{x}^{\mu} \dot{x}^{\nu} = m,
    \label{eq:Lagrangian}
\end{equation}
with the dot denoting differentiation with respect to the affine parameter $\tau$ and $m$ equals to $0$ or $-1$ for null geodesics and timelike geodesics, respectively. 
From Eq.~\eqref{eq:Lagrangian}, the canonical momentum conjugate to the coordinate $x^{\mu}$ is obtained as
\begin{equation}
    p_{\mu} = \frac{\partial \mathcal{L}}{\partial \dot{x}^{\mu}}.
    \label{eq:canonical_momentum}
\end{equation}
Evaluating this for an asymptotically symmetric spacetime (Eq.~\eqref{final}) yields the following components
\begin{align}\label{eq:ppt}
    p_t &= -f(r) \, \dot{t} = -E, \\ \label{eq:pr}
    p_r &= \frac{\dot{r}}{f(r)},  \\ 
    p_{\theta} &= r^2 \dot{\theta}, \label{eq:ptheta} \\ 
    p_{\phi} &= r^2 \sin^2\theta \, \dot{\phi} = L, \label{eq:pphi}
\end{align}
where $E$ and $L$ are constants of motion, identified respectively as the specific energy and specific angular momentum of the particle, and $f(r)$ denotes the lapse function $A(r)$ defined in Eq. \eqref{eq:lapse}. 
Using the conserved quantities in Eq.~\eqref{eq:ppt} and Eq.~\eqref{eq:pphi}, we derive the first--order equations of motion for timelike geodesics
\begin{align}
    &\dot{t} = \frac{E}{f(r)}, \label{eq:t_dot} \\
    &\dot{\phi} = \frac{L}{r^2}, \label{eq:phi_dot} \\
    &\dot{r}^{2}  +f(r)r^2\dot{\theta}^2+f(r)\frac{L^2}{r^2 \sin^2\theta}-E^2=-1 , \label{eq:rdot}
\end{align}
where Eq.~\eqref{eq:rdot} describes the radial dynamics. For motion in a spherically symmetric potential, we can restrict our attention to orbits confined to the equatorial plane without loss of generality. We therefore set $\theta = \frac{\pi}{2}$, so that $\dot{\theta} = 0$. Then, the radial motion reduces to the following form 
\ie
\dot{r}^2+V_\text{eff}=E^2,
\fe
where an effective potential $V_\text{eff}$ defined as 
\begin{align}
    V_\text{eff}&=f(r)\left(1+\frac{L^{2}}{r^{2}}\right)\\
    &=\left(1-\frac{2M}{\sqrt{r^2+g^2}}-\frac{4\pi\rho_{\text{s}}r_{\text{s}}^{3}}{r+r_{\text{s}}}\right)\left(1+\frac{L^2}{r^2}\right).
\end{align}
As a key quantity governing radial motion, the effective potential $V_\text{eff}$ is shown in Fig.~\ref{fig:Veff}, and its parametric dependence is analyzed. The upper panels show that increasing either of the \textit{Hernquist} halo parameters---the density $\rho_\mathrm{s}$ and the scaled radius $r_\text{s}$---leads to a decrease of the potential over all radius, especially the peak of potential suppresses more than other places. A direct comparison indicates that variations in the density parameter $\rho_\mathrm{s}$ produce a more significant change in $V_{\mathrm{eff}}$ than equivalent variations in $r_\text{s}$.

Conversely, as shown in the lower panel, the magnetic monopole charge parameter reinforces the potential structure, increasing the peak height in a subtle yet detectable manner. This result indicate the inverse effect of $g$ with both $r_\text{s}$ and $\rho_\mathrm{s}$. The right lower panel shows the impact of the particle's orbital angular momentum $L$, with the peak height increasing monotonically with $L$, being consistent with standard results in spherically symmetric systems. 
 
Since the spacetime is asymptotically flat $(f(r) \xrightarrow{r \to \infty} 1 )$, and consequently the effective potential satisfies $(V_{\rm eff}\xrightarrow{r \to \infty} 1)$, $E = 1$ can be indicated as a critical value of the conserved energy for massive test particles, based on Eq. \eqref{eq:rdot}. For particles with $E > 1$, the radial equation of motion implies $\dot{r}^2 = E^2 - V_{\rm eff}> 0 \quad \text{as} \quad r \to \infty$, indicating that the particle retains an unbound orbit. Consequently, bound motion is only possible for particles with $E \leq 1$.

For a massive particle orbiting a magnetically charged black hole embedded in a \textit{Hernquist} dark matter halo, bound orbits exist only within a finite region of parameter space, delimited by the marginally bound orbit (MBO) and the innermost stable circular orbit (ISCO). The conserved energy and angular momentum of a particle in a bound orbit must satisfy
\begin{equation}\label{eq:EL}
E_{\rm ISCO} \leq E \leq E_{\rm MBO} = 1,
\qquad
L \geq L_{\rm ISCO}.
\end{equation}
The upper bound $E_{\rm MBO} = 1$ corresponds to the MBO, for which the particle possesses exactly the minimum energy required to escape to spatial infinity. The lower bound $E_{\rm ISCO}$ is associated with the ISCO; particles with energies below this threshold are dynamically unstable and inevitably plunge into the black hole. Similarly, particles with angular momentum smaller than $L_{\rm ISCO}$ cannot sustain stable circular orbits and undergo direct infall toward the black hole.
To explore the periodic orbits, we first investigate the MBO and ISCO conditions in the following sections.
\subsection{Marginally stable orbit}
\begin{figure}
    \centering
      \includegraphics[scale=0.4]{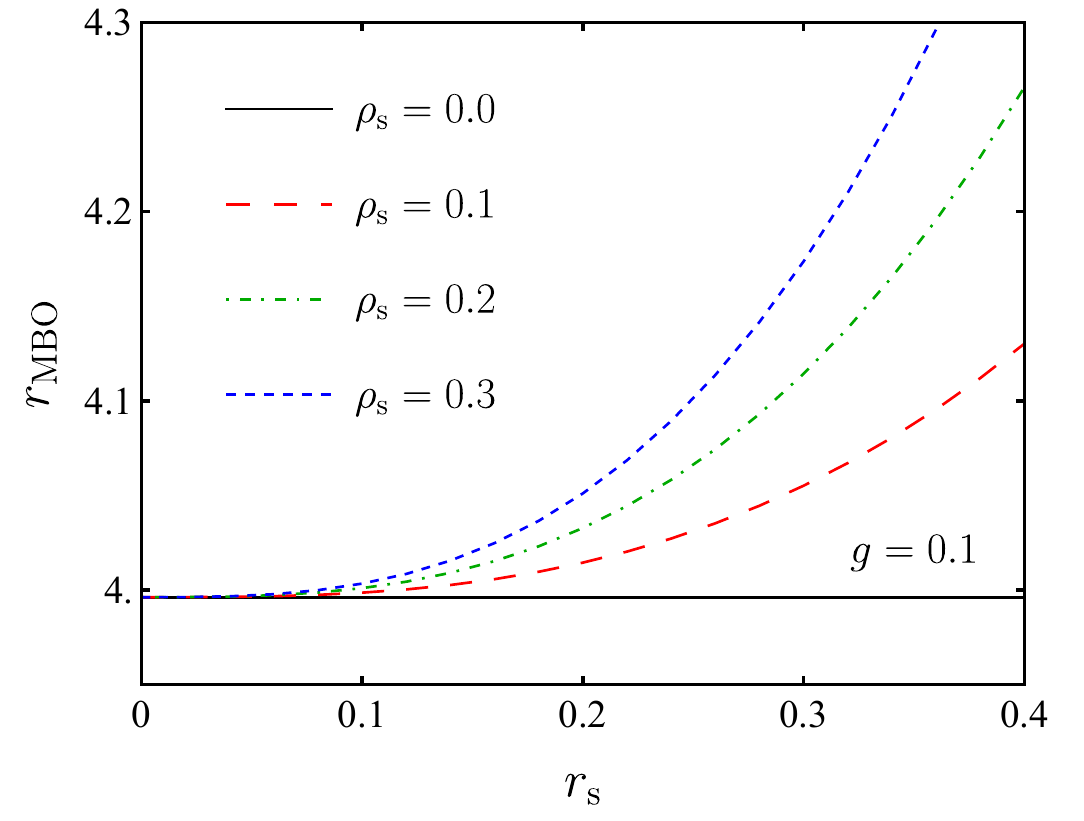}
       \includegraphics[scale=0.4]{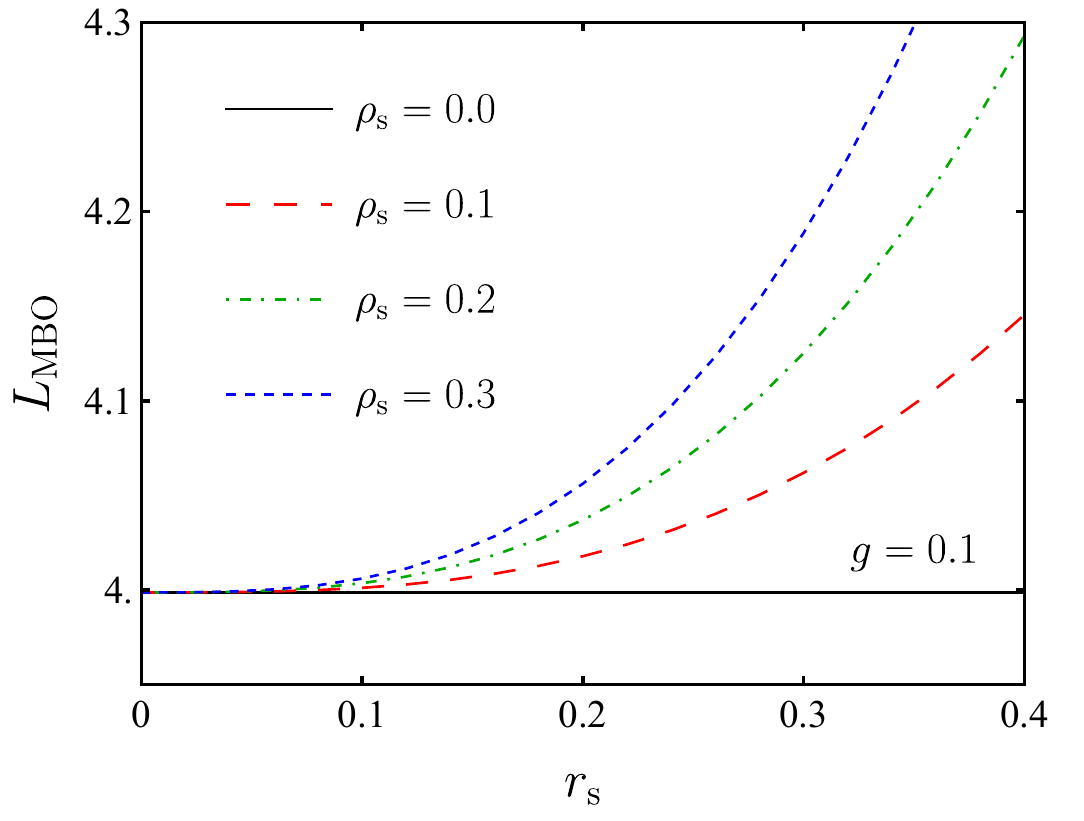}
        \includegraphics[scale=0.4]{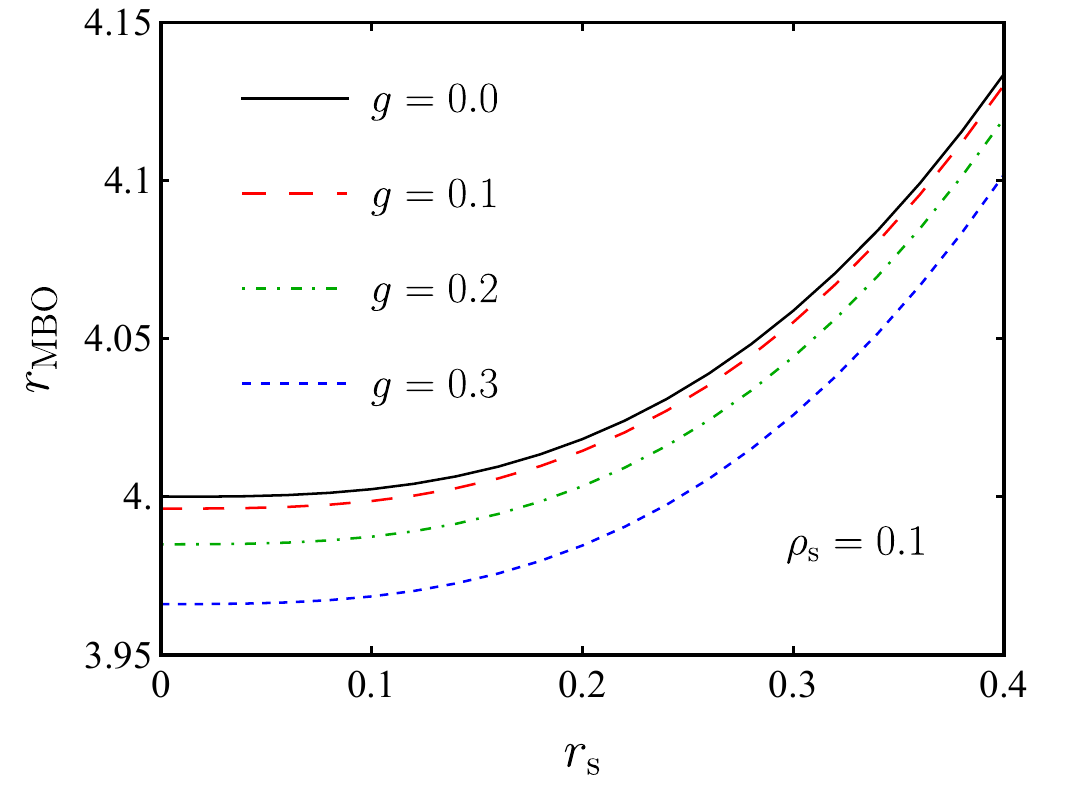}
        \includegraphics[scale=0.4]{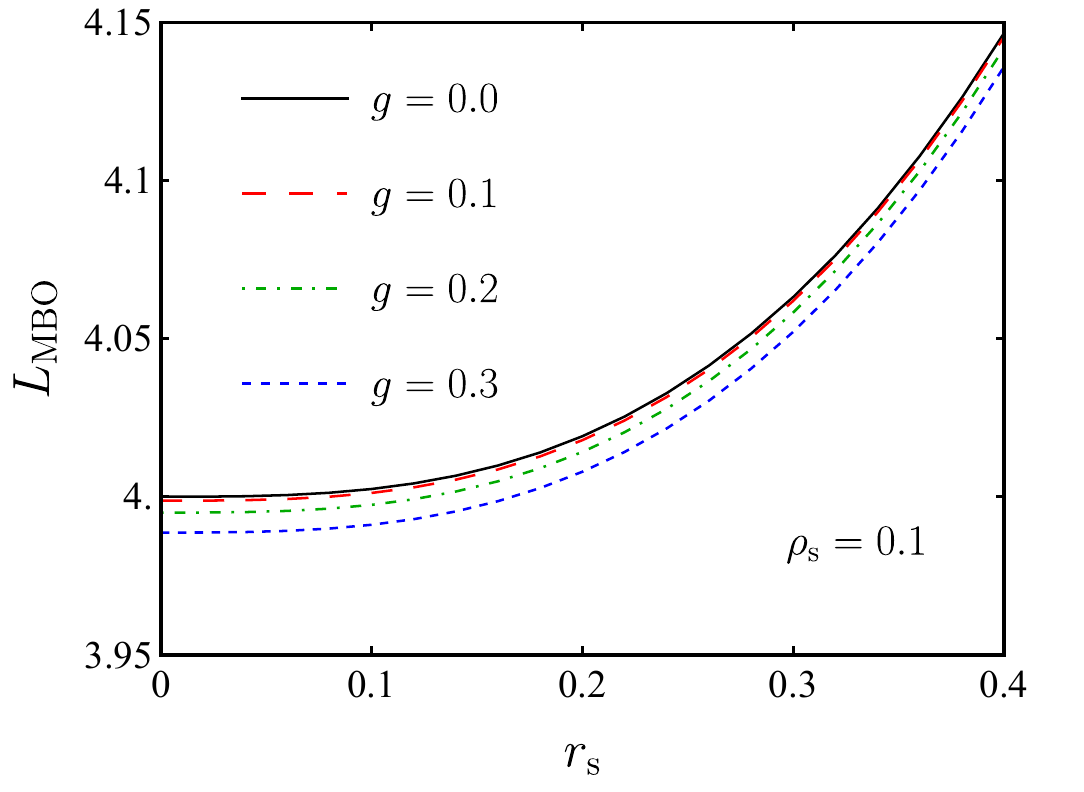}
    \caption{Variation of $r_\text{MBO}$ and $L_\text{MBO}$ as functions of the scaled radius $r_\text{s}$. In the upper panels, the magnetic charge is fixed at $g=0.1$, while different values of $\rho_\mathrm{s}$ are considered. In the lower panels, the halo density is fixed at $\rho_\mathrm{s}=0.1$, while $g$ is varied. }
	\label{fig:MBO1}
\end{figure}

\begin{figure}[ht!]
    \centering
    \includegraphics[width=80mm]{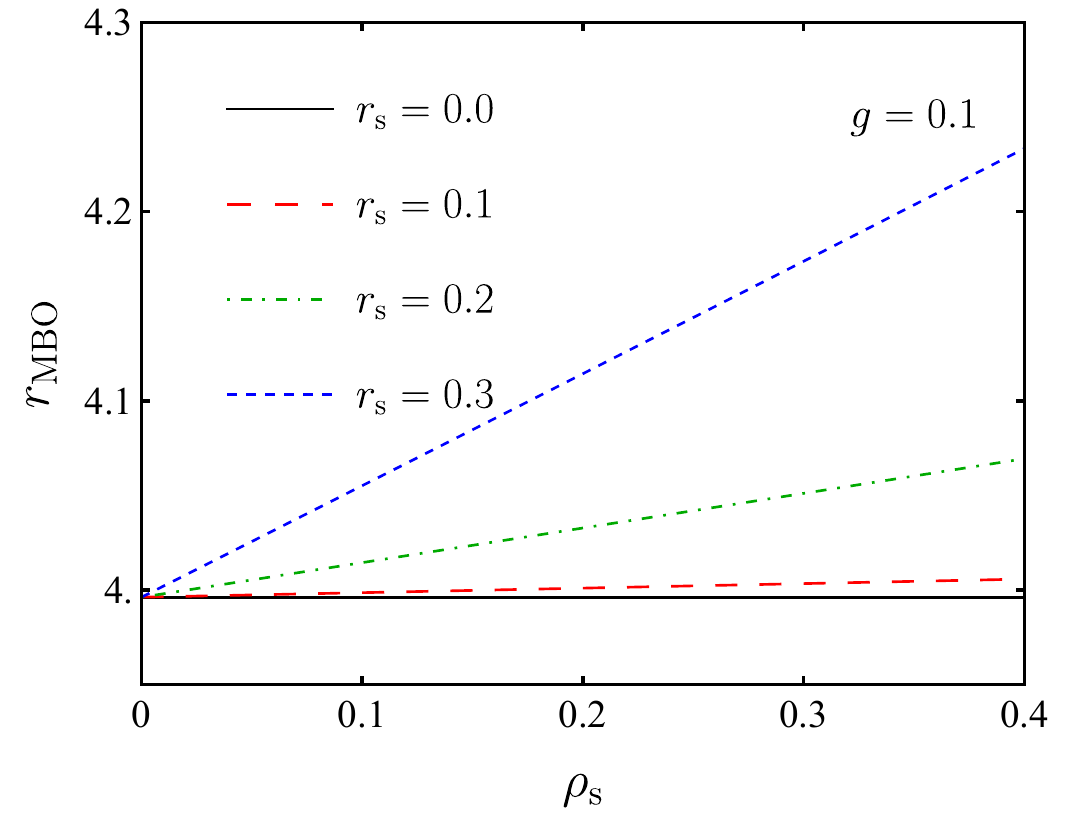} \hspace{2mm}
    \includegraphics[width=80mm]{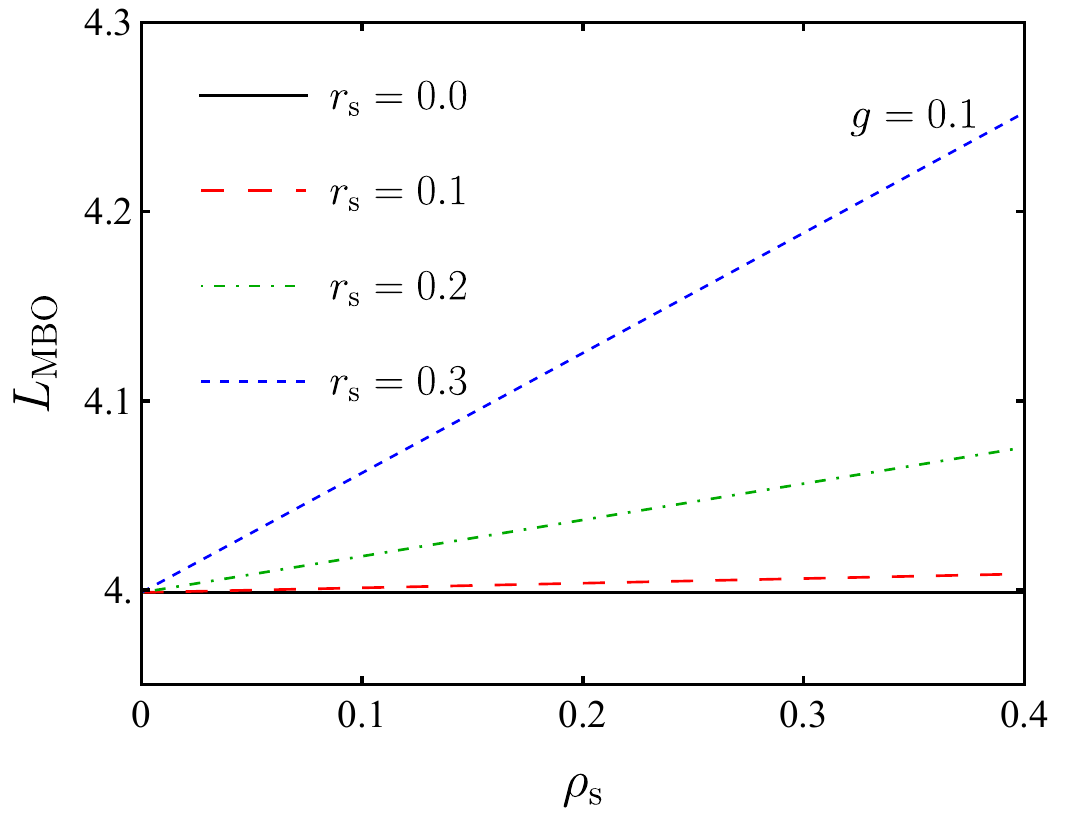}
    \includegraphics[width=82mm]{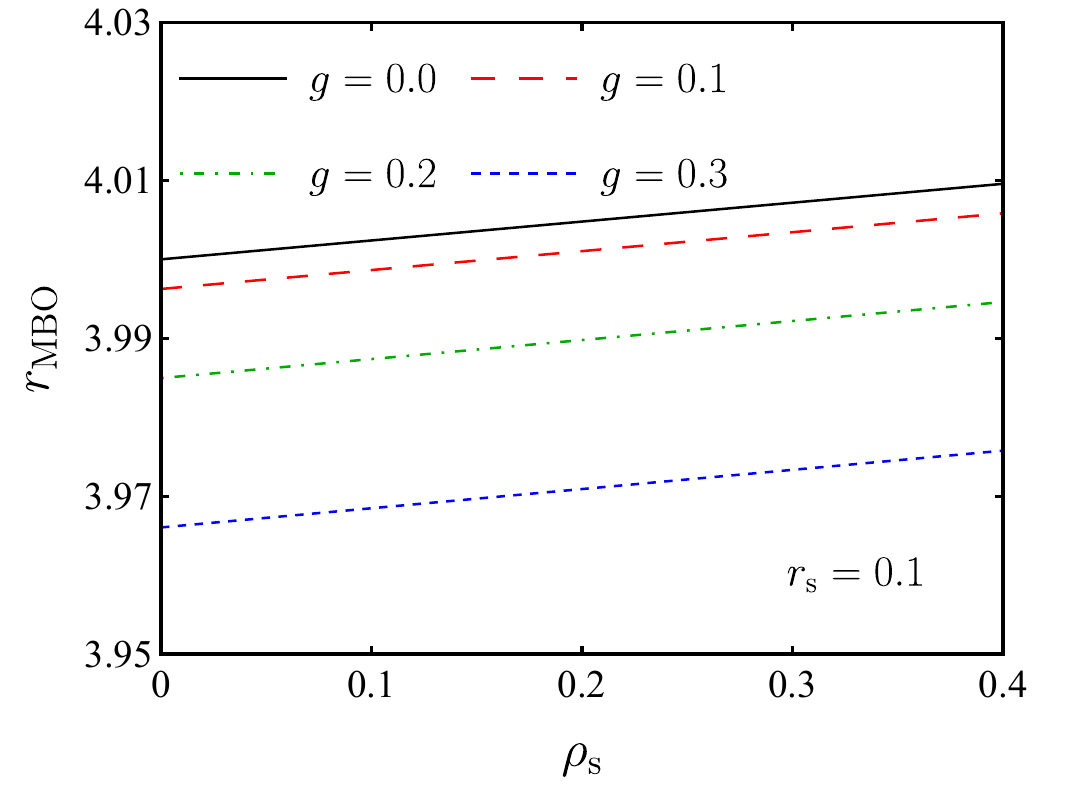} \hspace{2mm}
    \includegraphics[width=82mm]{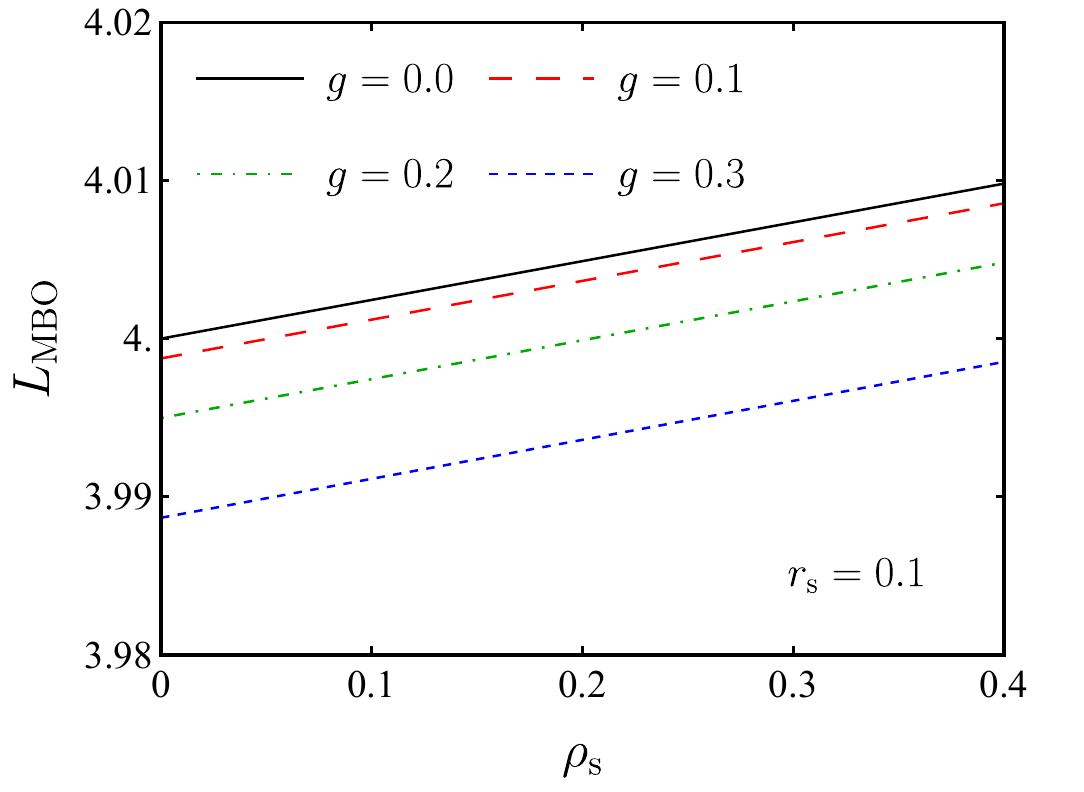} 
  \caption{Variation of $r_{\mathrm{MBO}}$ and $L_{\mathrm{MBO}}$ with respect to the density parameter $\rho_{\mathrm{s}}$. In the upper panels, $g$ is fixed at $0.1$ for different values of $\rho_{\mathrm{s}}$; in the lower panels, $r_{\mathrm{s}}$ is fixed at $0.1$ for various values of $g$. }
    \label{fig:MBO2}
\end{figure}

The MBO conditions can be described as 
\begin{align}
    V_\text{eff}=1 \quad\quad \text{and}\quad\quad\frac{\mathrm{d}V_\text{eff}}{\mathrm{d}r}=0.
\end{align}
which leads to the following expressions for radius and angular momentum for MBO
\begin{align}\label{eq:rMBO}
    r_{\mathrm{MBO}} &= \frac{2 f(r_{\mathrm{MBO}}) \left(1 - f(r_{\mathrm{MBO}})\right)}{f'(r_{\mathrm{MBO}})} , \\ \label{eq:LMBO}
    L_{\mathrm{MBO}} &= \sqrt{\frac{1 - f(r_{\mathrm{MBO}})}{f(r_{\mathrm{MBO}})}} \; r_{\mathrm{MBO}} .
\end{align}
Solving Eq.~\eqref{eq:rMBO} -- \eqref{eq:LMBO} numerically, we show the impact of \textit{MHDM} BH parameters on MBO's radius and momentum in Fig.~\ref{fig:MBO1}~--~\ref{fig:MBO2}.

We explore the behavior of the marginal bound orbit radius $r_{\mathrm{MBO}}$ and the corresponding angular momentum $L_{\mathrm{MBO}}$ as functions of the scale radius $r_{\mathrm{s}}$ in Fig.~\ref{fig:MBO1}. Both quantities exhibit a distinctly nonlinear dependence on $r_{\mathrm{s}}$, which shows a remarkable correlation between the dark matter halo scale and the orbital characteristics. In the upper panels of Fig.~\ref{fig:MBO1}, the magnetic monopole charge is fixed at $g = 0.1$, while different curves correspond to varying values of the scale density $\rho_{\mathrm{s}}$. All curves originate from a common point, indicating a universal behavior in the limit of small $r_{\mathrm{s}}$. As $r_{\mathrm{s}}$ increases, the nonlinearity becomes more pronounced, with higher densities yielding steeper curvature and more rapid growth in both $r_{\mathrm{MBO}}$ and $L_{\mathrm{MBO}}$. This suggests that for larger $\rho_{\mathrm{s}}$, the system becomes increasingly sensitive to variations in the halo scale radius.  The lower panels of Fig.~\ref{fig:MBO1} present the dependence on the magnetic monopole charge $g$, with the scale density fixed at $\rho_{\mathrm{s}} = 0.1$. Here again, the behavior is nonlinear. Increasing $g$ results in a systematic downward shift of the entire curve, indicating that a larger magnetic charge reduces both the orbital radius and angular momentum for a given $r_{\mathrm{s}}$. Notably, as $r_{\mathrm{s}}$ increases, the curves for different $g$ converge, suggesting that the influence of the magnetic charge becomes less significant for larger halo scales. 
Unlike the nonlinear trend observed with respect to $r_\mathrm{s}$, the behavior of the marginal bound orbit radius $r_{\mathrm{MBO}}$ and the corresponding angular momentum $L_{\mathrm{MBO}}$ with respect to the $\rho_\mathrm{s}$ is linear in Fig.~\ref{fig:MBO2}. In the upper panels, for a fixed magnetic monopole charge $g = 0.1$, all curves originate from a common point, indicating a universal threshold behavior independent of the density scale. The slope of these linear relations increases monotonically with $\rho_{\mathrm{s}}$, suggesting that higher densities enhance the sensitivity of both the orbital radius and angular momentum to variations in the halo parameters.  
The lower panels of Fig.~\ref{fig:MBO1} illustrate the dependence on the magnetic monopole charge $g$ while keeping the scale radius fixed at $r_{\mathrm{s}} = 0.1$. Here, the linear trends for both $r_{\mathrm{MBO}}$ and $L_{\mathrm{MBO}}$ persist; however, increasing $g$ results in a general downward shift of the entire linear profile. Importantly, the slopes remain almost parallel across different values of $g$, indicating that the magnetic charge primarily modifies the overall scale of the orbits without altering the functional dependence on $\rho_{\mathrm{s}}$.

\subsection{Innermost stable circular orbits}

The ISCO condition is determined by solving the system of equations.

\begin{align}
	V_{\mathrm{eff}}&= E^2, \quad \frac{\mathrm{d}V_{\mathrm{eff}}}{\mathrm{d}r} = 0,\quad \text{and}\quad \frac{\mathrm{d}^2V_{\mathrm{eff}}}{\mathrm{d}r^2} = 0.
\end{align}

The first two conditions yield the angular momentum $L$ and energy $E$ of a circular orbit, while the third condition selects the innermost stable orbit where the curvature of the potential changes sign. Substituting $f(r)$ into the ISCO conditions leads to an algebraic equation for $r_{\mathrm{ISCO}}$ 
\begin{align}
&r_{\mathrm{ISCO}}\, f(r_{\mathrm{ISCO}})\, f''(r_{\mathrm{ISCO}})
- 2 r_{\mathrm{ISCO}} f'(r_{\mathrm{ISCO}})^2
+ 3 f(r_{\mathrm{ISCO}})\, f'(r_{\mathrm{ISCO}}) = 0,\\
&L_{\mathrm{ISCO}} = \sqrt{
\frac{r_{\mathrm{ISCO}}^{3}\, f'(r_{\mathrm{ISCO}})}
{2 f(r_{\mathrm{ISCO}}) - r_{\mathrm{ISCO}} f'(r_{\mathrm{ISCO}})}
}\\
&E_{\mathrm{ISCO}} = \sqrt{
\frac{2\, f(r_{\mathrm{ISCO}})^2}
{2 f(r_{\mathrm{ISCO}}) - r_{\mathrm{ISCO}} f'(r_{\mathrm{ISCO}})}
}.
\end{align}

\begin{figure}
    \centering
    \includegraphics[width=53mm]{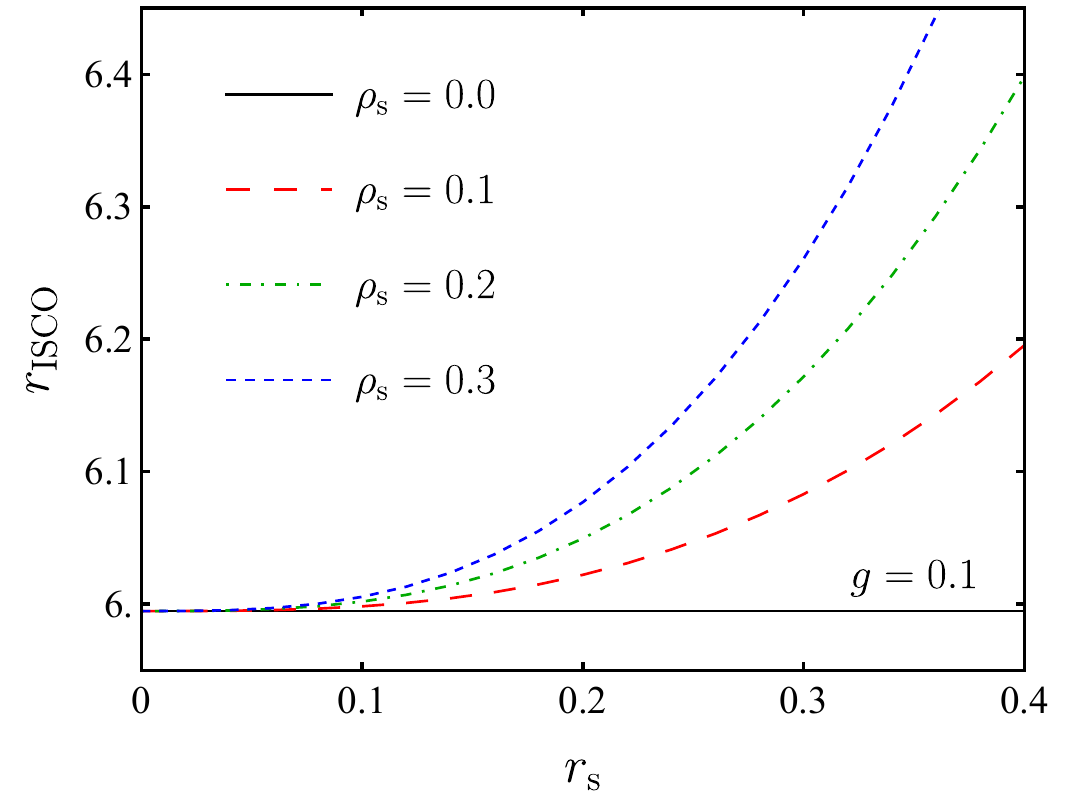}
    \includegraphics[width=53mm]{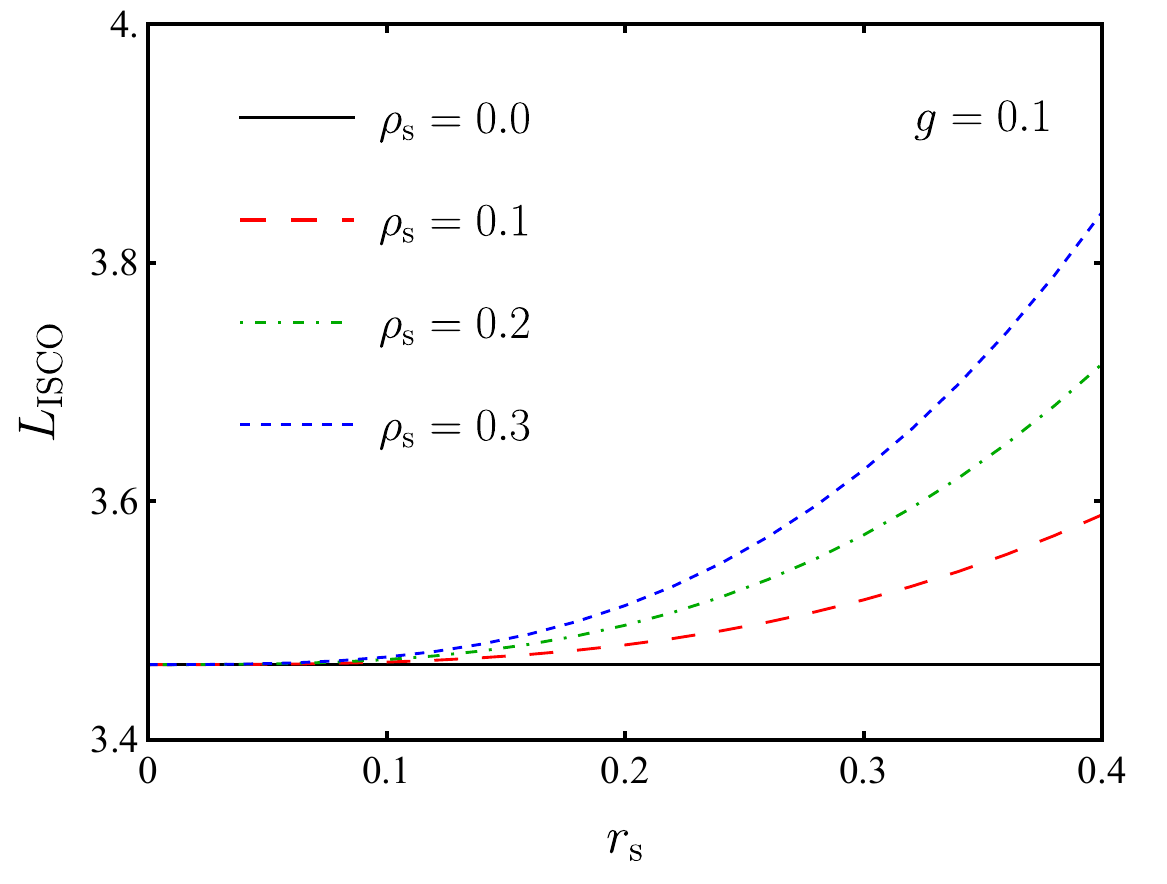}
    \includegraphics[width=60mm]{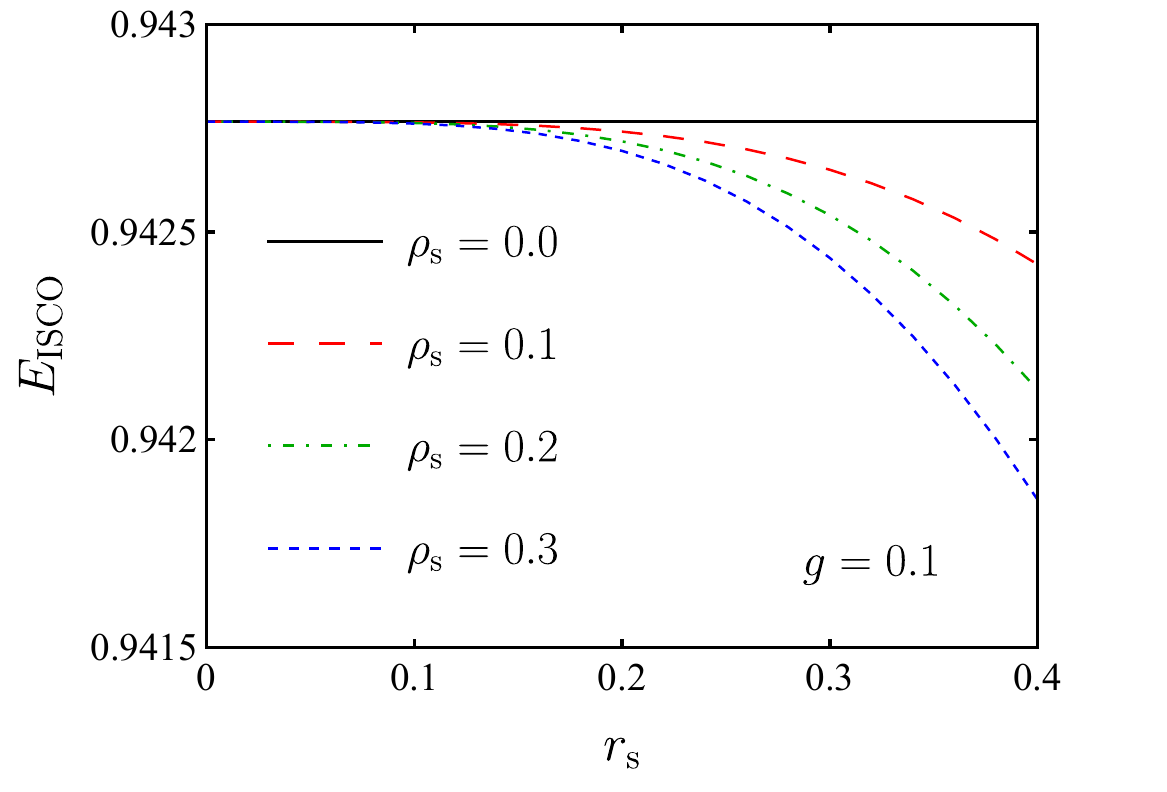} 
    \caption{Innermost stable circular orbit (ISCO) properties for the \textit{MHDM} BH as a function of $r_\mathrm{s}$, for $M = 1$, a fixed value of $g$ and different values of the density parameter $\rho_\mathrm{s}$. The panels display: (left) the ISCO radius, (middle) the orbital angular momentum, and (right) the orbital energy.}
    \label{fig:ISCO1}
\end{figure}

\begin{figure}
    \centering
      \includegraphics[scale=0.29]{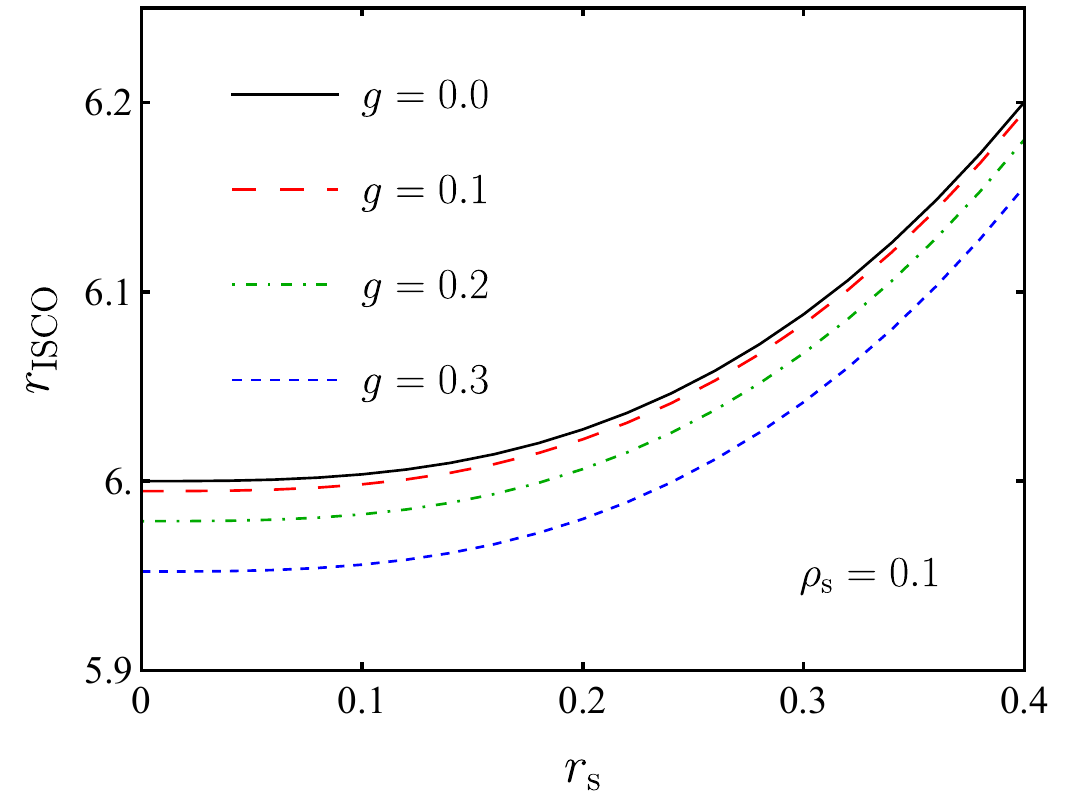}
       \includegraphics[scale=0.28]{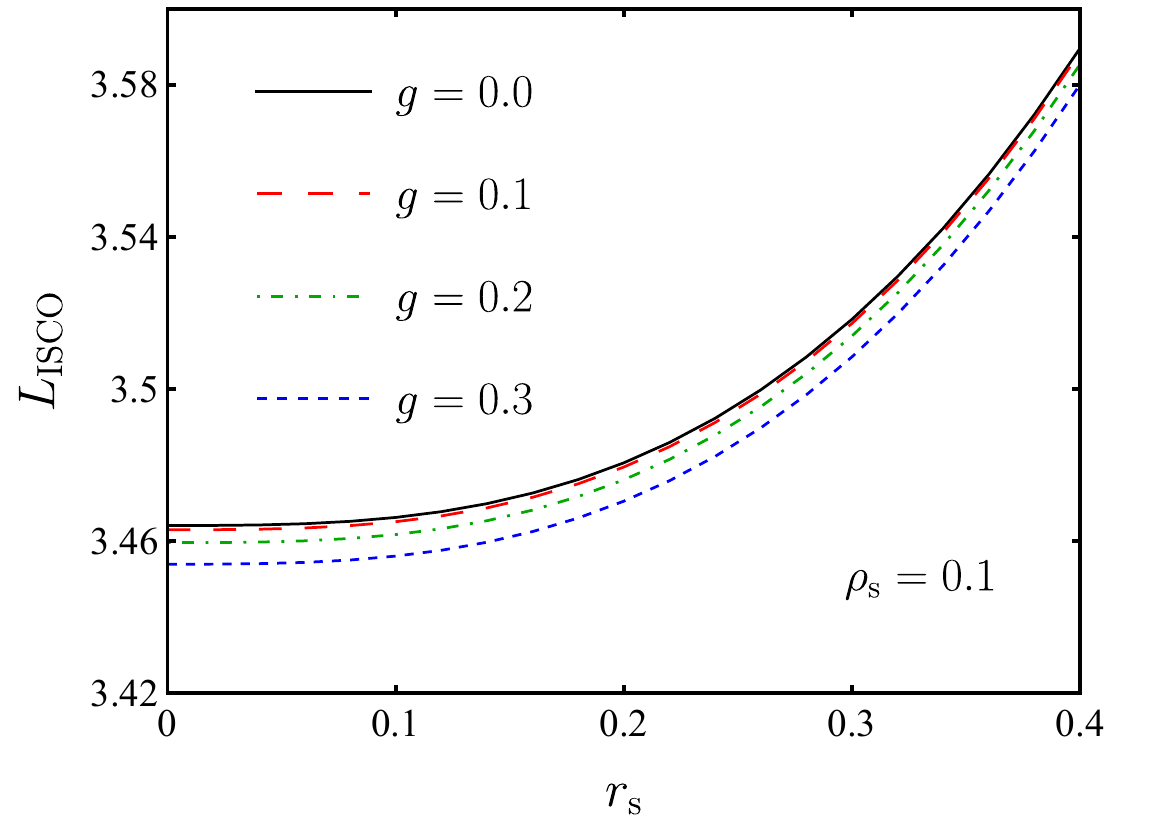}
        \includegraphics[scale=0.303]{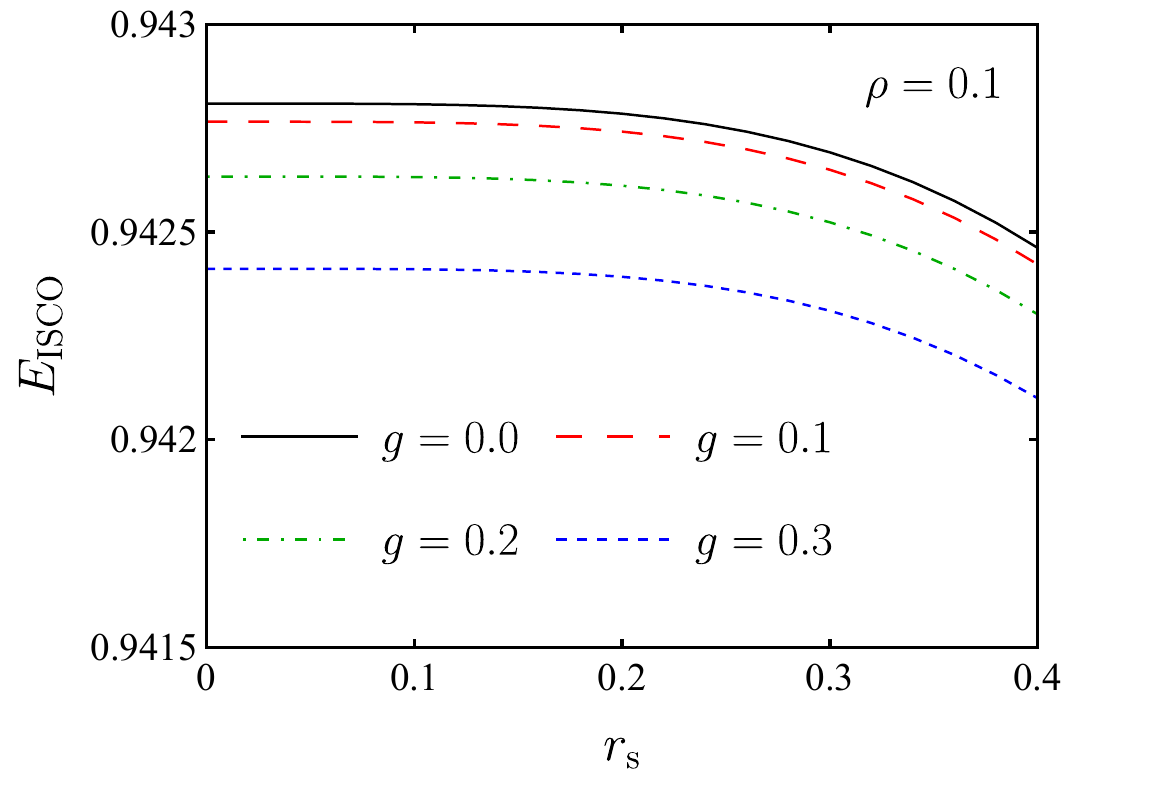}
    \caption{ISCO properties of the \textit{MHDM} BH as a function of $r_\mathrm{s}$ for fixed $\rho_\mathrm{s}$, and varying $g$. The left, middle, and right panels show the ISCO radius, orbital angular momentum, and orbital energy, respectively.}
    \label{fig:ISCO2}
\end{figure}

We solve this equation numerically to investigate how these parameters influence the ISCO radius, momentum, and energy. The results are presented in Fig.~\ref{fig:ISCO1}~--~\ref{fig:ISCO2}, where we explore the individual and combined effects of the halo parameters and the magnetic monopole charge. The analysis reveals that the presence of dark matter and nonlinear electrodynamics significantly shifts the ISCO compared to the standard Schwarzschild black hole (\textit{Sch} BH) case, which is $r^\textit{Sch}_{\mathrm{ISCO}} = 6M$.

In Fig.~\ref{fig:ISCO1}, for a fixed $g$, both ISCO radius and orbital angular momentum exhibit a nonlinear increasing dependence on $r_\mathrm{s}$, with larger values of the density parameter $\rho_\mathrm{s}$ yielding higher $r_{\mathrm{ISCO}}$ and $L_{\mathrm{ISCO}}$. This growth becomes more pronounced at larger $r_\mathrm{s}$. In contrast, the orbital energy $E_{\mathrm{ISCO}}$ shows the opposite behavior: increasing either $r_\mathrm{s}$ or $\rho_\mathrm{s}$ reduces the energy required for the innermost stable circular orbit. According to Fig.~\ref{fig:ISCO2}, $r_{\mathrm{ISCO}}$ and $L_{\mathrm{ISCO}}$ increase monotonically with $r_\mathrm{s}$. For a fixed value of $\rho_\mathrm{s}$, increasing the parameter $g$ results in lower values of both $r_{\mathrm{ISCO}}$ and $L_{\mathrm{ISCO}}$, effectively shifting the corresponding curves downward. Notably, this downward shift diminishes at larger $r_\mathrm{s}$, where the curves tend to converge. In contrast, the orbital energy $E_{\mathrm{ISCO}}$ decreases with increasing $r_\mathrm{s}$. A similar downward shift is observed for larger $g$, and as with $r_{\mathrm{ISCO}}$ and $L_{\mathrm{ISCO}}$, the differences become less pronounced at higher $r_\mathrm{s}$. 

\begin{figure}[ht!]
    \centering
    \includegraphics[width=55mm]{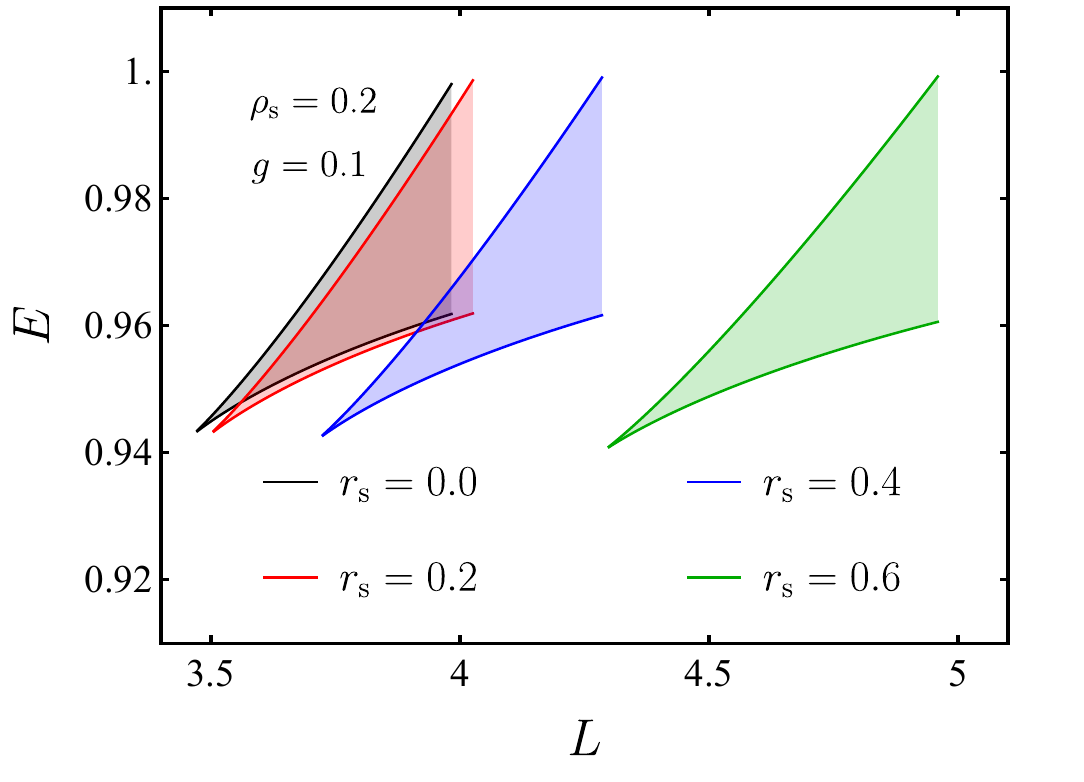} 
    \includegraphics[width=55mm]{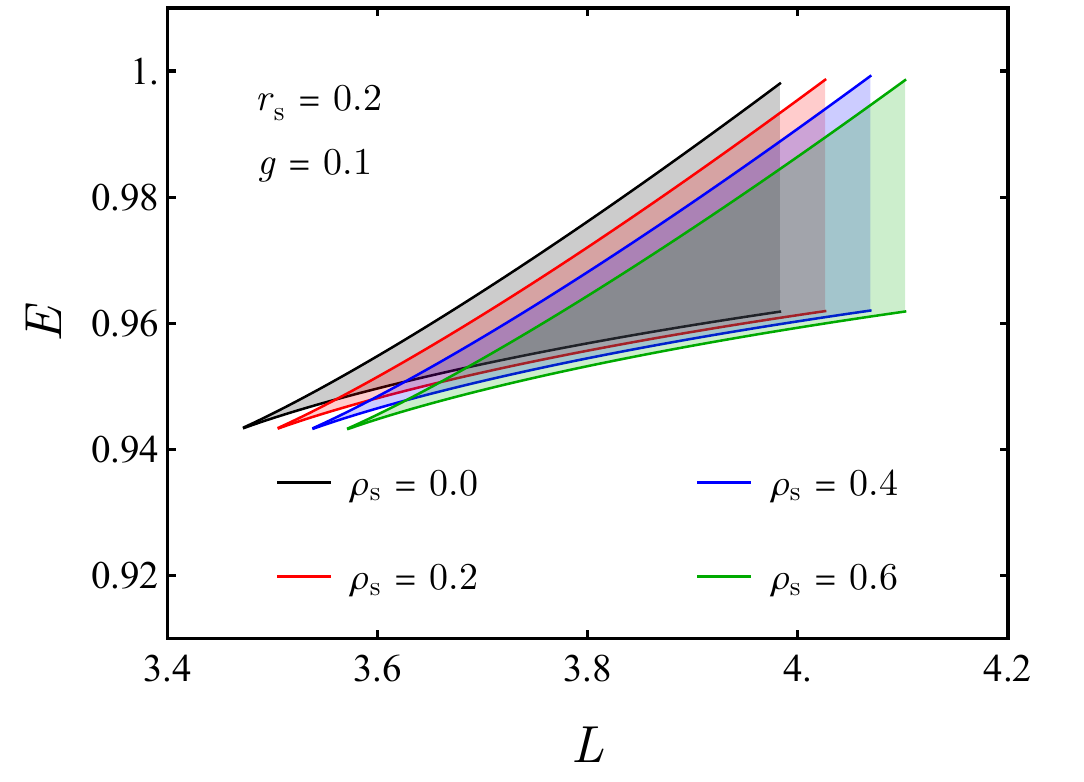}
     \includegraphics[width=55mm]{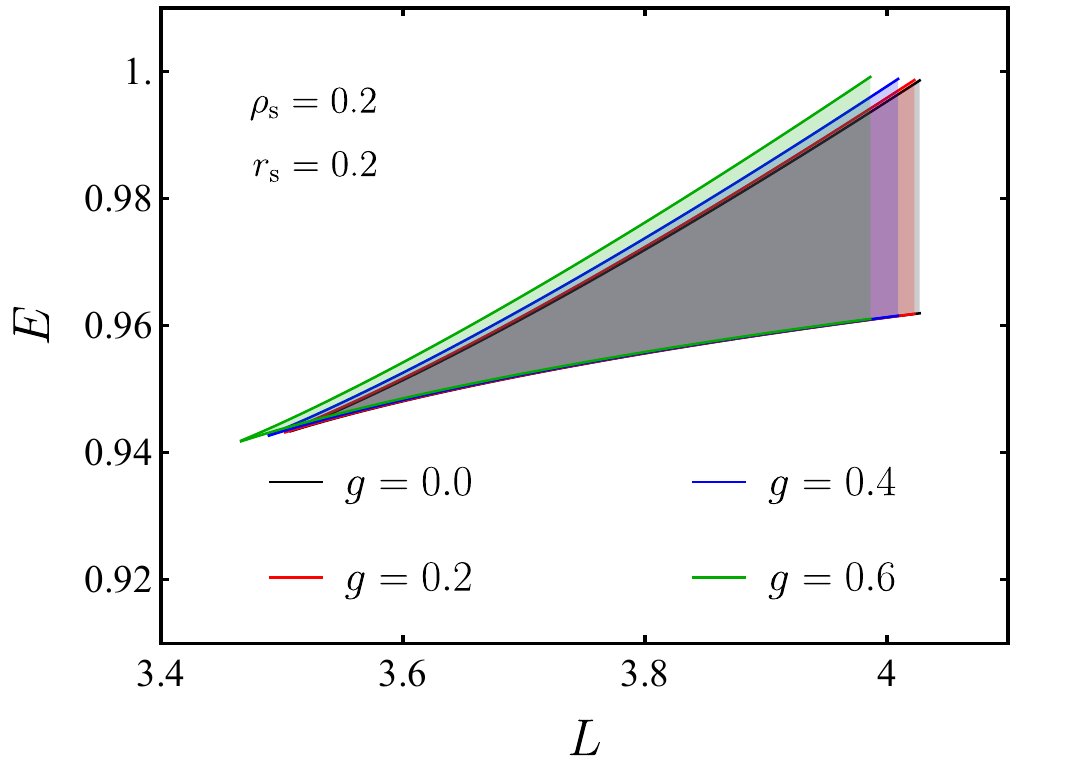}
    \caption{Allowed parameter space for bound orbits around the \textit{MHDM} BH in the $(E, L)$ plane. The left, middle, and right panels illustrate the effects of varying $r_\mathrm{s}$, the density parameter $\rho_\mathrm{s}$, and the parameter $g$, respectively.}
    \label{fig:EL1}
\end{figure}

\begin{figure}[ht!]
	\centering
	\includegraphics[width=85mm]{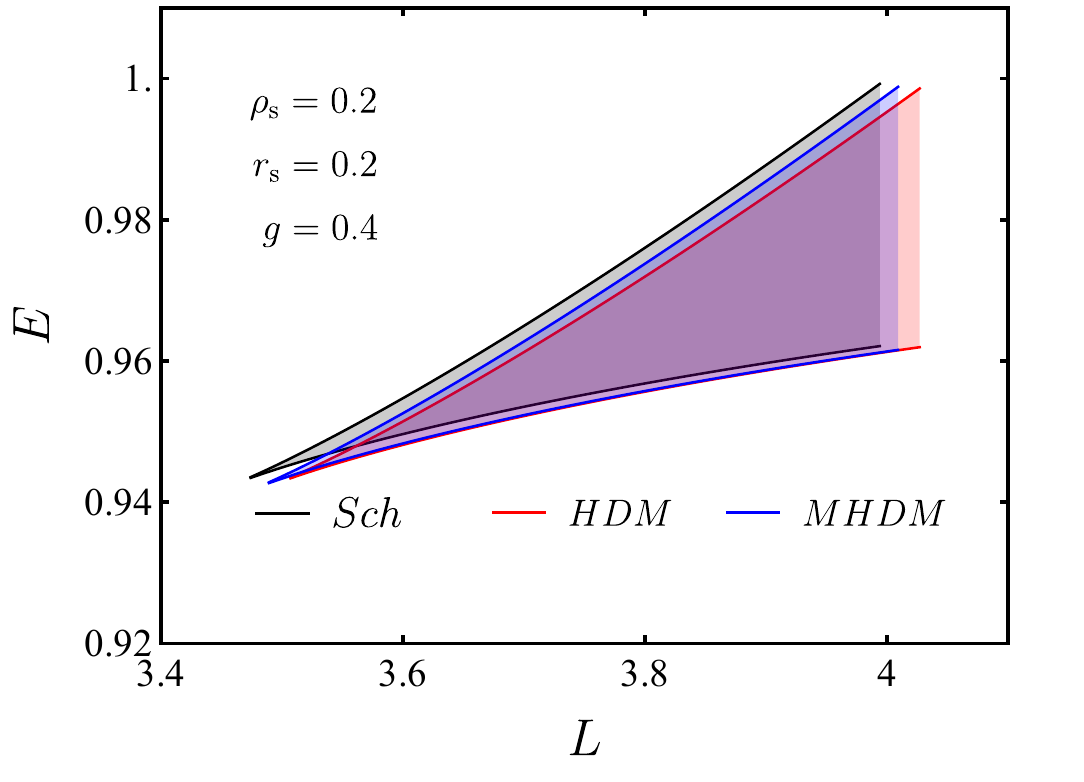} 
	\caption{Allowed region in the $(E, L)$ plane for bound orbits around \textit{Sch}, \textit{HDM}, and \textit{MHDM} BHs. The shaded regions indicate the permitted parameter space for each black hole configuration. In all cases, the mass is fixed at $M = 1$}
	\label{fig:EL2}
\end{figure}

For a better understanding, the allowed parameter space for bound orbits around the \textit{MHDM} BH, defined by the conditions in Eq.~\ref{eq:EL}, is illustrated in Fig.~\ref{fig:EL1}~--~\ref{fig:EL2}. The left, middle, and right panels represent the effects of varying $r_\mathrm{s}$, the density parameter $\rho_\mathrm{s}$, and the parameter $g$, respectively. For a fixed $\rho_\mathrm{s}$ and $g$, increasing $r_\mathrm{s}$ shifts the allowed region toward higher orbital angular momentum $L$, while the upper bound on energy remains fixed at $E_{\mathrm{MBO}} = 1$. In other words, this indicates that a more extended dark matter scaled radius requires larger angular momentum to maintain stable circular orbits. Similarly, increasing $\rho_\mathrm{s}$ at fixed $r_\mathrm{s}$ and $g$ also expands the accessible $L$ range, reflecting the enhanced gravitational influence of a denser dark matter halo. This impact of $\rho_\mathrm{s}$ is weaker than that of $r_\mathrm{s}$. In contrast, varying $g$ at fixed $r_\mathrm{s}$ and $\rho_\mathrm{s}$ shows a more subtle effect: larger $g$ reduces the minimum allowed $L$, slightly shrinking the permitted region. In all cases, the energy interval remains bounded between $E_{\mathrm{ISCO}}$ and $1$, with the lower bound $E_{\mathrm{ISCO}}$ decreasing as the dark matter parameters increase, thereby enlarging the accessible energy range for bound orbits.

Fig.~\ref{fig:EL2} presents a comparison of the allowed parameter space for bound orbits around \textit{Sch}, \textit{HDM}, and \textit{MHDM} BHs in the $(E, L)$ plane. Comparing the allowed region of the \textit{Sch} and \textit{HDM} case shows that the inclusion of \textit{HDM} shifts the allowed profile rightward, requiring larger orbital angular momentum and lowering the ISCO energy. Interestingly, the magnetized \textit{HDM} black hole lies between the \textit{Sch} and \textit{HDM} cases. In other words, this suggests that the magnetic monopole charge introduces a competing effect that reduces the influence of the \textit{Hernquist} dark matter halo on the bound orbit parameter space.


\section{Periodic orbits}

We now turn our attention to periodic orbits around the \textit{MHDM} BH, which play a crucial role in the study of particle dynamics near compact objects, as they serve as building blocks for understanding more complex bound trajectories. In generic spacetimes, the motion of a test particle is characterized by oscillations in the radial and azimuthal directions. When the ratio of the corresponding frequencies is rational, the trajectory closes upon itself after a finite number of cycles, giving rise to a periodic orbit. Conversely, an irrational frequency ratio results in a precessing orbit that never exactly repeats. Nevertheless, any generic bound orbit can be approximated arbitrarily closely by a periodic one, making the study of periodic orbits essential for interpreting the orbital dynamics and the associated gravitational wave emission from systems involving black holes surrounded by dark matter halos.

In the spherically symmetric geometry, we consider the motion of a test particle confined to an equatorial plane. Based on Eq. \ref{eq:rdot}, the radial motion is governed by the effective potential, and the orbit can be characterized by the periapsis $r_1$ and apoapsis $r_2$, which are the turning points where the radial velocity vanishes. Considering Eq. \ref{eq:phi_dot}~--~\ref{eq:rdot}, the accumulated azimuthal angle over one full radial oscillation, known as the apsidal angle, is given by

\begin{equation} \label{eq:deltaphi}
\Delta \phi =\oint \mathrm{d}\phi = 2 \int_{\phi_1}^{\phi_2} \mathrm{d}\phi= 2 \int_{r_1}^{r_2} \frac{\dot{\phi}}{\dot{r}} \, \mathrm{d}r = 2 \int_{r_1}^{r_2} \frac{L}{r^2 \sqrt{E^2 - V_{\mathrm{eff}}}} \, \mathrm{d}r,
\end{equation}
where the factor of 2 accounts for the outward and inward journey between the turning points.
Based on the classification scheme introduced in Ref.~\cite{levin2008periodic}, we characterize periodic orbits by three integers $(z, w, v)$, known as the zoom, whirl, and vertex numbers, respectively. The frequency ratio between the azimuthal frequency $\omega_\phi$ and the radial frequency $\omega_r$ is expressed as

\begin{equation}\label{eq:q1}
	q = \frac{\omega_\phi}{\omega_r} - 1 = w + \frac{v}{z},
\end{equation}
where $z$, $w$, and $v$ are positive integers with $v < z$ to avoid degeneracy. This ratio is directly related to the apsidal angle via
\begin{equation}\label{eq:q2}
\frac{\omega_\phi}{\omega_r} = \frac{\Delta \phi}{2\pi},
\end{equation}
and utilizing the Eq. (\ref{eq:deltaphi}), the rational number $q$ can be written explicitly as
\begin{equation}
q = \frac{\Delta \phi}{2\pi} - 1 = \frac{1}{\pi} \int_{r_1}^{r_2} \frac{L}{r^2 \sqrt{E^2 - V_{\mathrm{eff}}}} \, \mathrm{d}r - 1.
\end{equation}

\begin{figure}[ht!]
    \centering
    \includegraphics[width=55mm]{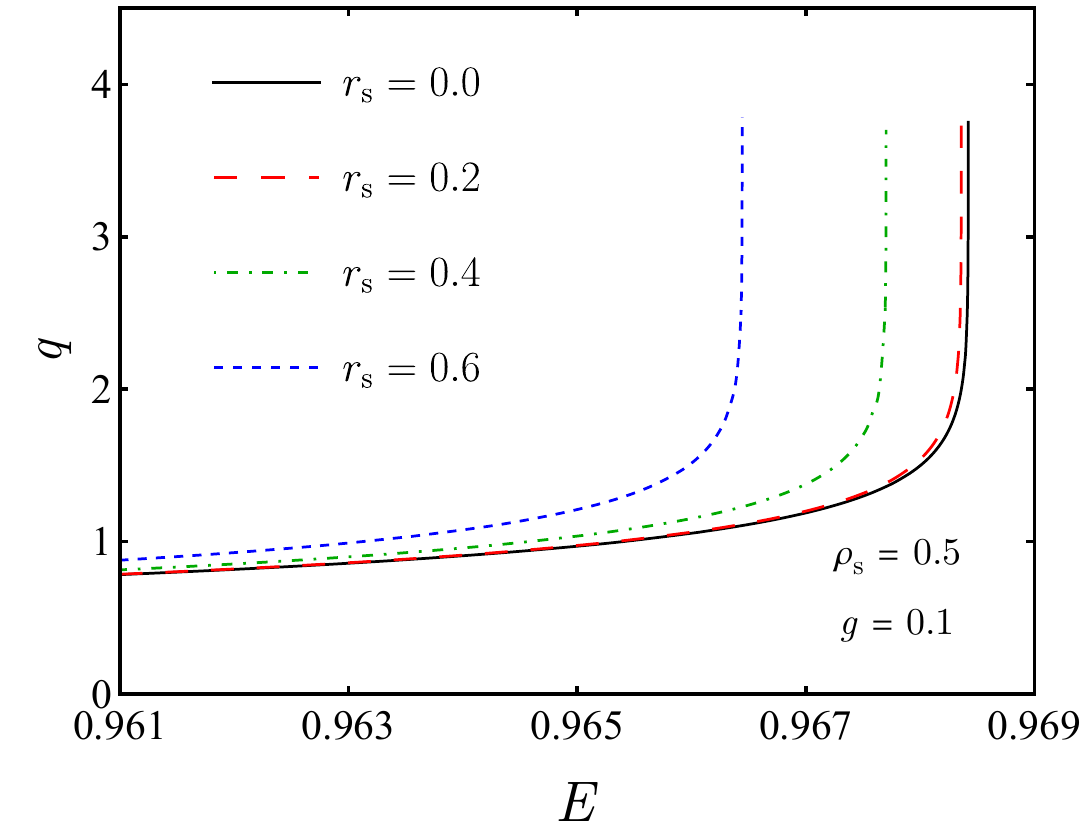}
    \includegraphics[width=55mm]{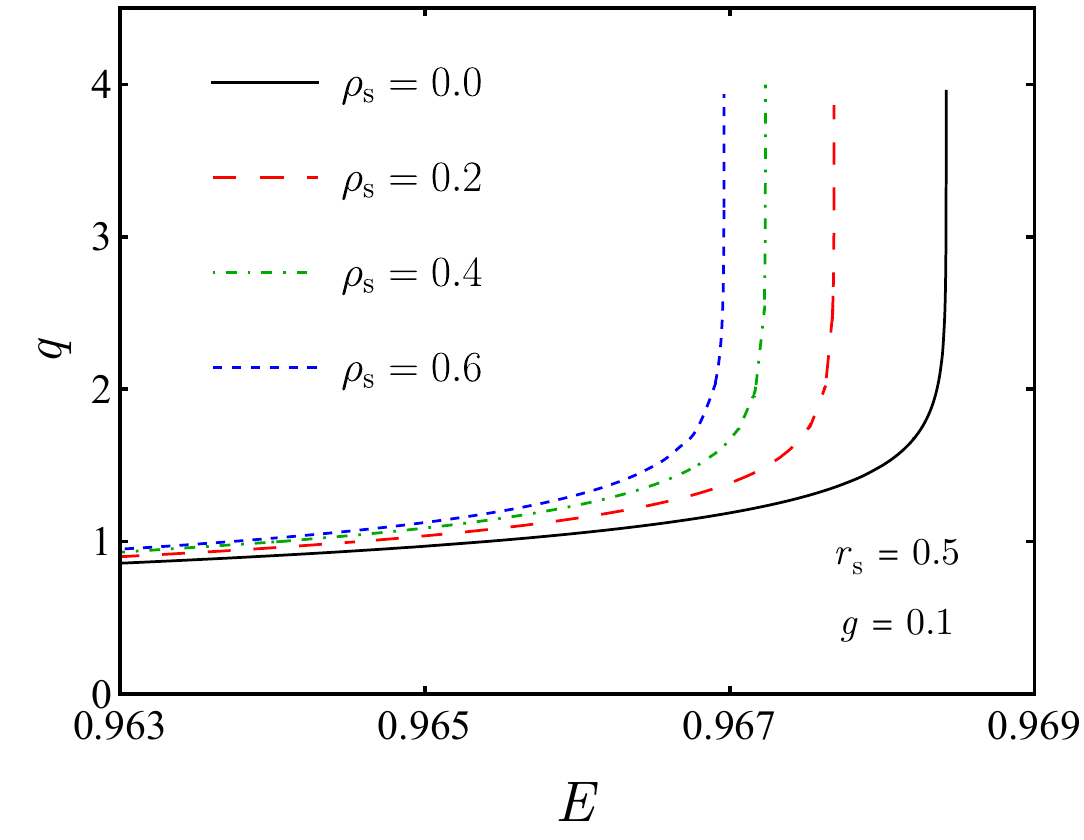}
    \includegraphics[width=55mm]{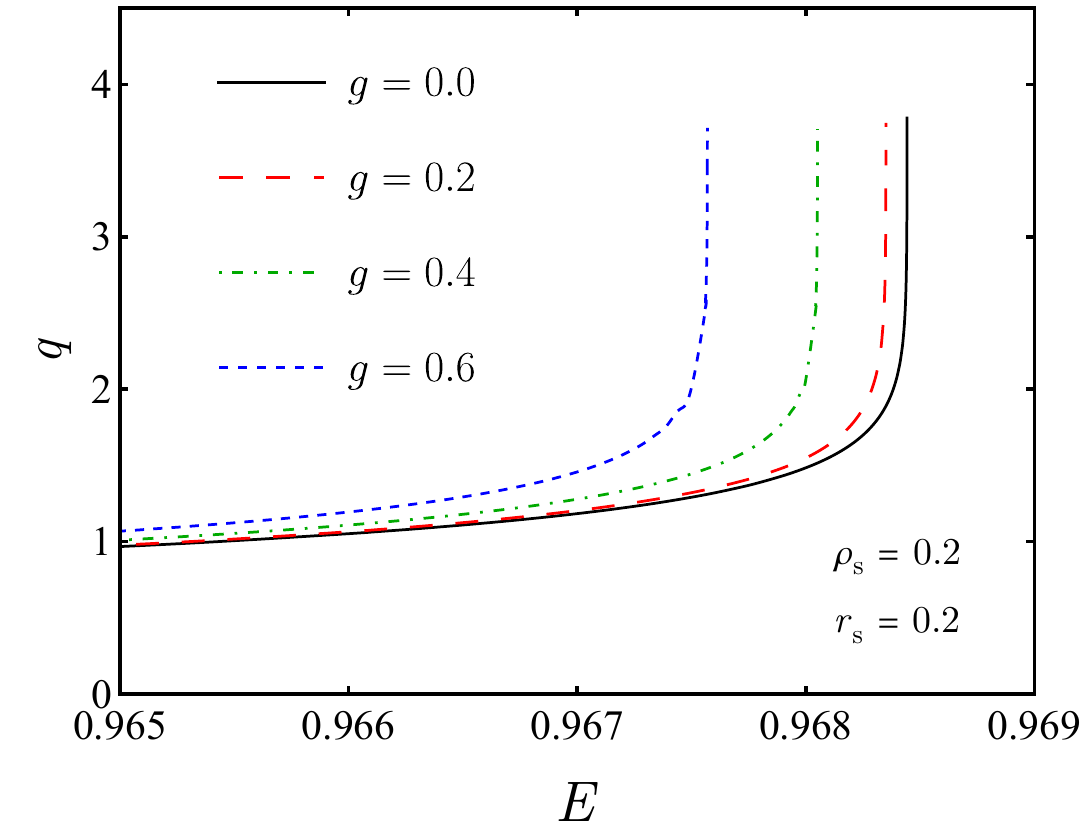}
    \caption{The rational number $q$ plotted against the energy $E$ of periodic orbits in a \textit{\textit{MDMH}} BH spacetime, for selected values of the parameters $\rho_\mathrm{s}$, $r_\mathrm{s}$, and $g$. The orbital angular momentum is fixed to $L_\text{av}$.}
    \label{fig:qE}
\end{figure}

This integral depends on the parameters of the black hole, the dark matter halo, and the magnetic field, as well as on the energy $E$ and angular momentum $L$ of the particle. For bound orbits, the angular momentum is constrained by $L_{\mathrm{ISCO}} \leq L \leq L_{\mathrm{MBO}}$. To actually explore the parameter space, it is convenient to parameterize $L$ as
\begin{equation}
L = L_{\mathrm{ISCO}} + \epsilon \, (L_{\mathrm{MBO}} - L_{\mathrm{ISCO}}),
\end{equation}
with $\epsilon \in [0,1]$, where $\epsilon = 0$ and $\epsilon = 1$ correspond to the ISCO and MBO, respectively.
We consider a fixed averaged angular momentum $L_\text{av}=( L_{\mathrm{ISCO}} + L_{\mathrm{ISCO}})/2$ where $\epsilon = 0.5$ to examine the dependence of the rational number $q$ on the energy $E$ for different values of the spacetime parameters in Fig. \ref{fig:qE}. 
In all cases, the rational number $q$ increases gradually with increasing energy, exhibiting a steep rise as $E$ approaches its maximum. 
Although the overall behavior of $q$ with respect to the spacetime parameters $r_\mathrm{s}$, $\rho_\mathrm{s}$, and $g$ shows a consistent trend --- shifting toward lower energies as these parameters increase --- the sensitivity of this shift differs among them. We perform a numerical analysis of the effects of the parameters $\rho_\mathrm{s}$, $r_\mathrm{s}$, and $g$ on the energy $E$ of periodic orbits. For fixed orbital angular momentum $L_\text{av}$, we consider orbits characterized by distinct $(z, w, v)$ values. The computed energies are listed in Tables \ref{tab:qE1}~--~\ref{tab:qE3}, with the mass parameter set to $M = 1$ for simplicity. 
Across all configurations $(z, w, v)$, increasing the \textit{Hernquist} halo parameters $r_{\mathrm{s}}$ and $\rho_{\mathrm{s}}$ leads to a systematic decrease in the energy required for periodic orbits, while the corresponding average orbital angular momentum $L_{\mathrm{av}}$ increases. A similar reduction in the required energy is observed with increasing magnetic monopole charge $g$; however, in this case, $L_{\mathrm{av}}$ exhibits a qualitatively different behavior. This suggests that the presence of the dark matter halo and magnetic monopole charge modifies the effective potential in such a way that periodic orbits become accessible at lower energies.

\begin{figure}[htbp]
    \centering
    \floatsetup{heightadjust=all, valign=c}
    
    \begin{tabular}{@{}c@{\hspace{0.03\textwidth}}c@{\hspace{0.03\textwidth}}c@{}}
        \includegraphics[width=0.3\textwidth]{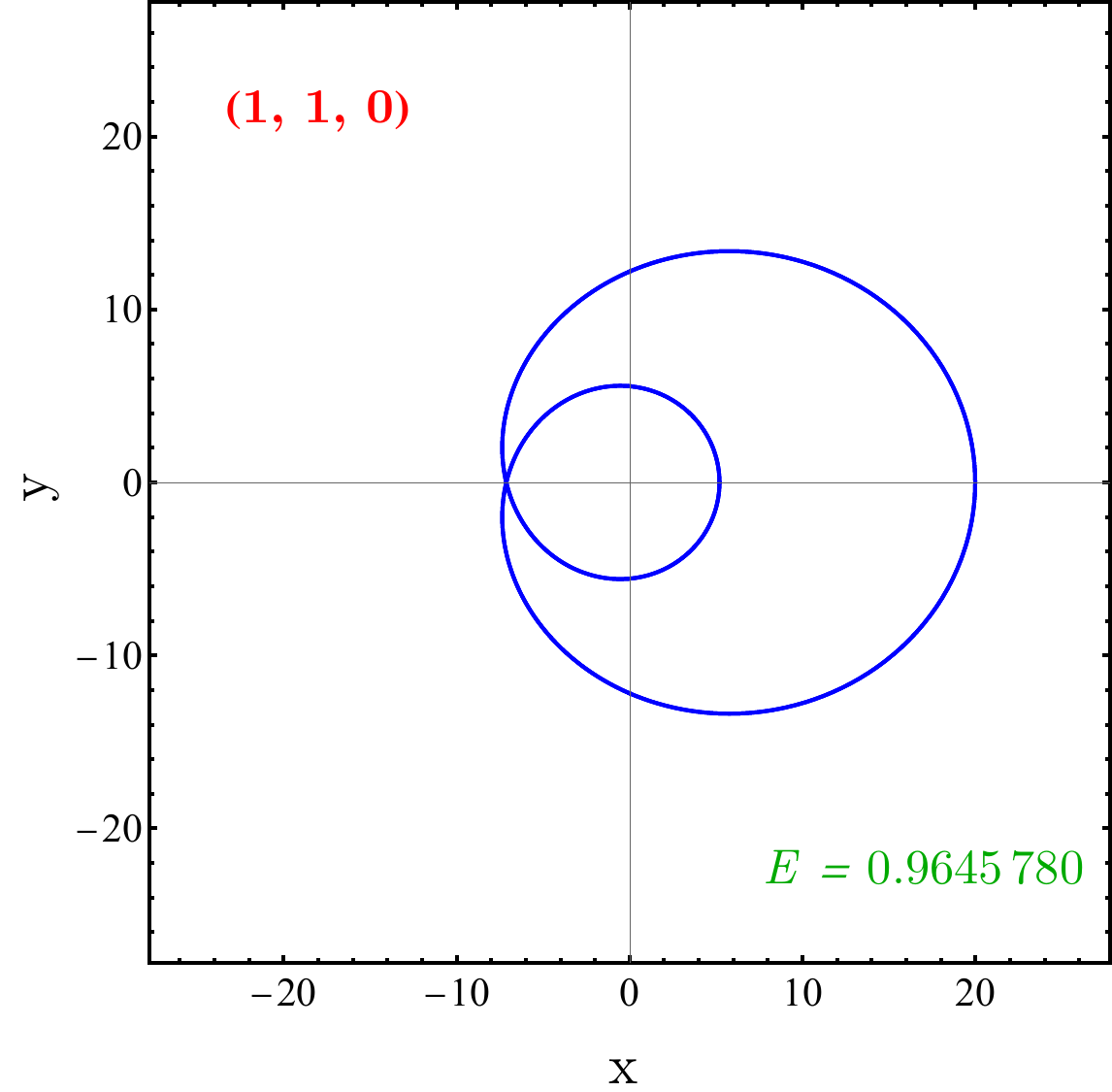} &
        \includegraphics[width=0.3\textwidth]{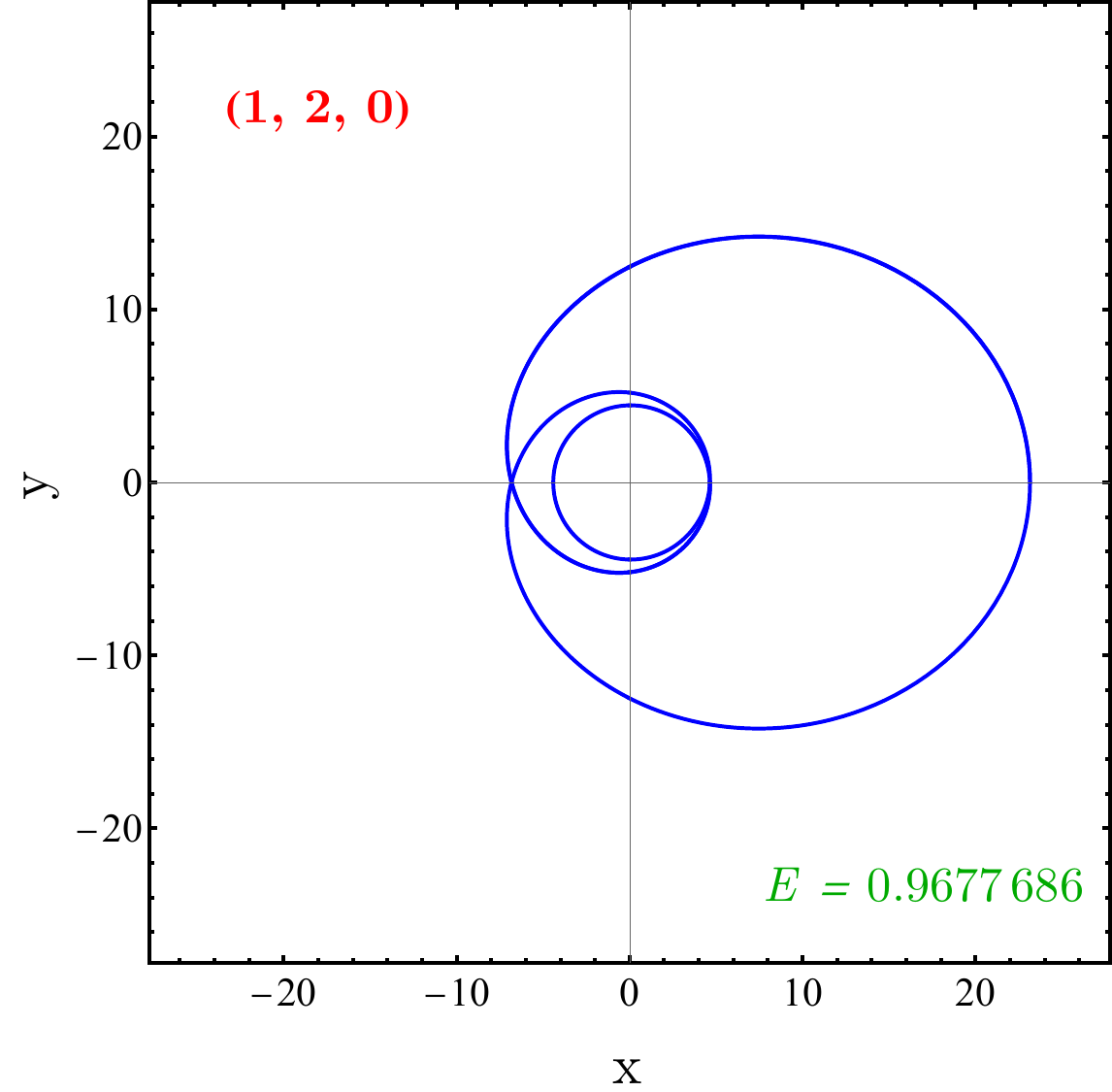} &
        \includegraphics[width=0.3\textwidth]{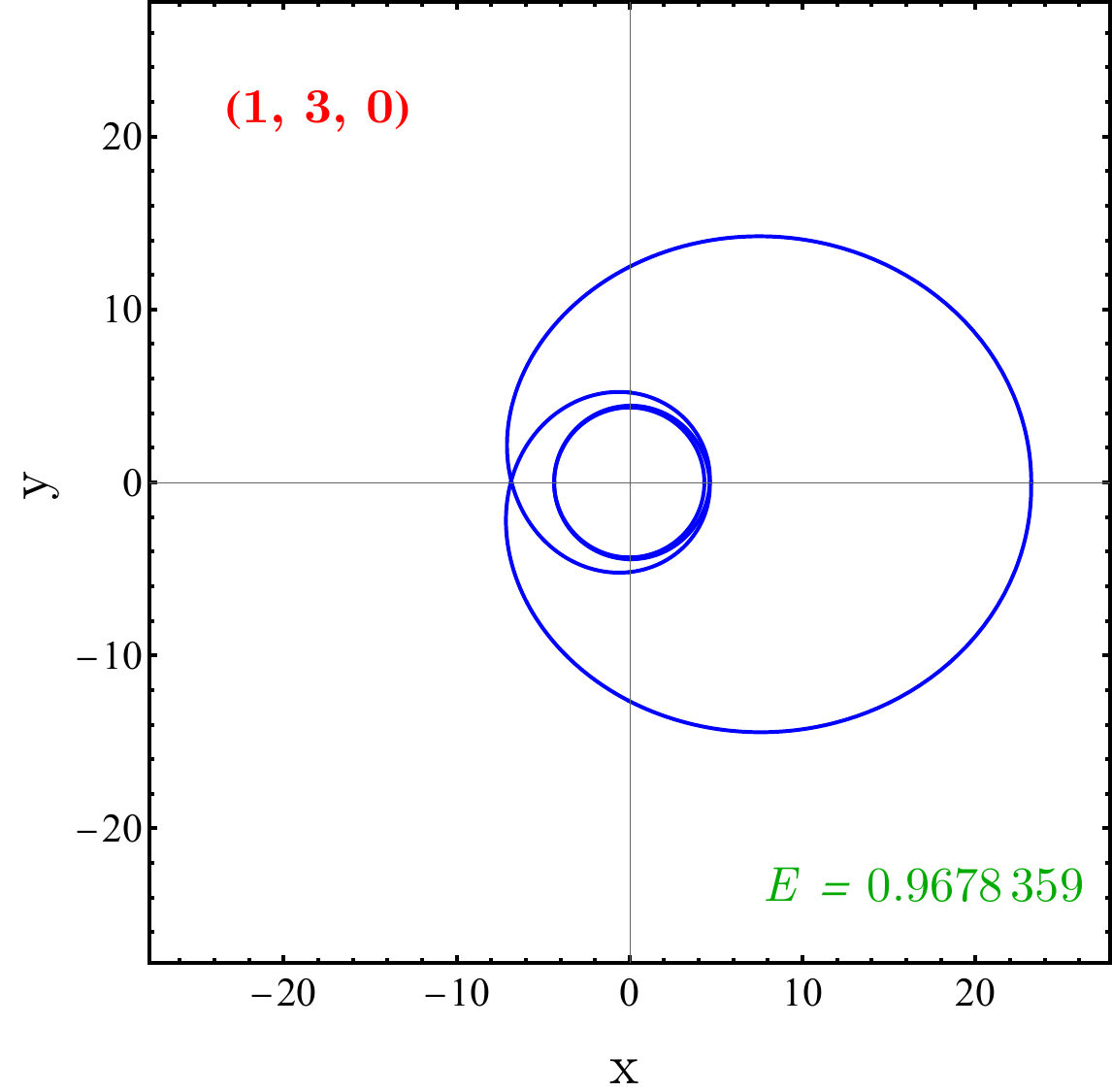} \\
        
        \includegraphics[width=0.3\textwidth]{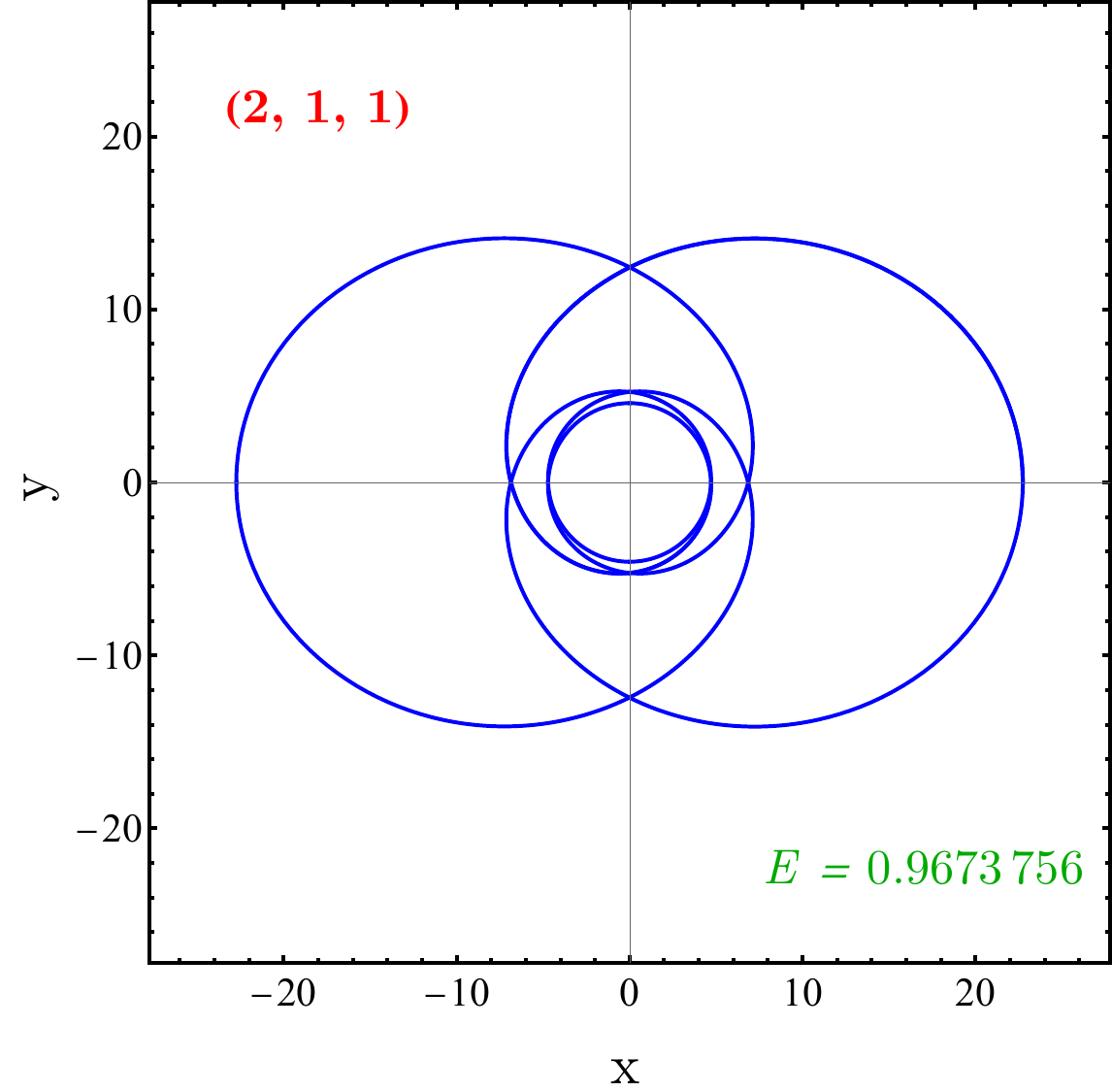} &
        \includegraphics[width=0.3\textwidth]{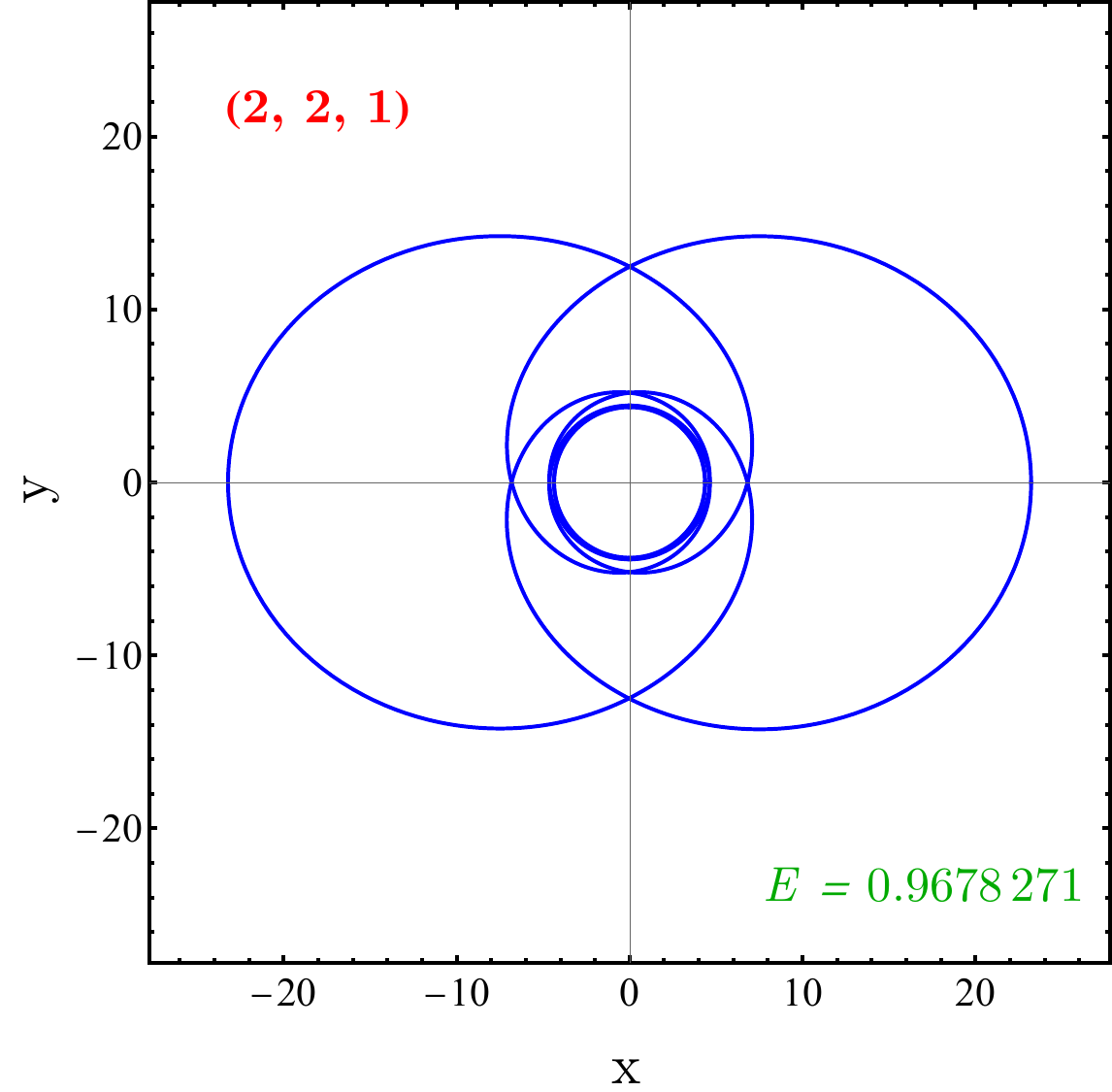} &
        \includegraphics[width=0.3\textwidth]{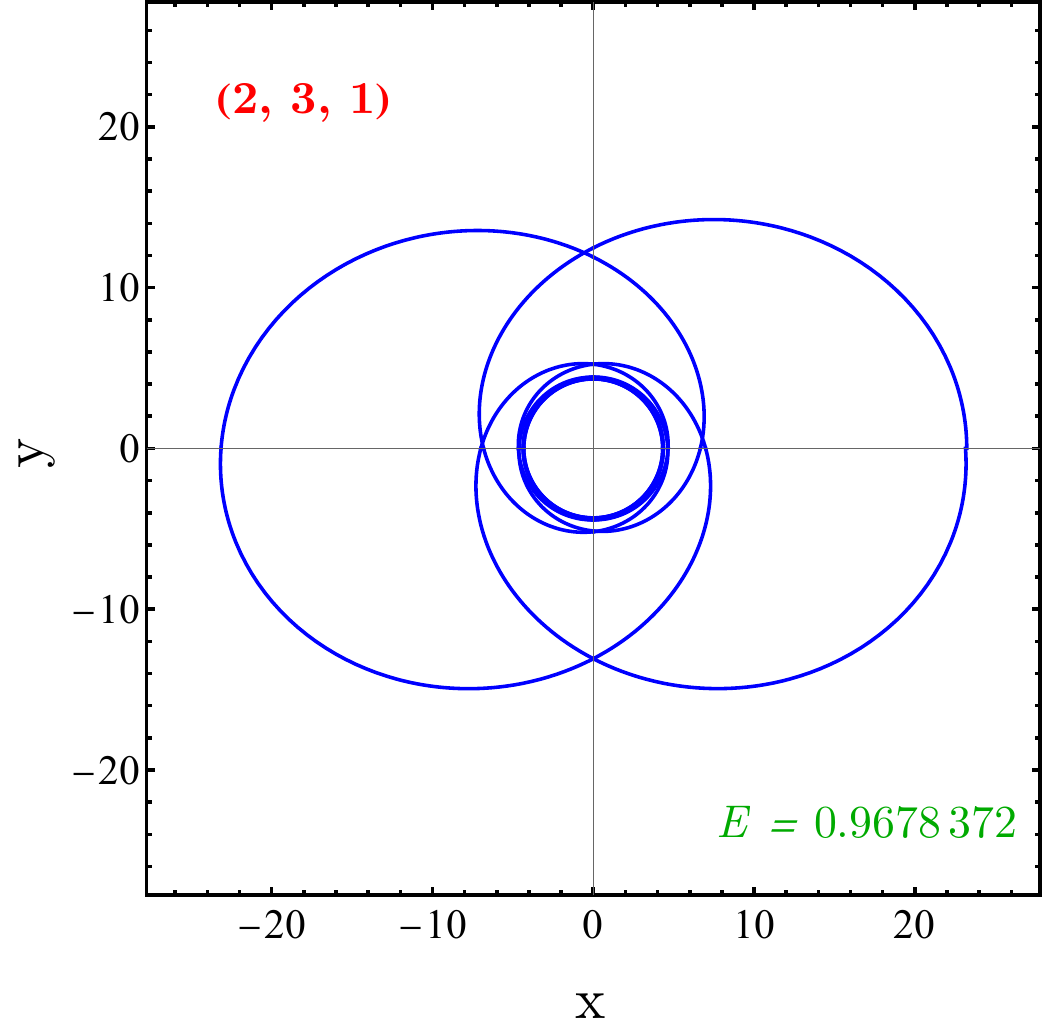} \\
        
         \includegraphics[width=0.3\textwidth]{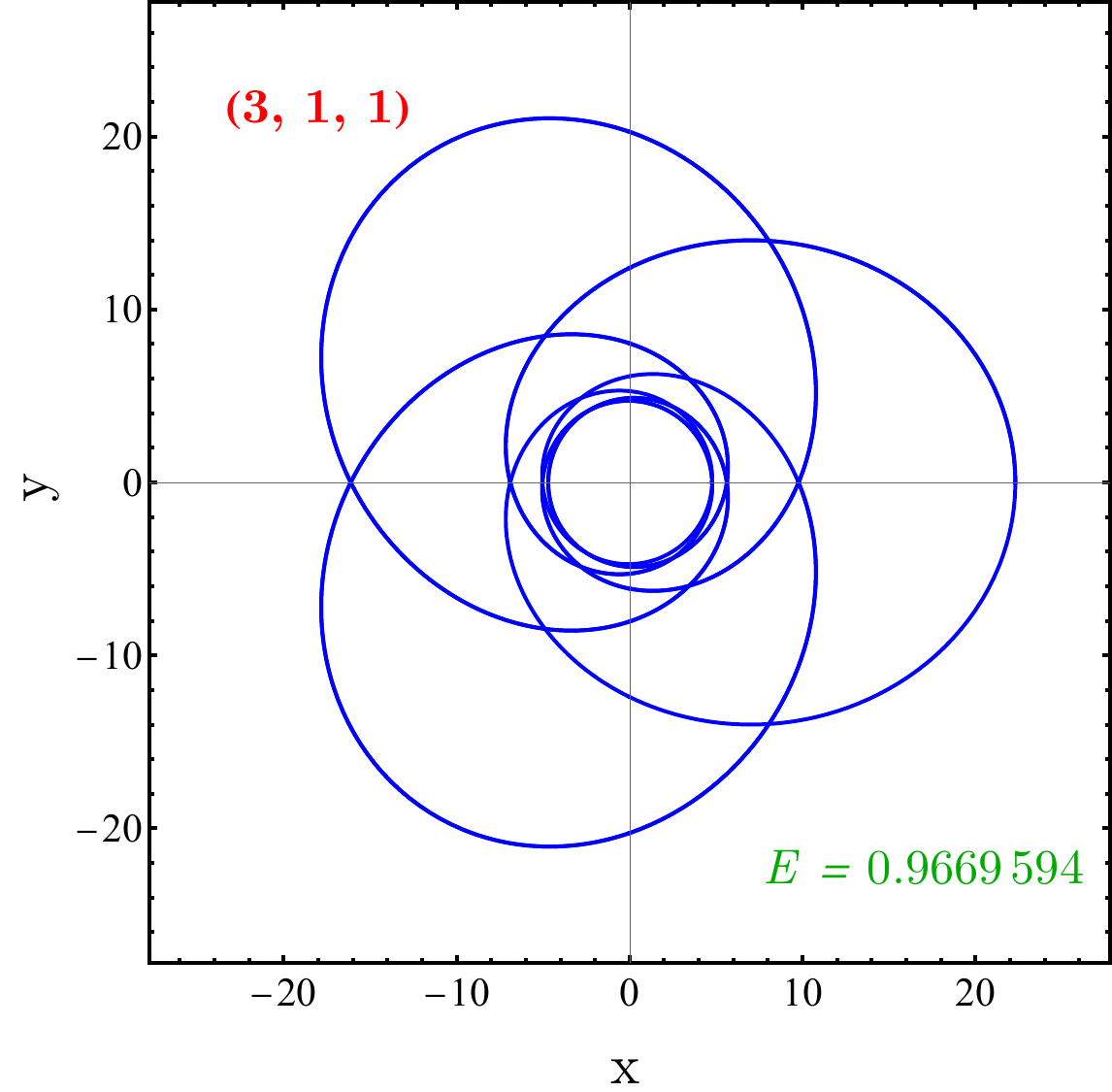} &
        \includegraphics[width=0.3\textwidth]{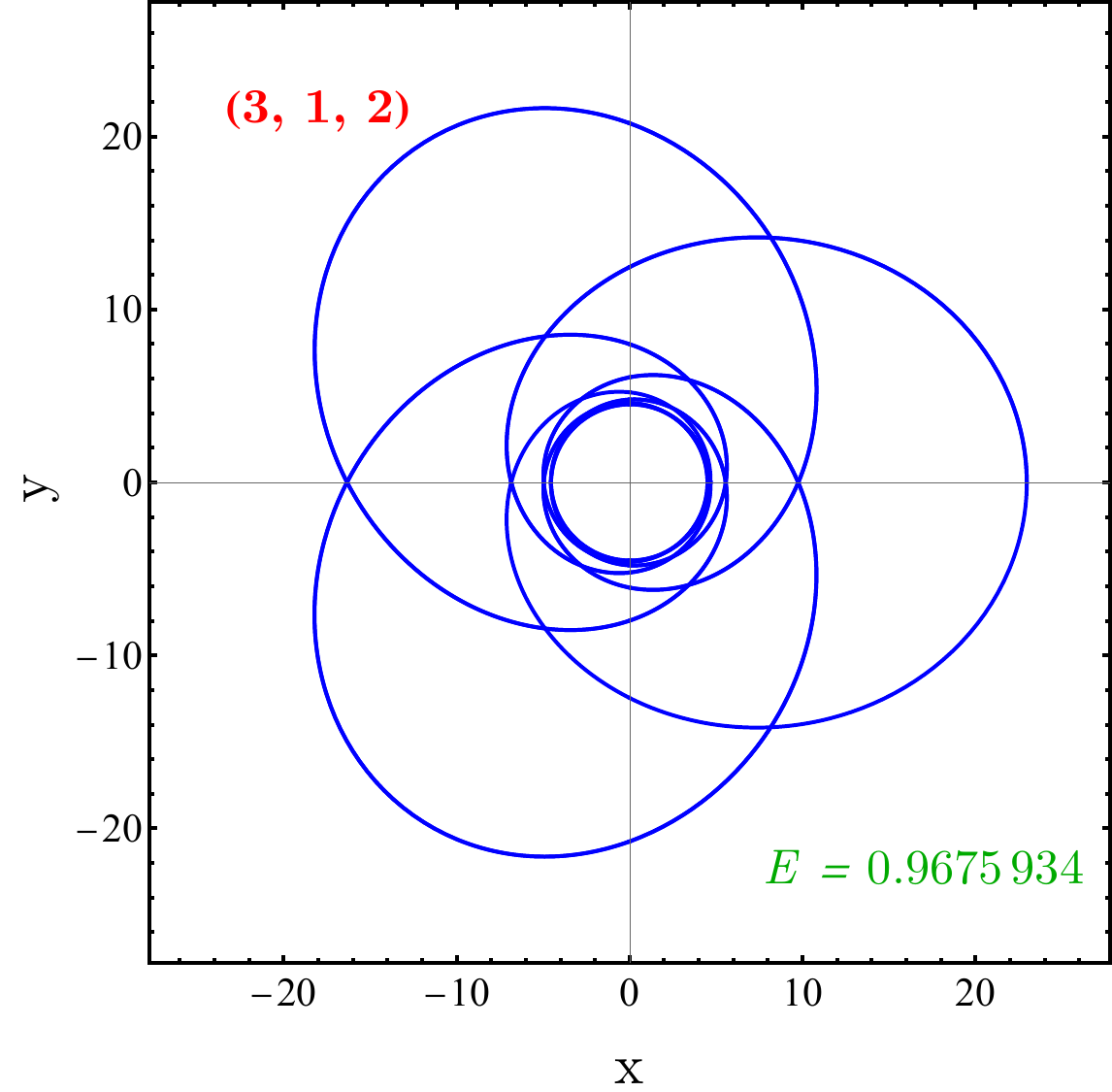} &
        \includegraphics[width=0.3\textwidth]{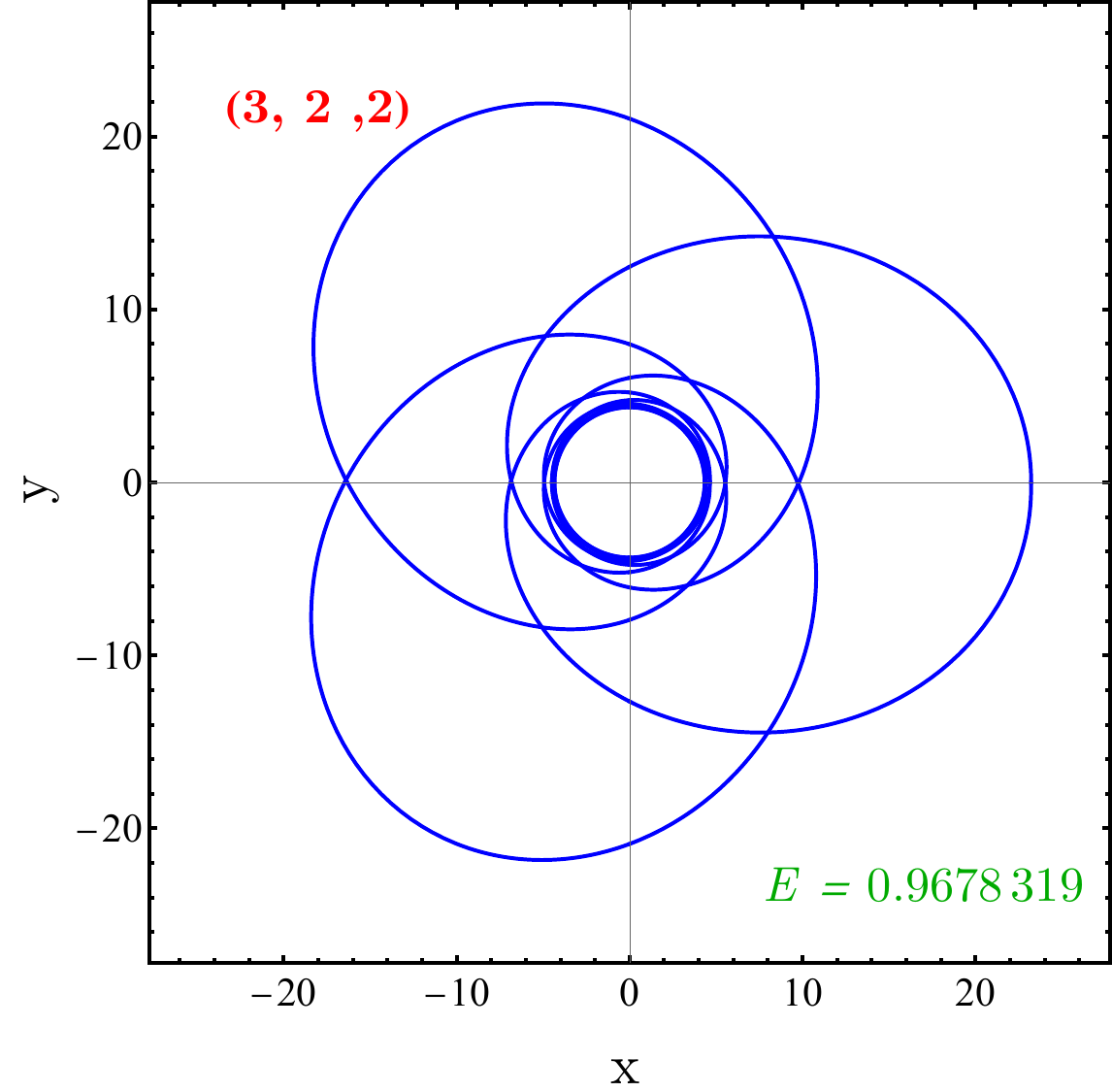} \\
        
         \includegraphics[width=0.3\textwidth]{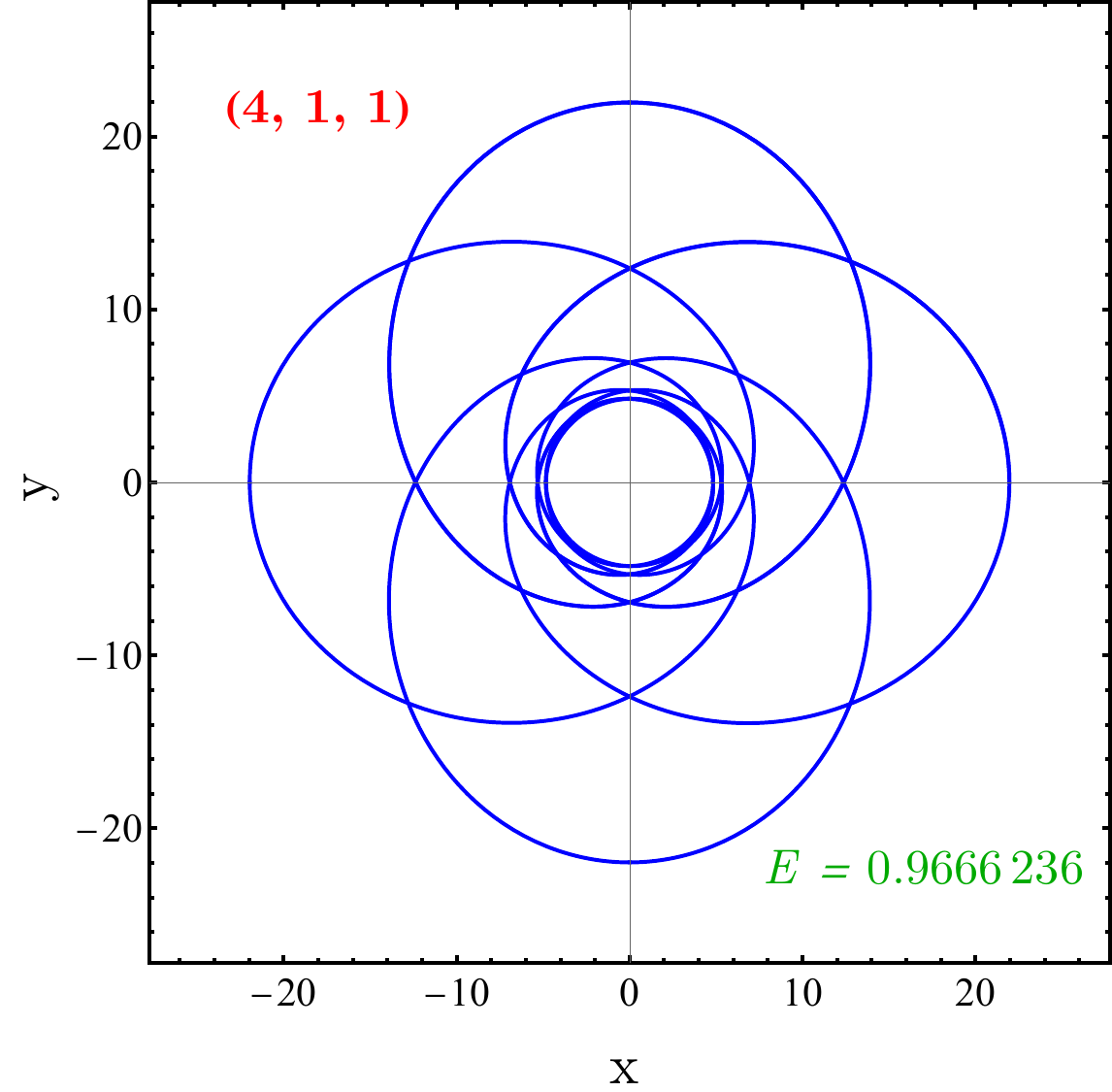} &
        \includegraphics[width=0.3\textwidth]{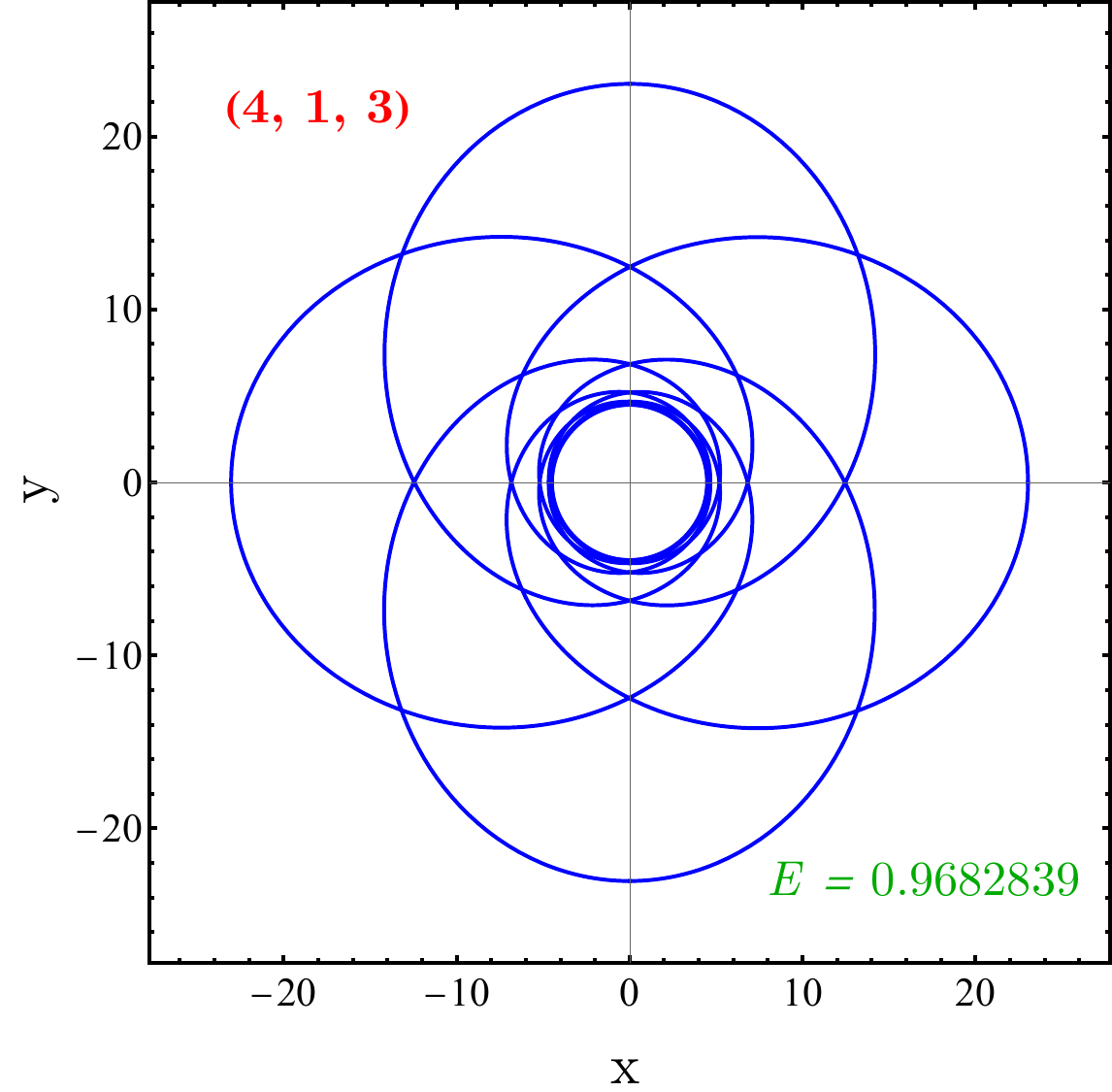} &
        \includegraphics[width=0.3\textwidth]{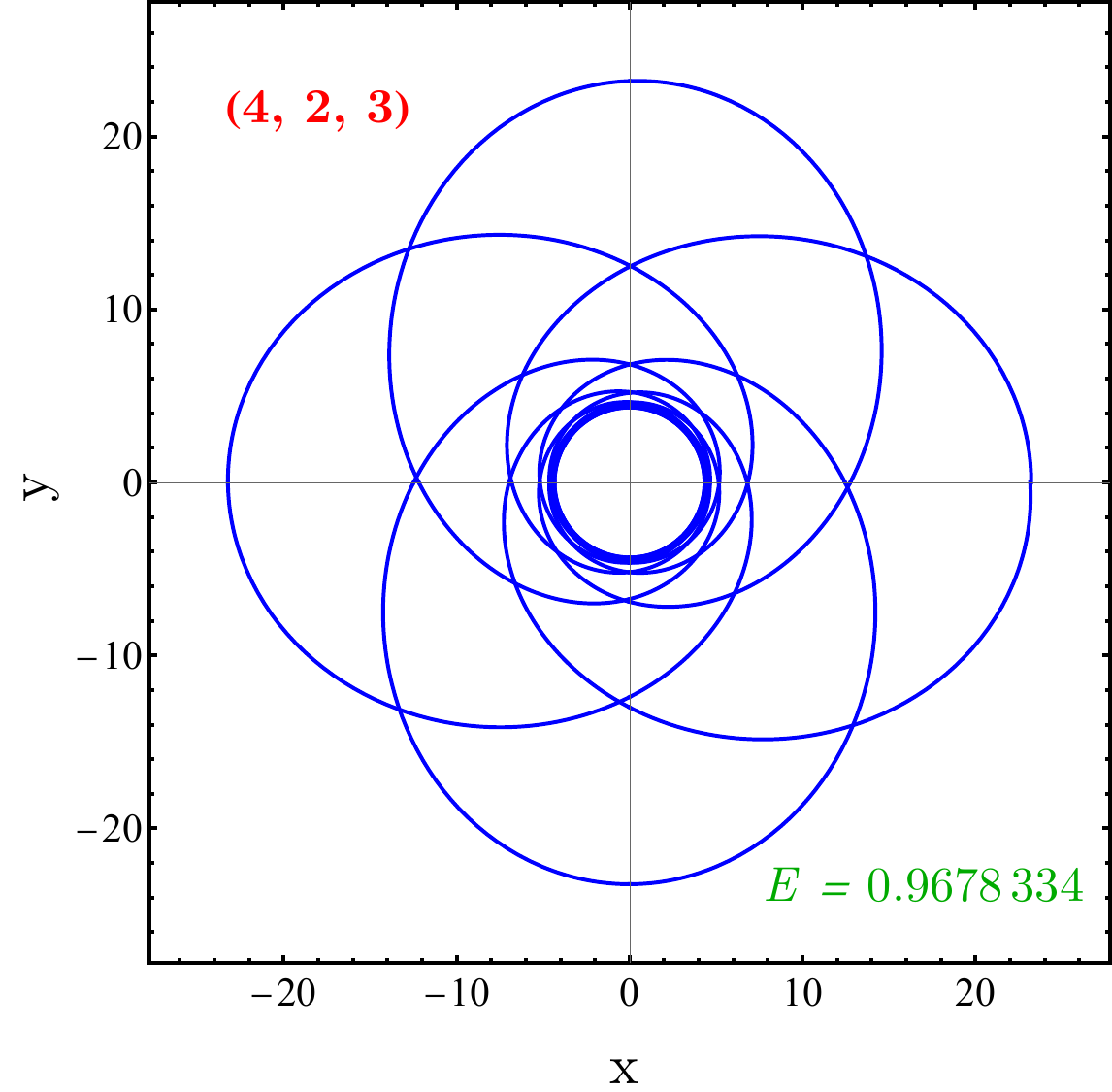} 
    \end{tabular}
    
   \caption{Periodic orbits characterized by different $(z, w, v)$ triples in the vicinity of an \textit{MHDM} BH. The spacetime parameters are fixed at $r_\mathrm{s} = \rho_\mathrm{s} = 0.2$ and $g = 0.5$.}
   
    \label{fig:allorbits}
\end{figure}

In Fig.~\ref{fig:allorbits}, we illustrate the trajectories of periodic orbits around the \textit{MDMH} black hole for various $(z, w, v)$ configurations, using the corresponding energies and averaged angular momenta. It can be noticed that in each column, increasing the zoom number $z$ yields progressively richer structural complexity in the periodic orbits. In contrast, a larger whirl number $w$, in each row, results in a higher number of revolutions around the central black hole between consecutive apoapsis points.

\begin{table}[htbp]
\centering
\caption{The values of the energy $E$ are tabulated for different periodic orbits characterized by $(z, w, v)$. The magnetic monopol charge $g$ varies from $0$ to $0.6$, and $r_\text{s}$ and $\rho_\mathrm{s}$ are fixed at $0.2$.}
\label{tab:qE1}
\small 
\setlength{\tabcolsep}{4pt} 
\begin{tabular}{|c|S[table-format=1.5]|*{8}{S[table-format=1.6]|}}
\hline
$g$ & {$L_\text{av}$} & {$E(1,1,0)$} & {$E(1,2,0)$} & {$E(2,1,1)$} & {$E(2,2,1)$} & {$E(3,1,2)$} & {$E(3,2,2)$} & {$E(4,1,3)$} & {$E(4,2,3)$} \\
\hline \hline
0.0   &{ 3.76770} & {0.965395} & {0.968356} &{0.968356}&{0.968408}&{0.968198}&{0.968412}&{0.968258}&{0.9684139} \\
\hline

0.2 & {3.76303} & {0.965295} & {0.968288} & {0.967926} & {0.968341} & {0.968128} & {0.968345} & {0.968189} & {0.968346}\\
\hline

0.4 & {3.74882}& {0.964880}& {0.967985}& {0.967605}& {0.968041}& {0.967816}& {0.968045}& {0.967880} & {0.968047}     \\
\hline
0.6   & {3.72453} & {0.964194} & {0.967496} & {0.967085} & {0.967558} & {0.967312} & {0.967563} & {0.967382} & {0.967564}     \\
\hline
\end{tabular}
\end{table}

\begin{table}[htbp]
\centering
\caption{Energy $E$ of periodic orbits characterized by different $(z, w, v)$ triples as a function of the scaled radius $r_\mathrm{s}$. The scaled radius varies while the \textit{Hernquist} dark matter halo density and the parameter $g$ are fixed at $\rho_\mathrm{s} = 0.5$ and $g = 0.1$, respectively. }
\label{tab:qE2}
\small 
\setlength{\tabcolsep}{4pt} 
\begin{tabular}{|c|S[table-format=1.5]|*{8}{S[table-format=1.6]|}}
\hline
$r_\text{s}$ & {$L_\text{av}$} & {$E(1,1,0)$} & {$E(1,2,0)$} & {$E(2,1,1)$} & {$E(2,2,1)$} & {$E(3,1,2)$} & {$E(3,2,2)$} & {$E(4,1,3)$} & {$E(4,2,3)$} \\
\hline \hline
0.0   & {3.73086} &{0.965393} & {0.968359} & {0.968002} & {0.968411} & {0.968201} & {0.968415} & {0.968261} & {0.968417} \\
\hline
0.2   & {3.82011} & {0.965323} & {0.968298} & {0.967939} & {0.968350} & {0.968139} & {0.968354} & {0.968199} & {0.968355}  \\
\hline
0.4   & {4.41967} & {0.964572} & {0.967637} & {0.967266} & {0.967692} & {0.967472} & {0.967697} & {0.967535} & {0.967698}    \\
\hline
0.6  & {6.03445} & {0.963120} & {0.966372} & {0.965973} & {0.966432} & {0.966194} & {0.966437} & {0.966262} & {0.966438}     \\
\hline
\end{tabular}
\end{table}

\begin{table}[htbp]
	\centering
	\caption{Energy $E$ of periodic orbits characterized by different $(z, w, v)$ triples as a function of the \textit{Hernquist} dark matter halo density parameter $\rho_\mathrm{s}$. The density varies from $0$ to $0.6$, while the scaled radius and the parameter $g$ are fixed at $r_\mathrm{s} = 0.5$ and $g = 0.1$, respectively. }
	\label{tab:qE3}
	\small
	\setlength{\tabcolsep}{4pt} 
	\begin{tabular}{|c|S[table-format=1.5]|*{8}{S[table-format=1.6]|}}
		\hline
		$\rho_\mathrm{s}$ & {$L_\text{av}$} & {$E(1,1,0)$} & {$E(1,2,0)$} & {$E(2,1,1)$} & {$E(2,2,1)$} & {$E(3,1,2)$} & {$E(3,2,2)$} & {$E(4,1,3)$} & {$E(4,2,3)$} \\
		\hline \hline
		0.0   &{3.73086} & {0.965393} & {0.968359} & {0.968002} & {0.968411} & {0.968201} & {0.968415} & {0.968261} & {0.968417} \\
		\hline
		
		0.2 & {4.25598} & {0.964557} & {0.967621} & {0.967249} & {0.967675} & {0.967456} & {0.967680} & {0.967519} & {0.967681}\\
		\hline
		
		0.4 & {4.79326} & {0.964038} & {0.967169} & {0.966787} & {0.967225} & {0.966999} & {0.967229} & {0.967063} & {0.967231}   \\
		\hline
		0.6   & {6.03445} & {0.963721} & {0.966895} & {0.966507} & {0.966953} & {0.966724} & {0.966957} & {0.966789} & {0.966959}     \\
		\hline
	\end{tabular}
\end{table}


\section{Precession of Periodic Orbits}
\begin{figure}[ht!]
	\centering
	\includegraphics[width=80mm]{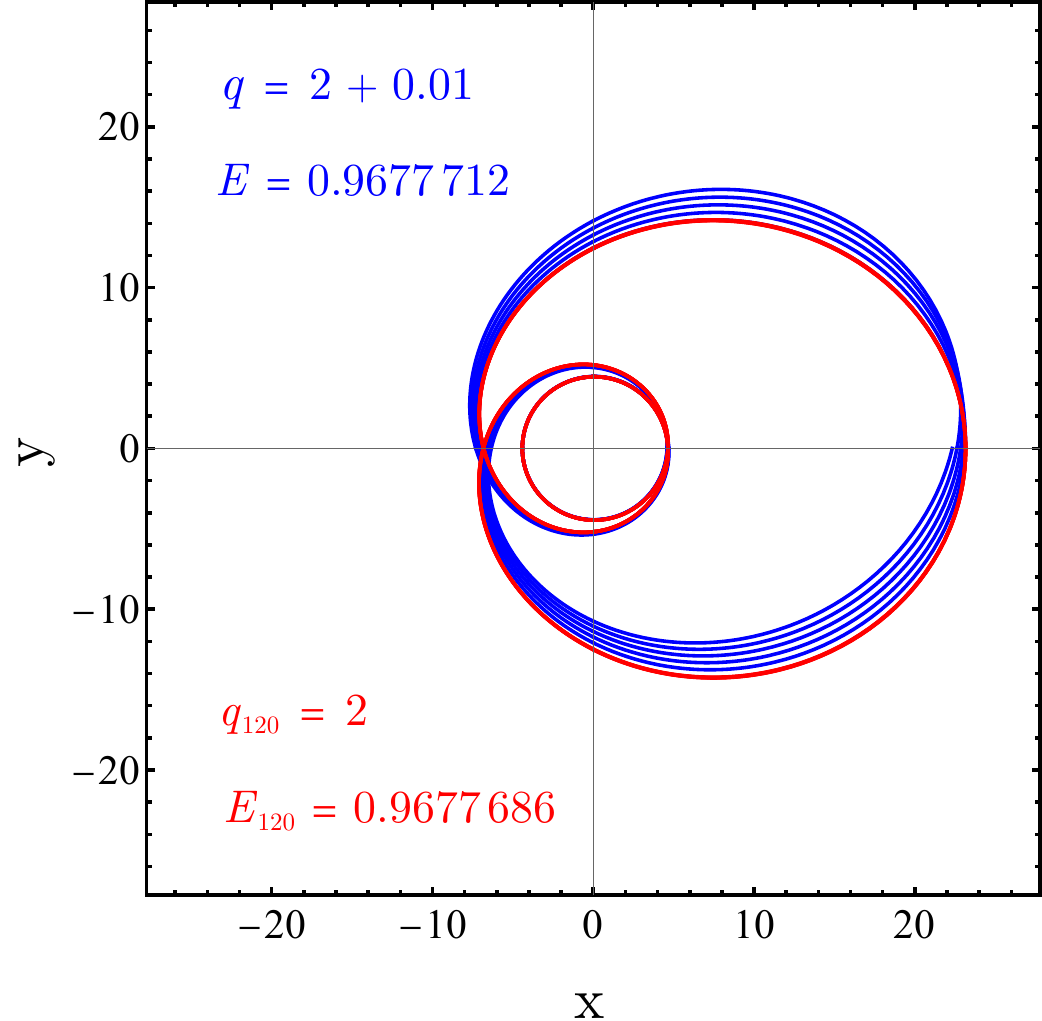}  \hspace{2mm}
	\includegraphics[width=80mm]{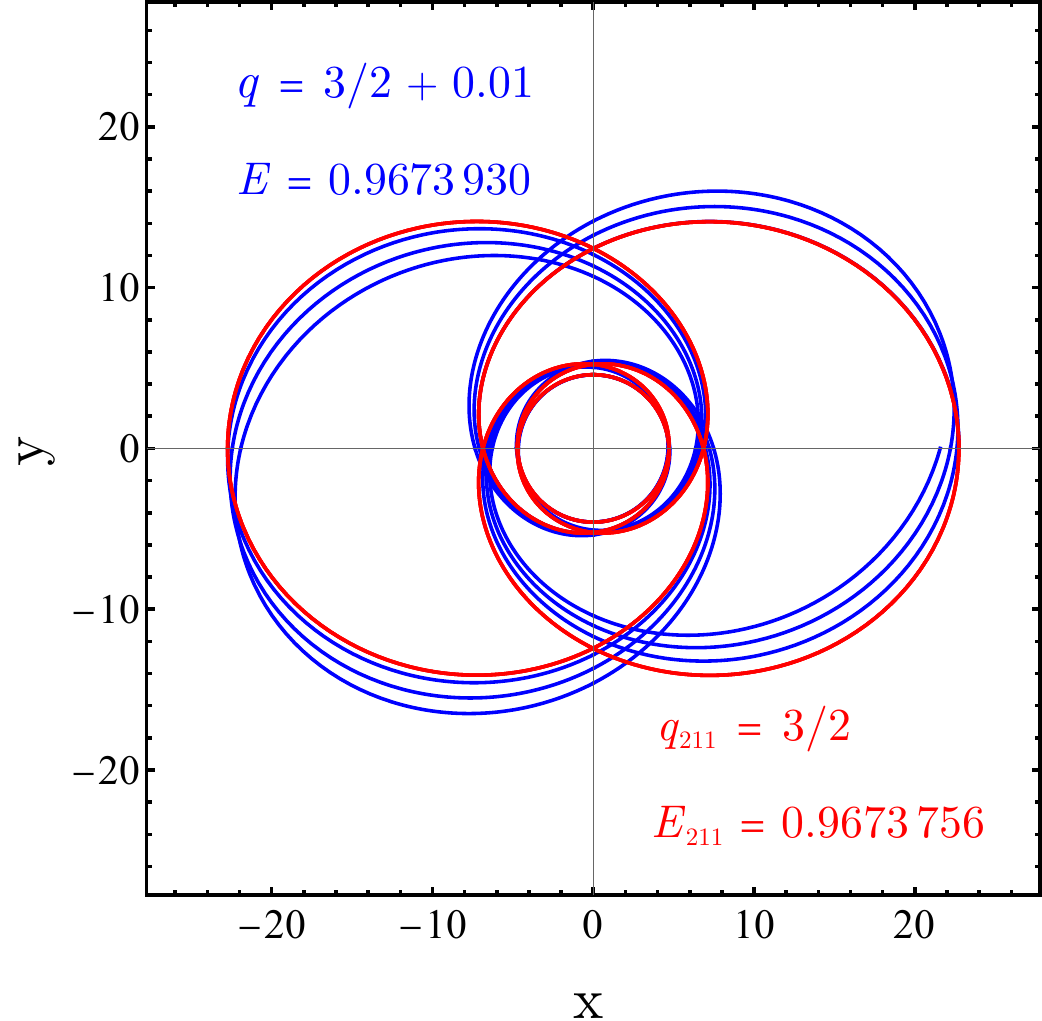}\newline \newline
	\includegraphics[width=80mm]{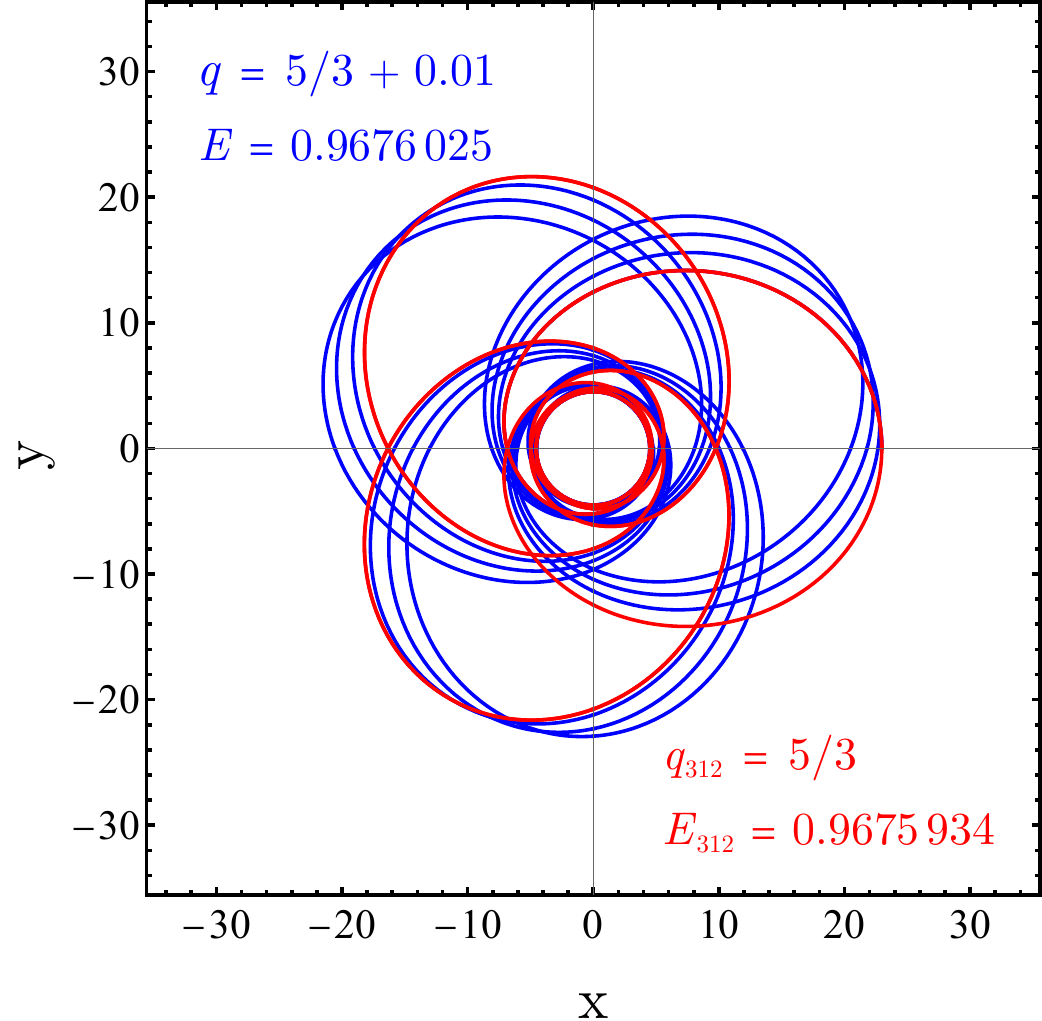} \hspace{2mm}
	\includegraphics[width=80mm]{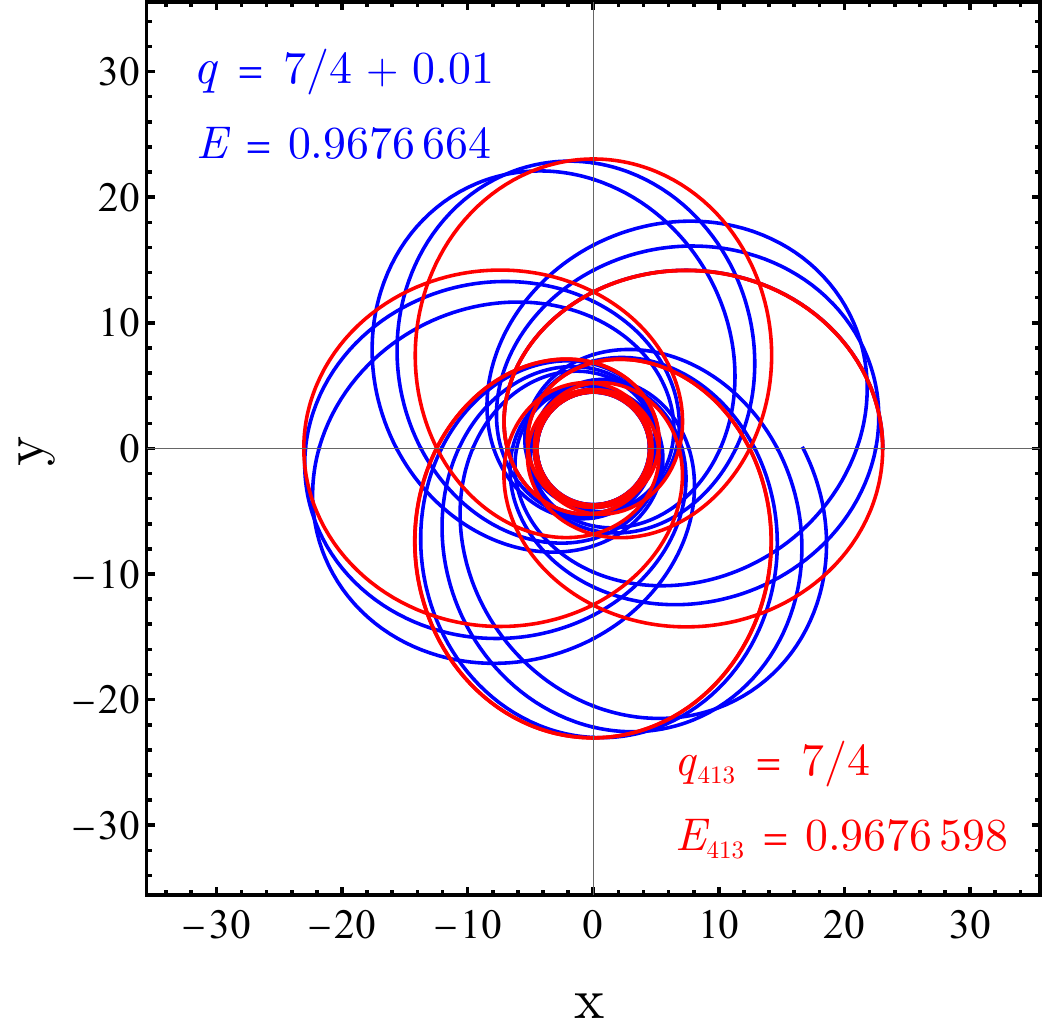}
	\caption{Precessing orbits (in blue) in the vicinity of periodic orbits (in red), computed for $r_\mathrm{s} = \rho_\mathrm{s} = 0.5$ and $g = 0.5$, illustrating the gradual secular shift in orbital orientation induced by small deviations from the exact periodic configuration.}
	\label{fig:prec}
\end{figure}

A fundamental distinction exists between periodic and generic orbits: periodic orbits yield rational values of $q$, whereas generic orbits yield irrational values and are thus non-periodic. Following the reasoning of Ref.~\cite{healy2009zoom}, one can approximate any generic orbit by a suitably chosen periodic orbit, drawing an analogy to the rational approximation of irrational numbers. When subjected to small perturbations, periodic orbits undergo precession, leading to a gradual shift in their orientation. In line with the treatment in Ref.~\cite{wang2022periodic,shabbir2025periodic}, such nearly periodic orbits can be expressed as $q = w +\frac{v}{z} \pm \delta$, where $\delta$ is a small positive parameter. We apply this formalism to generate precessing orbits near periodic solutions for different energies, with sample trajectories illustrated in Fig.~\ref{fig:prec}.
	
The orbital trajectories for four distinct periodic orbit configurations, namely $(1,2,0)$, $(2,1,1)$, $(3,1,2)$, and $(4,1,3)$ are presented in Fig.~\ref{fig:prec}. For each configuration, we examine the effect of small deviations from the exact periodic energy, with perturbations as small as $\sim 10^{-4}$ in magnitude. These infinitesimal energy variations give rise to precessing orbits, depicted in blue, while the exact periodic orbits and their associated parameters are shown in red. In all cases, the angular momentum is held fixed at the average value $L_{\text{av}}$ corresponding to each periodic orbit. The results clearly demonstrate that even minute departures from the precisely tuned energy lead to a gradual rotation of the orbital ellipse, which shows the sensitivity of the system to initial conditions, as we should expect.

\section{Gravitational waveform from periodic orbits}

In this section, we investigate the gravitational waveforms generated by periodic orbits around a magnetically charged black hole immersed in the \textit{Hernquist} dark matter halo. The system under consideration constitutes an extreme mass--ratio inspiral (EMRI), wherein a stellar--mass compact object follows a periodic trajectory around a supermassive black hole embedded within a dark matter distribution. Gravitational waves emitted by such EMRI systems encode valuable information regarding both the orbital dynamics and the properties of the central black hole and its surrounding dark matter environment \cite{cardoso2022gravitational,rahman2024probing}. To model the gravitational radiation, we adopt the adiabatic approximation, a widely employed approach in EMRI waveform calculation, where the orbital parameters evolve on a timescale significantly longer than the orbital period, allowing the motion to be treated as geodesic over a limited number of cycles \cite{hughes2000evolution,hughes2001evolution}. Consequently, the energy and angular momentum can be regarded as approximately constant over a single orbital period. For the purpose of computing waveforms over one complete cycle, we therefore neglect the back--reaction effects of gravitational radiation on the orbital motion.

\subsection{Time domain analysis}
\label{sec:time-domain}

We employ the numerical kludge waveform methodology~\cite{babak2007kludge}, which proceeds in two stages. First, we numerically integrate the equations of motion governing the small object's trajectory, as given in Eq.~\eqref{eq:t_dot}~--~\eqref{eq:rdot}. Subsequently, we construct the gravitational waveforms using the symmetric and trace--free (STF) mass quadrupole formalism. The quadrupole formula, expressed to second order, takes the form~\cite{zhao2025periodic}
\begin{eqnarray}
	h_{ij} = \frac{4\mu M}{D_L} \left( v_i v_j - \frac{m}{M} n_i n_j \right),
	\label{eq:quadrupole}
\end{eqnarray}
where $m$ and $M$ denote the masses of the small object and the central black hole, respectively, with $m \ll M$ characteristic of the EMRI regime. The luminosity distance is represented by $D_L$, and $\mu = Mm/(M+m)^2$ is the symmetric mass ratio. The vectors $n_i$ and $v_i$ correspond to the unit position vector and velocity components of the small object. This expression encapsulates the fundamental quadrupolar emission properties of gravitational waves in the far--zone radiation field.

To align the computed signal with realistic detector responses, we project the gravitational waveform onto a detector--adapted coordinate system. Following Refs.~\cite{poisson2014gravity,zhao2025periodic,yang2025gravitational}, we introduce an orthonormal basis $(e_X, e_Y, e_Z)$ oriented with respect to the detector as
\begin{eqnarray}
	e_X &=& [\cos \zeta,\; -\sin \zeta,\; 0], \nonumber\\
	e_Y &=& [\cos \iota \sin \zeta,\; \cos \iota \cos \zeta,\; -\sin \iota], \nonumber\\
	e_Z &=& [\sin \iota \sin \zeta,\; \sin \iota \cos \zeta,\; \cos \iota].
	\label{eq:basis}
\end{eqnarray}
Here, $\zeta$ denotes the longitude of the periastron, while $\iota$ represents the inclination angle between the orbital plane and the line of sight. Within this basis, the two independent polarization modes $h_+$ and $h_\times$ are expressed as~\cite{maselli2022detecting,liang2023probing}
\begin{eqnarray}
	h_+ &=& -\frac{2\mu M^2}{D_L r} \left(1 + \cos^2 \iota\right) \cos\left(2\phi + 2\zeta\right),
	\label{eq:hplus}\\
	h_\times &=& -\frac{4\mu M^2}{D_L r} \cos \iota \sin\left(2\phi + 2\zeta\right),
	\label{eq:hcross}
\end{eqnarray}
where $\phi$ is the orbital phase angle. These expressions provide the time--domain gravitational waveforms that would be detected by a laser interferometer such as LISA.

For our numerical simulations, we adopt geometrized units with $G = c = 1$. In these units, the luminosity distance is set to $D_L = 4.18 \times 10^{14}$, which corresponds to approximately $200\ \text{Mpc}$ for a $10^6 M_\odot$ black hole. The small object mass is taken as $m = 10^{-6}M$, consistent with the extreme mass--ratio condition. The angular parameters are fixed at $\zeta = \iota = \pi/4$, representing a moderately inclined orbit with respect to the line of sight. The gravitational waveforms are computed over one complete period of the periodic orbit, with the orbital phase $\varphi$ obtained from the numerical integration of the geodesic equations.

\begin{figure}[ht!]
    \centering
    \begin{tabular}[t]{cc}
        \includegraphics[width=0.48\textwidth]{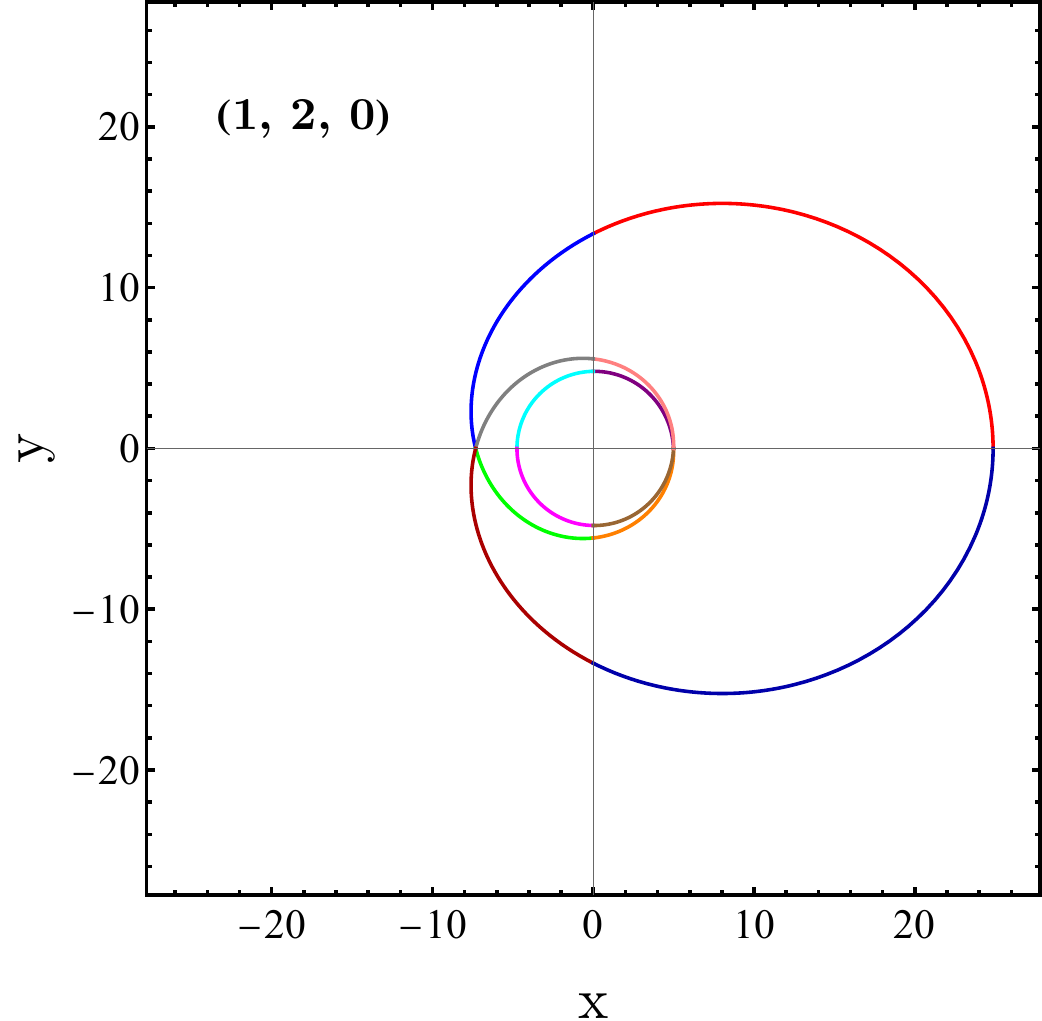}
        &
        \raisebox{4.4cm}{
        \begin{tabular}[t]{c}
            \includegraphics[width=0.48\textwidth]{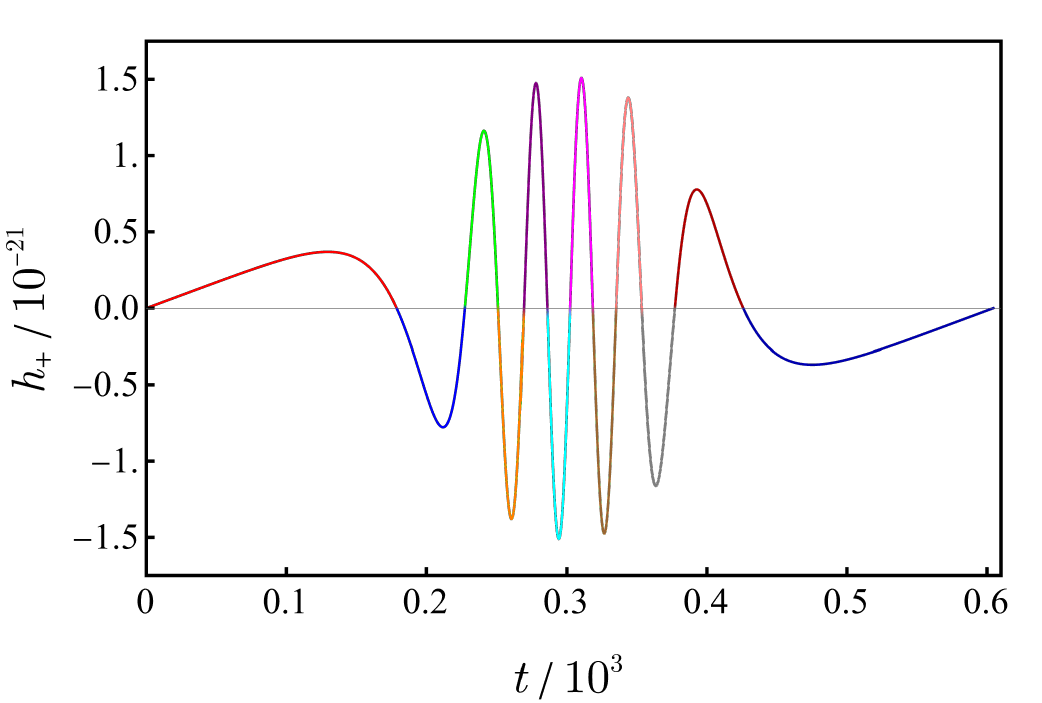} \\
            \vspace{-1cm} \\
            \includegraphics[width=0.48\textwidth]{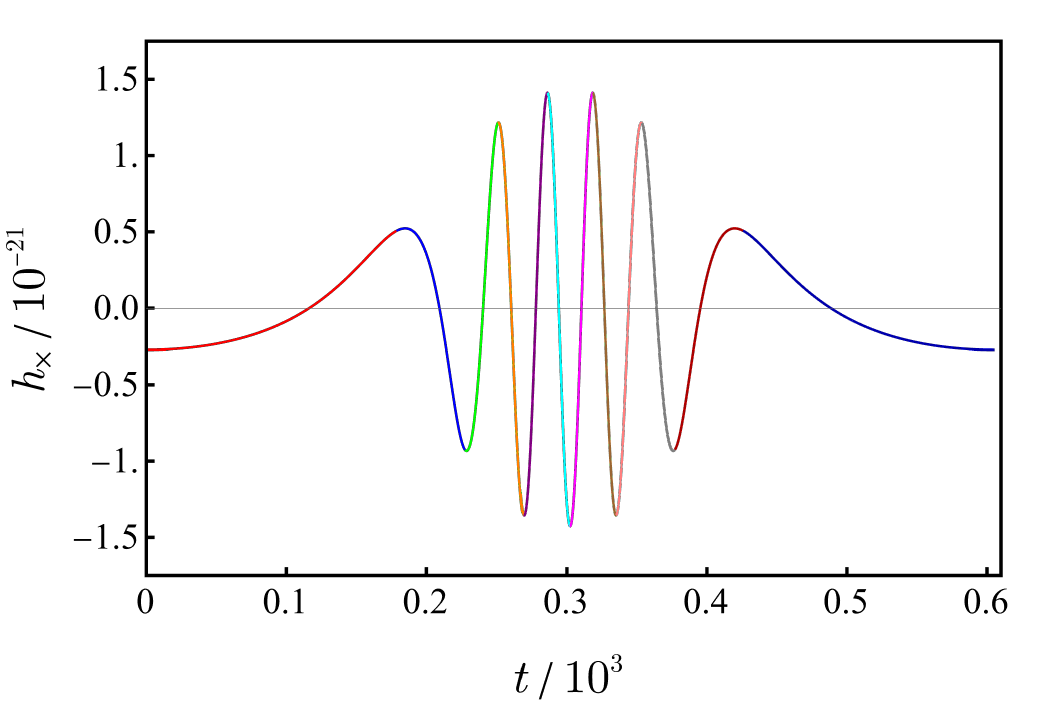}
        \end{tabular}}
    \end{tabular}
    \caption{The $(1,2,0)$ periodic orbit in the EMRI system, together with its associated gravitational waveform for $M = 1$, $r_{\mathrm{s}} = 0.5$, $\rho_{\mathrm{s}} = 0.1$, and $g = 0.1$. Distinct segments of the orbit are highlighted using different colors.\label{figss:colored}}
\end{figure}

Fig.~\ref{figss:colored} presents a phase--resolved visualization of the orbital motion and the associated gravitational wave emission for the \textit{MHDM} BH configuration. The adopted color scheme, based on successive $\pi/2$ intervals of the orbital phase, provides a direct mapping between segments of the trajectory and their corresponding contributions to the waveforms $h_+$ and $h_\times$, which shows a clear correlation between the gravitational wave and the particle’s position along the orbit. In particular, the amplitude is enhanced when the particle approaches the central region, where the gravitational field is strongest, and decreases as the particle moves toward larger radii. This results in a sequence of pronounced peaks associated with close passages, interspersed with lower amplitude intervals during the outward motion.
This behavior produces a characteristic modulation in both amplitude and frequency across a single orbital cycle. As the waveforms encode the essential features of the orbital dynamics in the \textit{MHDM} spacetime, a detailed comparison with other background geometries is represented in the following figure.

\begin{figure}[ht!]
	\centering
	\begin{minipage}{0.48\textwidth}
		\centering
		\includegraphics[width=\textwidth]{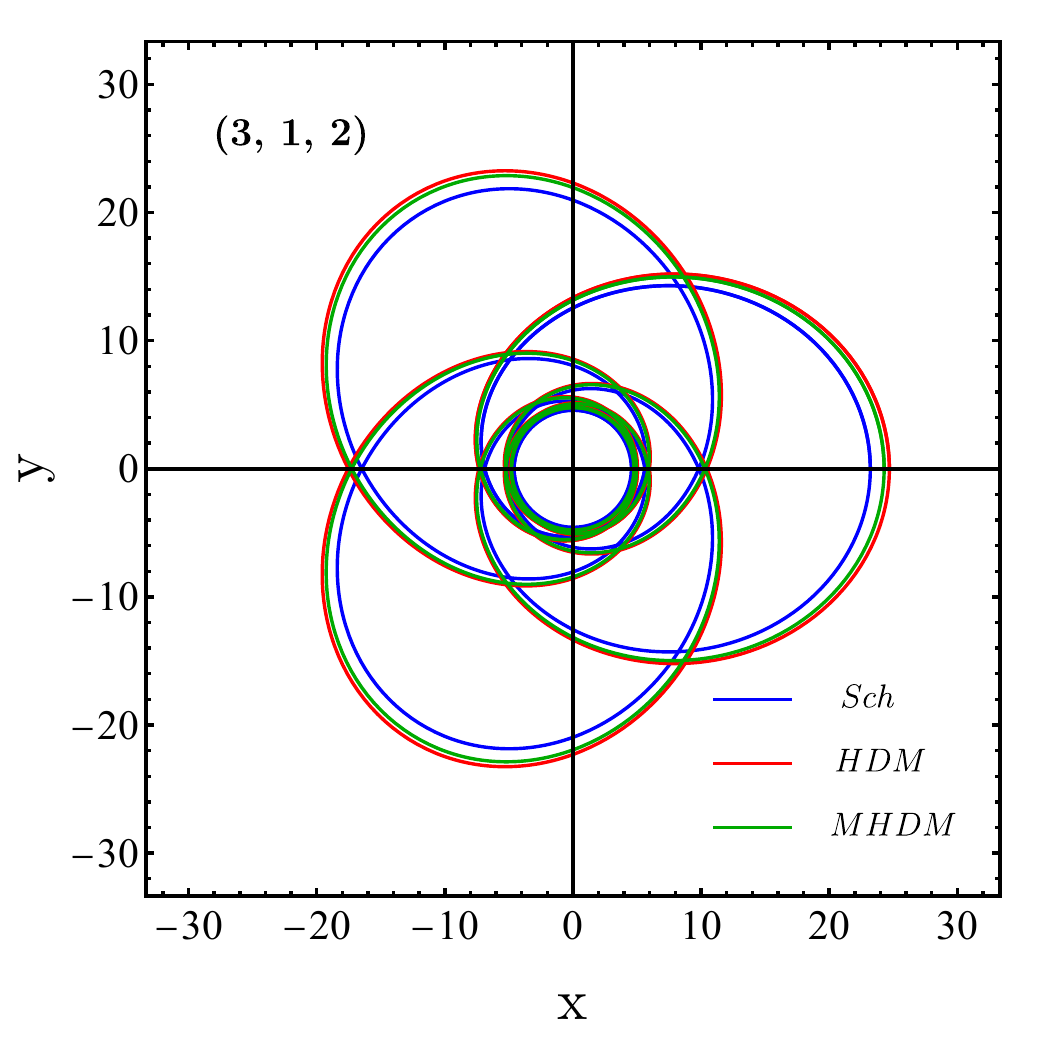}
	\end{minipage}
	\hfill
	\begin{minipage}{0.48\textwidth}
		\centering
		\includegraphics[width=\textwidth]{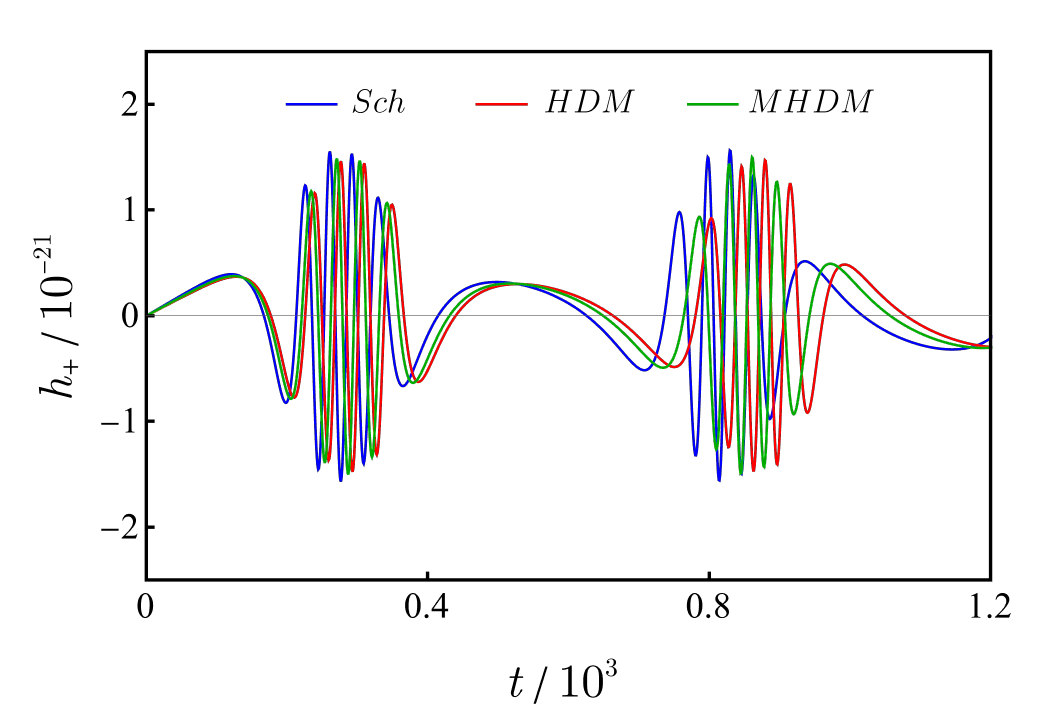}\\[0.5cm]
		\includegraphics[width=\textwidth]{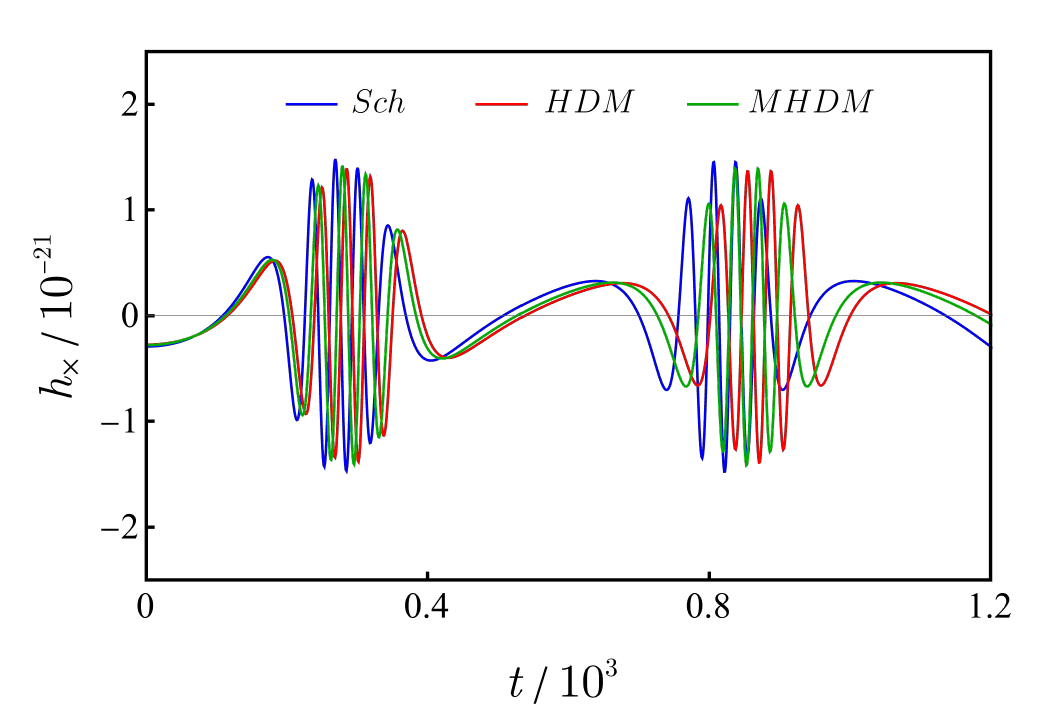}
	\end{minipage}
	\caption{Orbital trajectories (left) and gravitational waveforms $h_+$ (top right) and $h_\times$ (bottom right) for \textit{Sch} (blue), \textit{HDM} (red), and \textit{MHDM} (green) black holes. Each model is uniformly color-coded across trajectories and waveforms. Parameters: $M = 1$, $r_\mathrm{s} = 0.5$, $\rho_\mathrm{s} = 0.1$, $g = 0.5$.}
	\label{fig:comp}
\end{figure}

The results displayed in Fig.~\ref{fig:comp} provide a direct comparison between orbital dynamics and the associated gravitational wave emission in \textit{Sch}, \textit{HDM}, and \textit{MHDM} spacetimes. The left panel shows the particle trajectories, while the right panels present the corresponding waveform polarizations $h_+$ and $h_\times$. The comparison between Schwarzschild and \textit{Hernquist} spacetimes reveals a noticeable modification of the orbital structure. In particular, the periodic orbits in the \textit{Hernquist} background exhibit larger spatial extension compared to the Schwarzschild case. This behavior can be attributed to the modification of the effective gravitational potential induced by the surrounding matter distribution.
These changes in the orbital dynamics are directly reflected in the emitted gravitational waveforms. In the presence of dark matter, both polarization modes, $h_+$ and $h_\times$, display a reduction in amplitude relative to the Schwarzschild case. Furthermore, the temporal structure of the waveforms is also affected. The oscillation period of the gravitational waves increases in the \textit{Hernquist} spacetime, reflecting the longer orbital timescale associated with the enlarged trajectories. On the other hand, when the black hole carries a magnetic charge, the impact of the dark matter distribution is partially suppressed. In this case, the orbital configuration becomes more compact compared to the pure \textit{Hernquist} scenario, approaching the Schwarzschild limit. Consequently, the gravitational wave amplitude increases relative to the \textit{HDM} case, indicating a partial restoration of the emission efficiency.
A similar trend is observed in the waveform periodicity. The presence of the magnetic charge reduces the oscillation period, shifting it closer to the Schwarzschild value. This behavior suggests that the parameter $g$ counteracts the dispersive effect introduced by the dark matter halo, effectively driving the system back toward the vacuum geometry.

\begin{figure}[ht!]
    \centering
    \includegraphics[width=55mm]{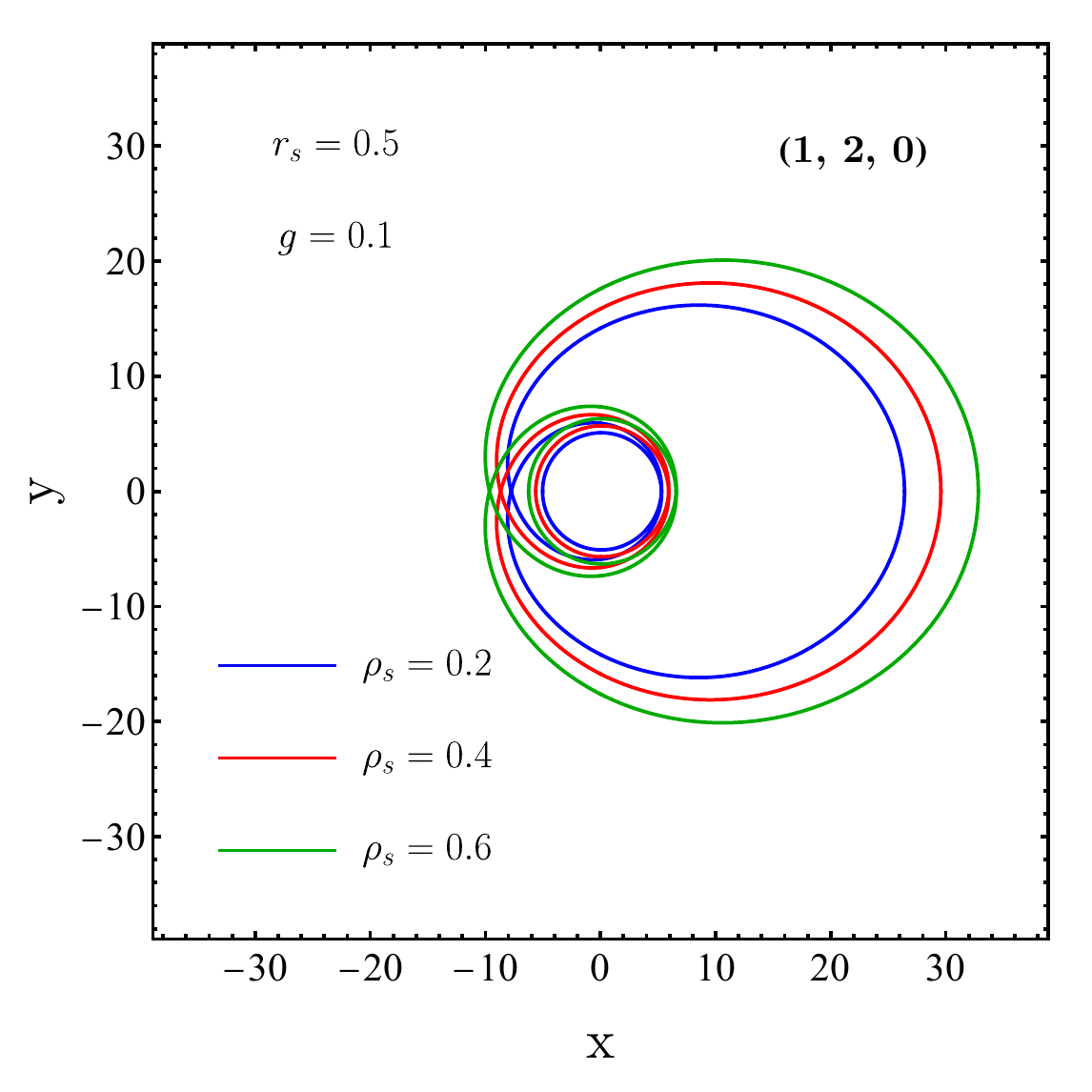}
    \includegraphics[width=55mm]{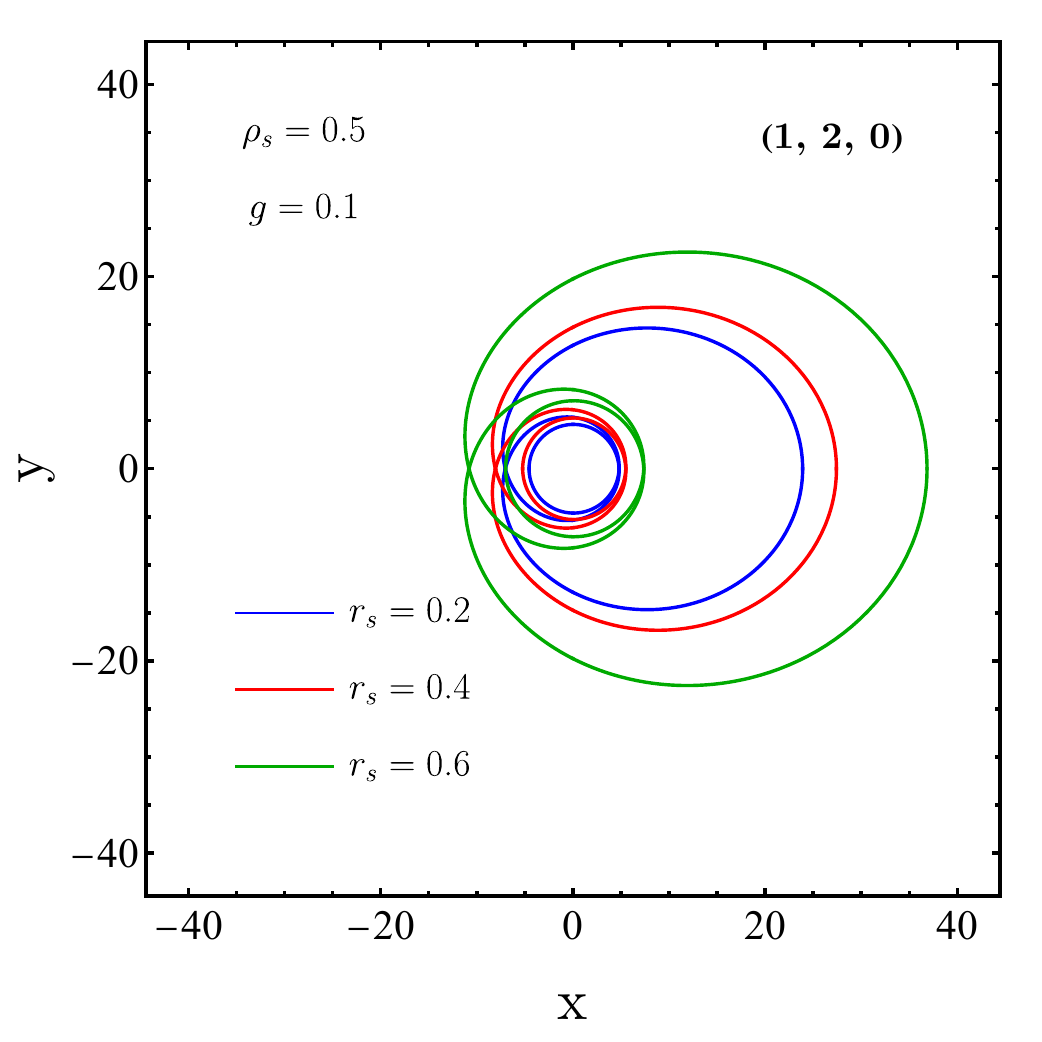}
    \includegraphics[width=55mm]{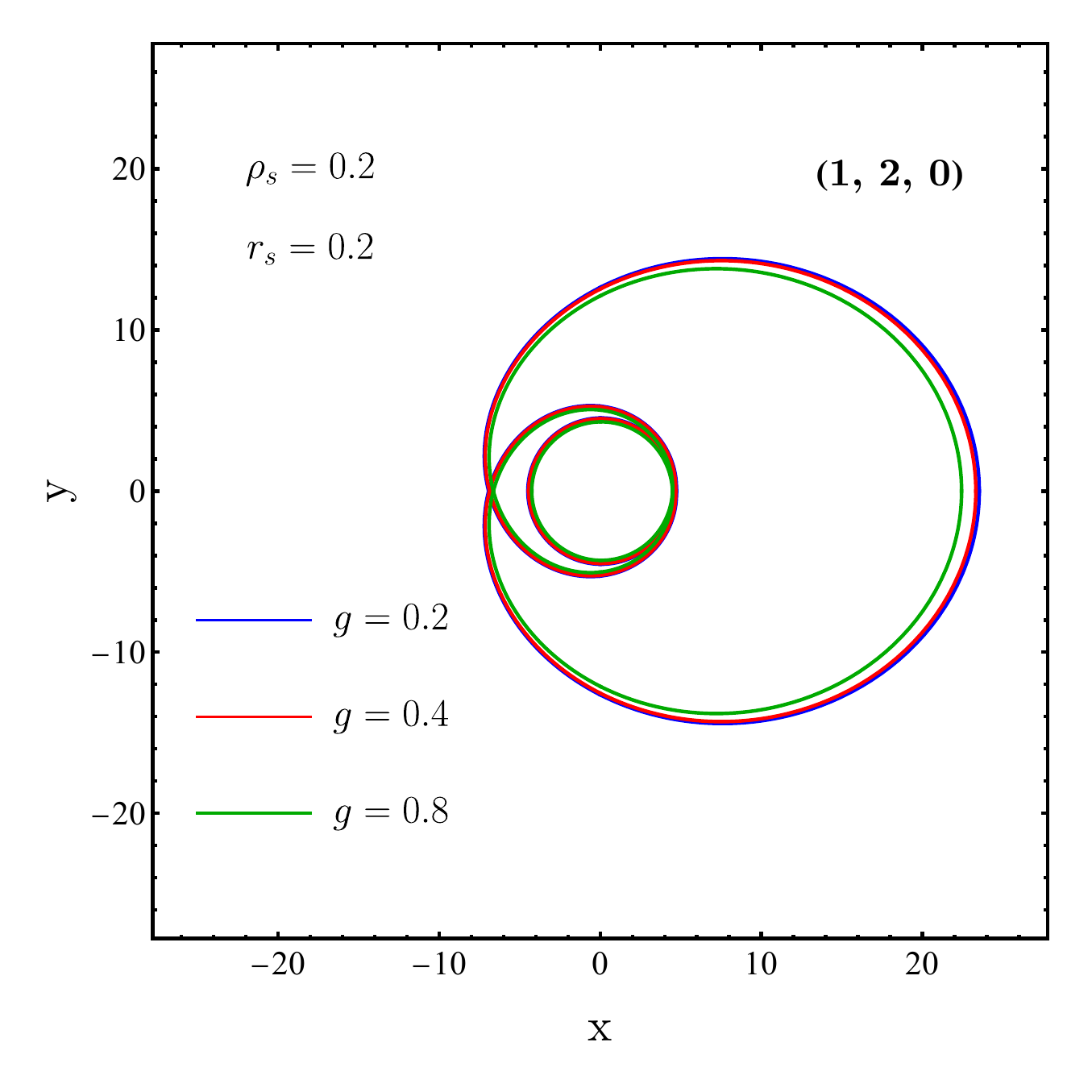}
    \caption{Periodic orbits in the spacetime of \textit{MHDM} BH, illustrating the influence of the black hole parameters for $(z, w, v) = (1, 2, 0)$. The left panel shows the effect of the dark matter halo density parameter $\rho_\mathrm{s}$, the central panel corresponds to the scale radius $r_\mathrm{s}$, and the right panel displays the impact of the magnetic monopole charge parameter $g$.}
    \label{fig:comporbit}
\end{figure}

\begin{figure}[ht!]
    \centering
    \includegraphics[width=55mm]{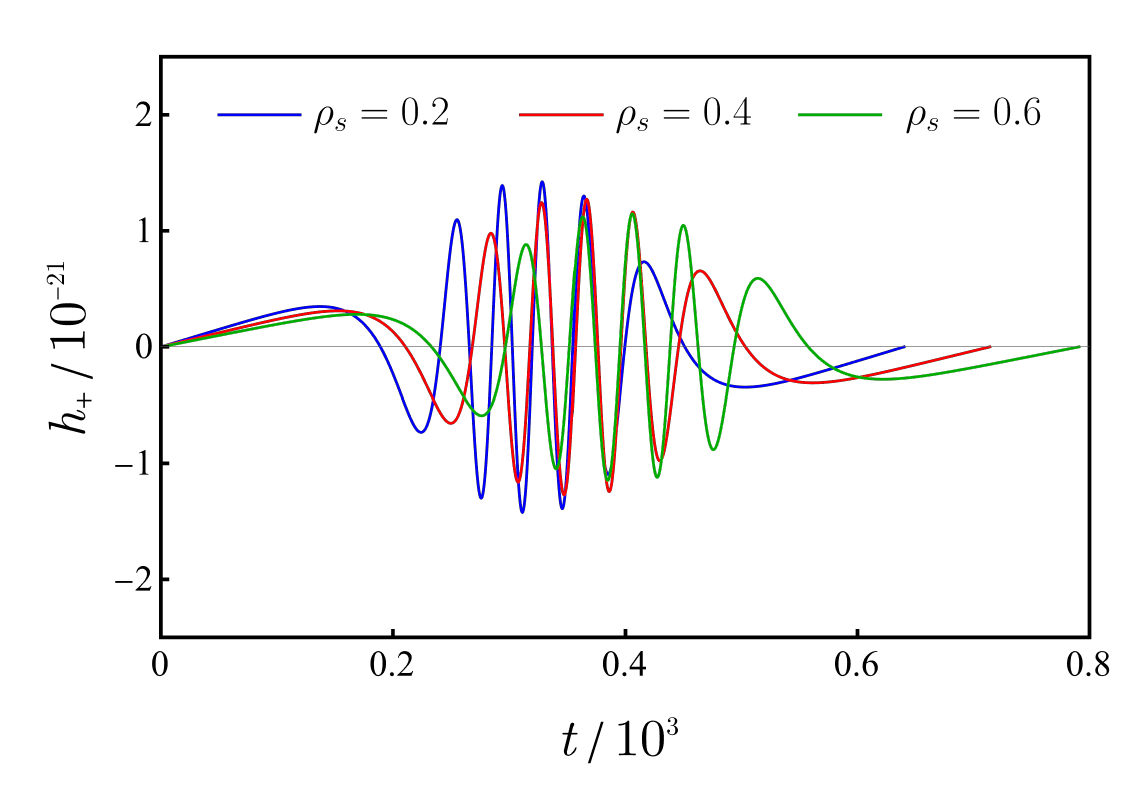}
    \includegraphics[width=55mm]{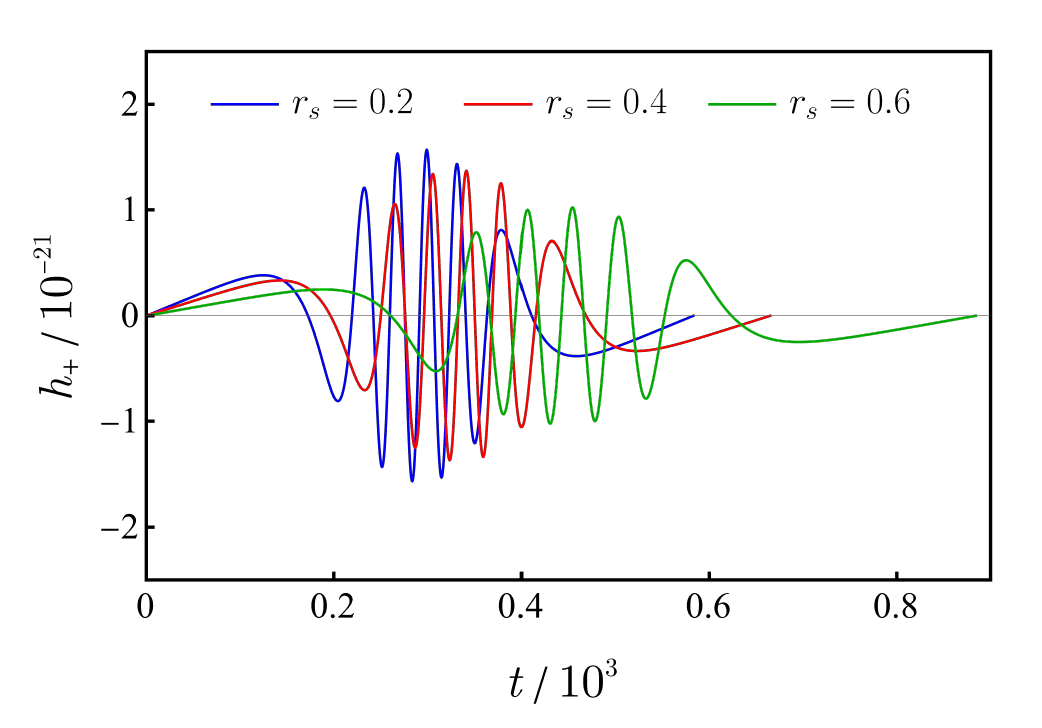}
    \includegraphics[width=55mm]{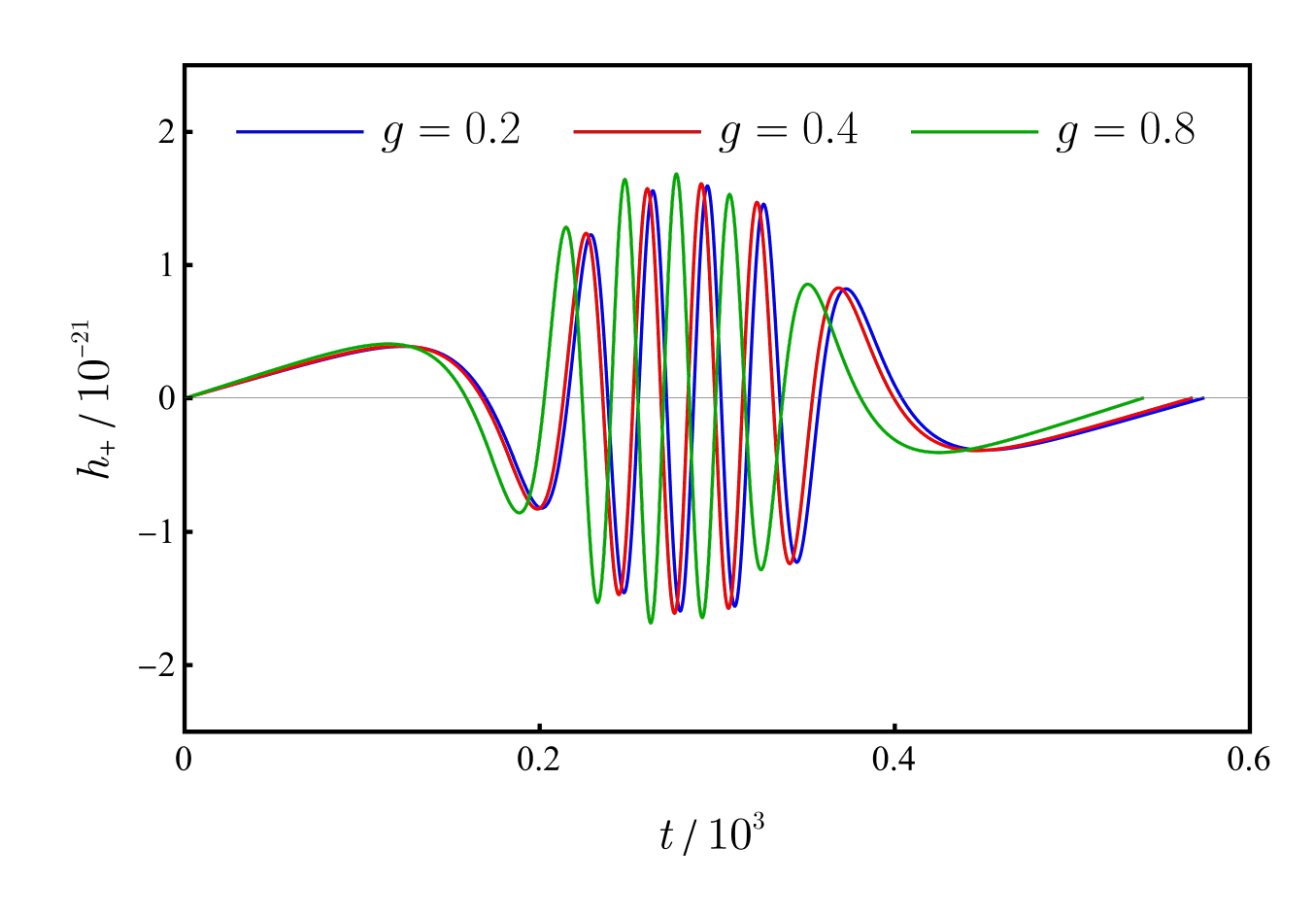}
    \caption{Plus polarization of the gravitational waveform $h_{+}$ in periodic
orbits characterized by $(z, w, v) = (1, 2, 0)$.}
    \label{fig:comphplus}
\end{figure}

\begin{figure}[ht!]
    \centering
    \includegraphics[width=55mm]{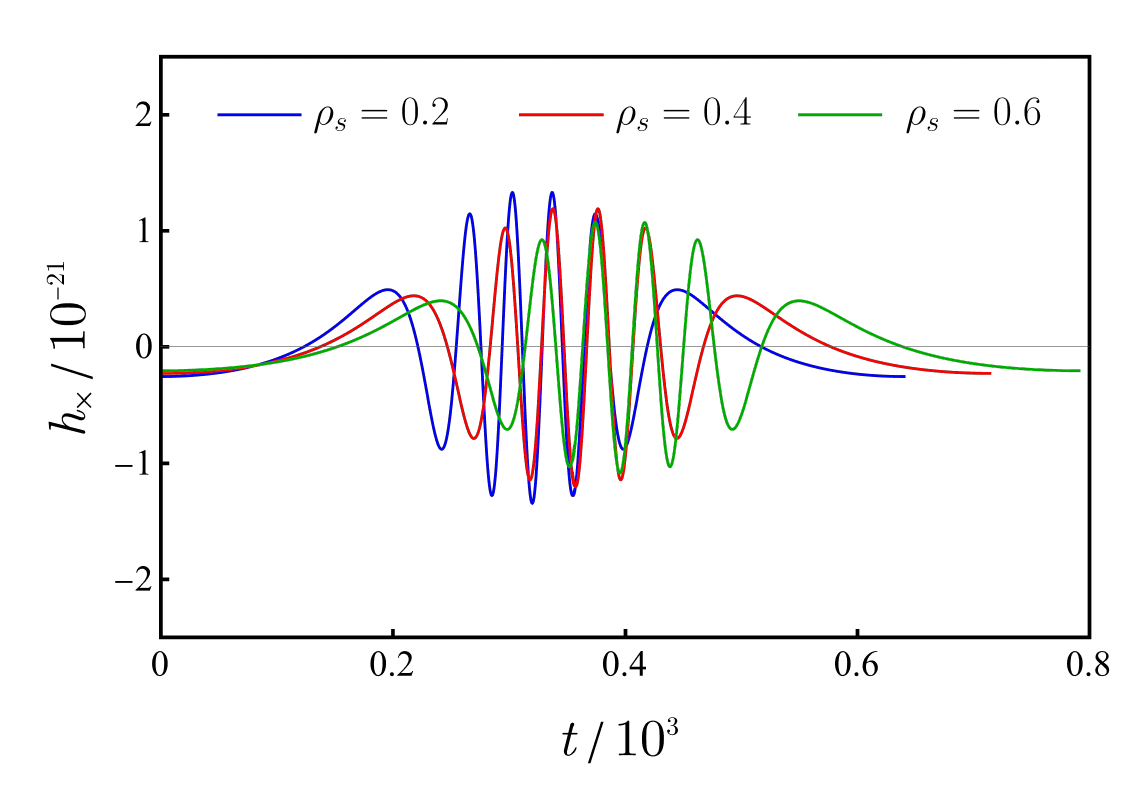}
    \includegraphics[width=55mm]{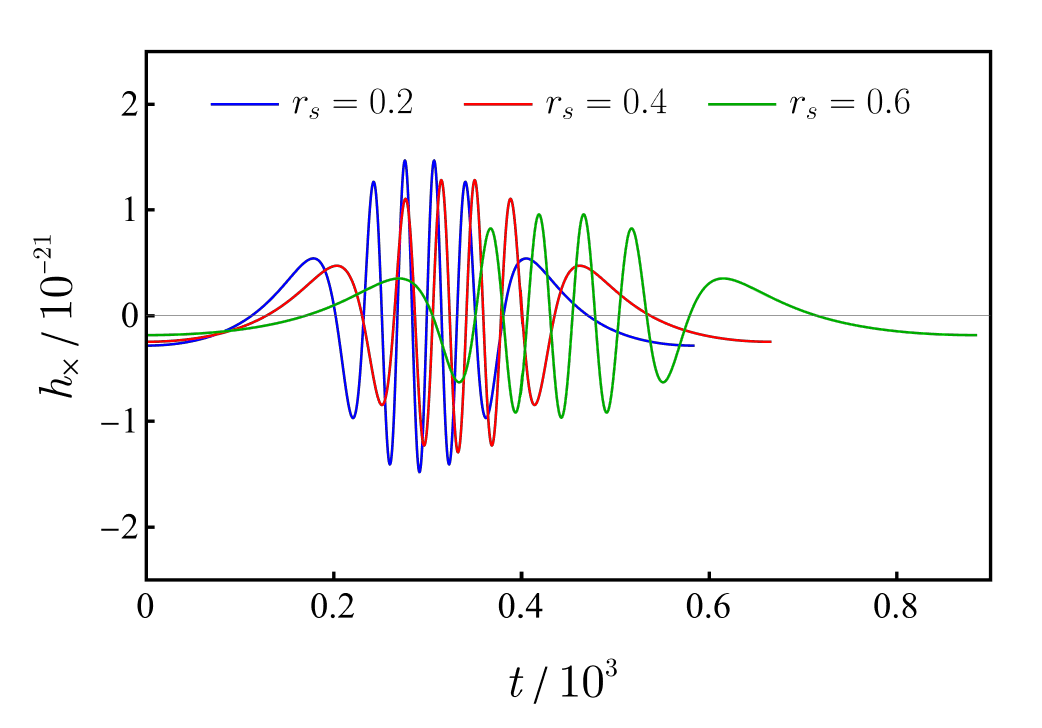}
    \includegraphics[width=55mm]{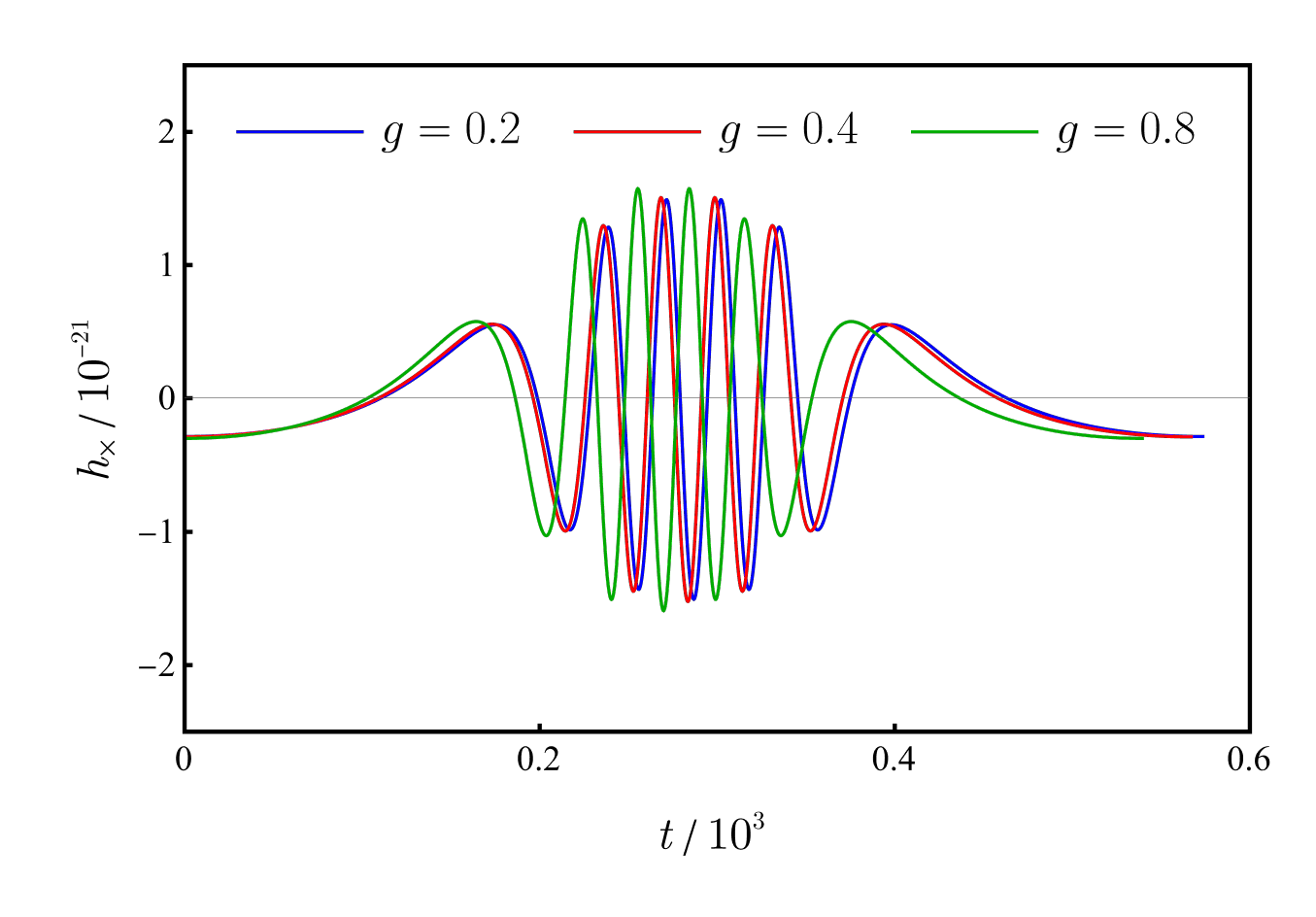}
    \caption{Cross polarization of the gravitational waveform $h_{\times}$ for periodic orbits characterized by $(z, w, v) = (1, 2, 0)$.}
    \label{fig:comphcross}
\end{figure}
This interplay highlights the sensitivity of both orbital dynamics and gravitational wave signals to the combined structure of the spacetime, providing a potential observational avenue to probe the distribution and properties of dark matter in the vicinity of compact objects, as well as the black hole's characteristics.



\subsection{Comparison with the LISA sensitivity}
\label{sec:LISA}
To compare the gravitational wave signals generated by periodic orbits with the sensitivity of LISA \cite{amaro2017laser}, we first transform the time domain polarization components, $h_{+}(t)$ and $h_{\times}(t)$, into the frequency domain using a discrete Fourier transform. The resulting Fourier components, $\tilde{h}_{+}(f)$ and $\tilde{h}_{\times}(f)$, are then used to calculate the characteristic strain. Following Refs.~\cite{michele2007gravitational,kumar2026probing}, the characteristic
strain is defined as
\begin{equation}
h_c(f) =2f \sqrt{
|\tilde{h}_{+}(f)|^2+|\tilde{h}_{\times}(f)|^2}.
\label{eq:characteristic-strain}
\end{equation}

\begin{figure}[ht!]
    \centering
    \includegraphics[height=70mm]{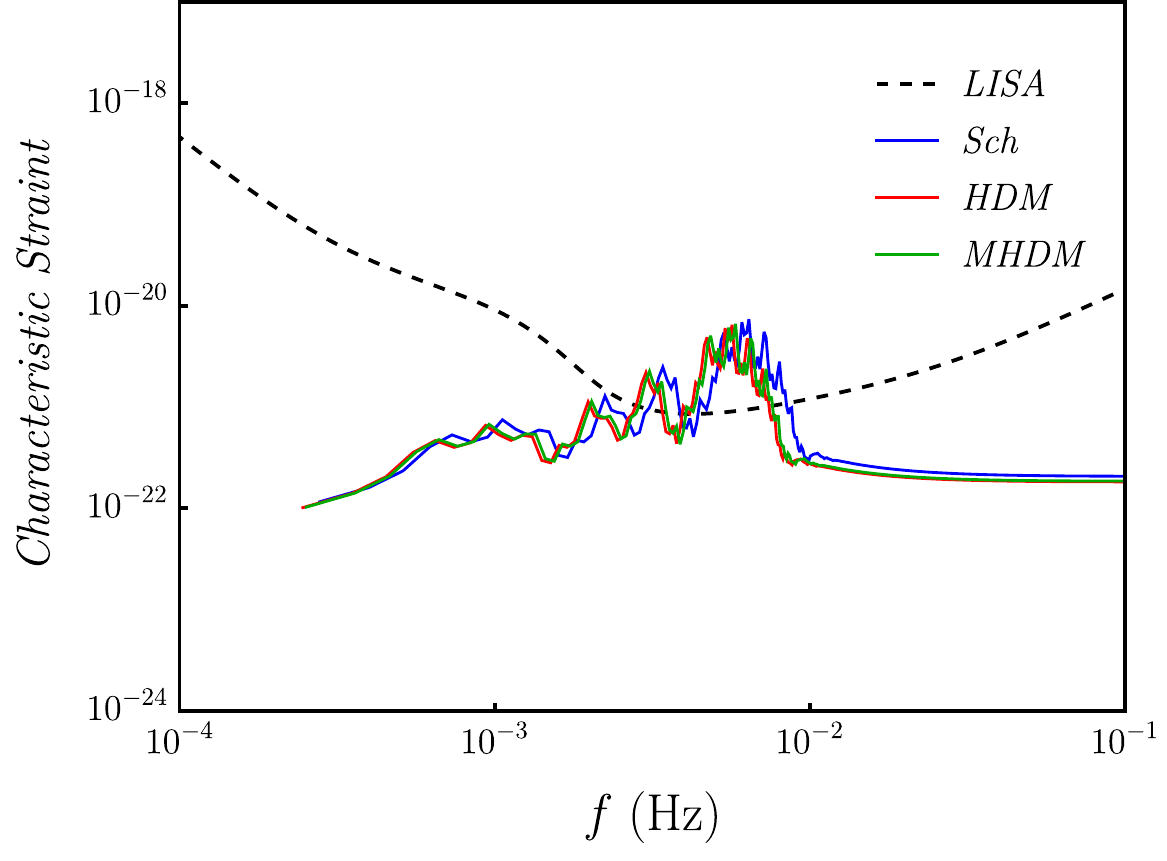}
    \caption{The characteristic straint $h_{c}$ for the periodic orbits $(z, w, v) = (3,1,2)$ with respect to frequency for \textit{Sch}, \textit{HDM} and \textit{MHDM} black holes. All remaining parameters are fixed at $r_s=0.4$, $\rho_s=0.4$, and $g=0.6$. The dashed line represents the sensitivity curve of space based GW detectors LISA.}
    \label{fig:LISA1}
\end{figure}

For the numerical analysis, we consider a central black hole of mass $M=10^{6}M_{\odot}$, a compact object with mass ratio $\mu/M=10^{-5}$, and a luminosity distance of $D_{L}=2\,\mathrm{Mpc}$. In Fig.~\ref{fig:LISA1}, the results of the characteristic strain spectra are compared to the LISA characteristic sensitivity. It is worth to mention that the characteristic strain is sensitive to all spacetime parameters; however, for these specific parameters, all top amplitudes of gravitational waves in \textit{Sch}, \textit{HDM}, and \textit{MHDM} black hole spacetimes are in the sensitive range of the LISA sensitivity curve, around $10^{-3}$ to $10^{-2}$. The presence of the \textit{Hernquist} dark matter causes a shift in the peak of the gravitational wave to lower frequencies. On the other side, the influence of the magnetic pole is so subtle that it is not easily seen. Still, in an exact observation, it elevates the effect of the dark matter and brings the gravitational wave closer back to the \textit{Sch}'s case.

\begin{figure}[ht!]
    \centering
    \includegraphics[height=60mm]{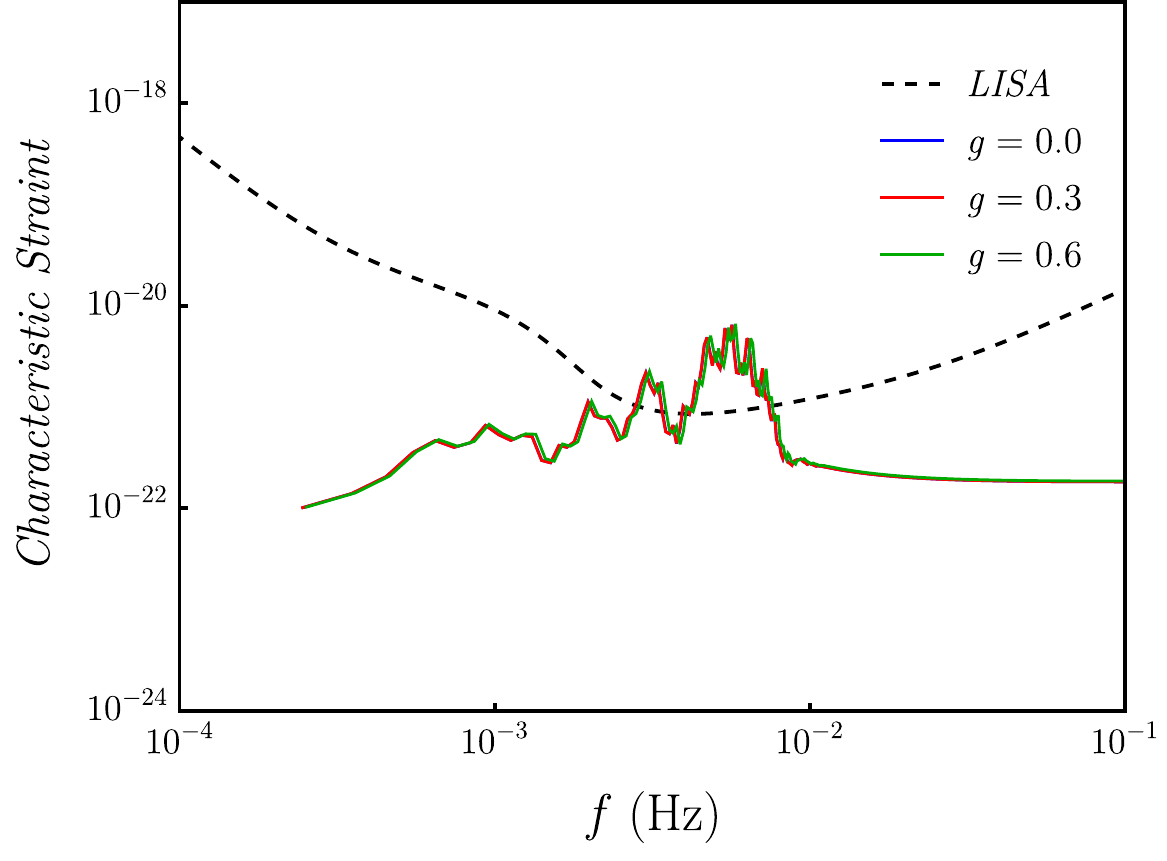}
    \includegraphics[height=60mm]{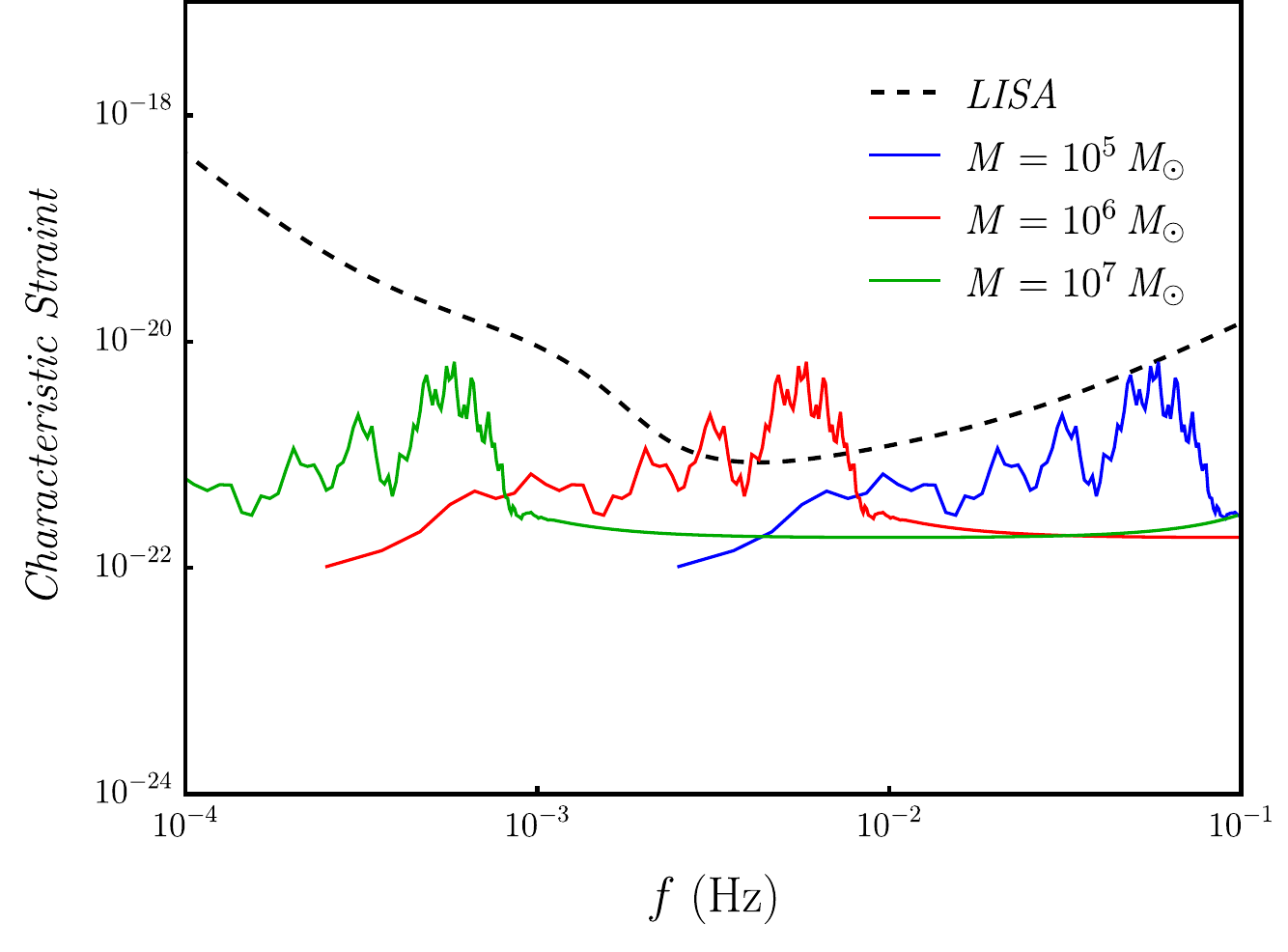}
    \caption{The LISA sensitivity curve and the characteristic straints for $r_s=0.4$, $\rho_s=0.4$ and small orbiting mass $m=10M_\odot$ and periodic orbit $(3,1,2)$. In the left panel, the effect of the magnetic charge $g$ is illustrated for $M=10^6M_\odot$, and in the right panel, the black hole mass is varied while $g=0.6$.}
    \label{fig:LISA2}
\end{figure}

In Fig.~\ref{fig:LISA2}, the impacts of black hole mass and magnetic monopole charge are explored. In the right panel, increasing the black hole mass keeps the overall spectral profile, including the relative heights and structure of the dominant peaks; howerver, it shifts the entire spectrum toward lower frequencies, consistent with the approximate inverse scaling $(f\propto M^{-1})$. The influence of the magnetic monopole charge $g$ displayed in the left panel does not lead to a considerable modification of the overall strain; nevertheless, a very small shift toward slightly higher frequencies can be observed as parameter $g$ increases.

\section{Conclusion}

In this work, we investigated the motion of massive test particles and the associated gravitational wave emission around a magnetically charged black hole immersed in a \textit{Hernquist} dark matter halo. Our analysis began with the effective potential and the corresponding conditions for bound motion, with particular attention to the marginally bound orbit and the innermost stable circular orbit. We showed that the dark matter halo parameters, namely the density $\rho_s$ and the scale radius $r_s$, enlarged the domain of bound motion and generally shifted the characteristic radii and angular momenta toward larger values. In contrast, the magnetic charge $g$ acted in the opposite direction, partially compensating for the effect produced by the halo. These features were also reflected in the allowed region of the $(E, L)$ plane, where the magnetically charged Hernquist configuration interpolated between the Schwarzschild and pure Hernquist cases.

We then examined the periodic sector of the timelike geodesics by means of the rational number $q$, which related the azimuthal and radial frequencies. For fixed average angular momentum, the periodic orbit energies varied systematically with the halo and magnetic parameters, while the corresponding trajectories displayed the expected progression from simpler closed patterns to rich zoom--whirl structures as the integers $(z,w,v)$ increased. We also studied small deviations from the exact periodic energies and showed that they generated precessing trajectories, indicating that even tiny perturbations were sufficient to produce a secular shift in the orbital orientation.

Moreover, within the adiabatic and numerical kludge framework, we computed the gravitational wave polarizations generated by these periodic motions in the extreme mass--ratio regime. The waveforms exhibited a clear correlation with the orbital motion: the signal was amplified during close passages to the central object and weakened when the particle moved farther away. A comparative analysis between the Schwarzschild, \textit{Hernquist}, and magnetically charged \textit{Hernquist} backgrounds showed that the presence of the dark matter halo tended to enlarge the orbit, reduce the waveform amplitude, and increase the oscillation period, whereas the magnetic monopole charge partially suppressed the influence of the dark matter and drove the signal closer to the Schwarzschild behavior. Finally, We have analyzed periodic orbits and the corresponding gravitational wave polarizations, $h_{+}$ and $h_{\times}$, and found a clear dependence on the dark matter halo parameters $\rho_{\mathrm{s}}$, $r_{\mathrm{s}}$, and the magnetic monopole charge $g$. In particular, the parameters $\rho_{\mathrm{s}}$ and $r_{\mathrm{s}}$ produce qualitatively similar effects: increasing either parameter leads to an elongation of the orbital period, which manifests as a longer periodicity in the gravitational wave signal, together with a suppression of its amplitude. In contrast, the magnetic charge $g$ induces a comparatively weaker effect with an opposite tendency. An increase in $g$ results in a slight reduction of the orbital scale, leading to shorter periodicity and a modest enhancement of the waveform amplitude. These findings are consistent with the trends identified in other dynamical and orbital characteristics discussed throughout this work. At the end, the characteristic strain has been calculated to compare the Schwarzschild, \textit{Hernquist}, and magnetically charged \textit{Hernquist} backgrounds with the LISA sensitivity. The presence of the dark matter results in almost no change in the amplitude but a small shift in the frequency range of the spectrum. For the chosen parameters, the characteristic strain has been in the LISA sensitivity range. The impact of magnetic monopole charge has led to only a weak shift toward higher frequencies with negligible variation in the strain amplitude.

\section*{Acknowledgments}
\hspace{0.5cm} 
N. H is supported by the Conselho Nacional de Desenvolvimento Científico e Tecnológico (CNPq), grant No. 152891/2025-0. N. H. also acknowledges the networking support provided by COST Action CA22113 -- Fundamental challenges in theoretical physics (Theory and Challenges), CA21106 -- COSMIC WISPers in the Dark Universe: Theory, astrophysics and experiments (CosmicWISPers), CA21136 -- Addressing observational tensions in cosmology with systematics and fundamental physics (CosmoVerse), and CA23130 -- Bridging high and low energies in search of quantum gravity (BridgeQG). A.A.A.F. is supported by Conselho Nacional de Desenvolvimento Cient\'{\i}fico e Tecnol\'{o}gico (CNPq) with the project number 150223/2025-0. I. P. L. was partially supported by the National Council for Scientific and Technological Development - CNPq, grant 312547/2023-4 and to acknowledge networking support by the COST Action BridgeQG (CA23130), the COST Action RQI (CA23115) and the COST Action FuSe (CA24101) supported by COST (European Cooperation in Science and Technology).


	\bibliography{main}
	\bibliographystyle{unsrt}
	
\end{document}